\documentclass[aps,10pt,prd,notitlepage]{revtex4-1}
\usepackage[utf8]{inputenc}
\usepackage{amsmath,amssymb,bm,epsfig}
\usepackage{color}
\usepackage{hyperref} 
\usepackage{natbib}
\usepackage{ulem}
\usepackage{graphicx}
\usepackage{xcolor}
\usepackage{pifont}
\usepackage{amsfonts}
\usepackage{dsfont,mathrsfs}
\usepackage{cancel}
\usepackage{bm}
\usepackage{bigints}
\newcommand{\nn}{\nonumber}

\newcommand{\MB}[1]{\left|#1\right|}
\newcommand{\FB}[1]{\left(#1\right)}
\newcommand{\SB}[1]{\left\{#1\right\}}
\newcommand{\TB}[1]{\left[#1\right]}

\newcommand{\scrL}{\mathscr{L}}

\newcommand{\sign}[1]{\text{sign}\left(#1\right)}
\newcommand{\munu}{{\mu\nu}}
\newcommand{\numu}{{\nu\mu}}
\newcommand{\alphabeta}{{\alpha\beta}}
\newcommand{\IM}{\text{Im}}
\newcommand{\RE}{\text{Re}}
\newcommand{\Tr}{\text{Tr}}
\newcommand{\rhopipi}{{\rho\pi\pi}}
\newcommand{\rhoNN}{{\rho NN}}

\newcommand{\tN}{\text{N}}
\newcommand{\tp}{\text{p}}

\newcommand{\tpn}{\text{p,n}}

\newcommand{\etatilde}{\tilde{\eta}}
\newcommand{\Pibar}{\overline{\Pi}}
\newcommand{\Pibarbar}{\overline{\Pibar}}
\newcommand{\kpll}{k_\parallel}
\newcommand{\qpll}{q_\parallel}
\newcommand{\kper}{k_\perp}
\newcommand{\qper}{q_\perp}
\newcommand{\gpll}{g_\parallel}
\newcommand{\gper}{g_\perp}

\newcommand{\utilde}{\tilde{u}}
\newcommand{\btilde}{\tilde{b}}

\newcommand{\qdotu}{q\cdot u}
\newcommand{\qdotb}{q\cdot b}
\newcommand{\bdotut}{b\cdot\utilde}

\newcommand{\Pimumu}{\Pi^\mu_{~\mu}}
\newcommand{\omegatilde}{\tilde{\omega}}
\newcommand{\Omegatilde}{\tilde{\Omega}}

\begin{document}
\title{General structure of the neutral $\rho$ meson self-energy and its spectral properties in a hot and dense magnetized medium}

\author{Snigdha Ghosh$^{a,b}$}
\email{snigdha.physics@gmail.com, snigdha.ghosh@saha.ac.in}
\author{Arghya Mukherjee$^{b,d}$}
\email{arghya.mukherjee@saha.ac.in}
\author{ Pradip Roy$^{b,d}$}
\email{pradipk.roy@saha.ac.in}
\author{Sourav Sarkar$^{c,d}$}
\email{sourav@vecc.gov.in}
\affiliation{$^a$Indian Institute of Technology Gandhinagar, Palaj, Gandhinagar 382355, Gujarat, India}
\affiliation{$^b$Saha Institute of Nuclear Physics, 1/AF Bidhannagar, Kolkata - 700064, India}
\affiliation{$^c$Variable Energy Cyclotron Centre, 1/AF Bidhannagar, Kolkata 700 064, India}
\affiliation{$^d$Homi Bhabha National Institute, Training School Complex, Anushaktinagar, Mumbai - 400085, India}

\begin{abstract}
The one loop self energy of the neutral $\rho$ meson 
is obtained for the effective $\rho\pi\pi$ and $\rho NN$ interaction  at finite temperature 
and density in the presence of  a constant background magnetic field of arbitrary strength. 
In our approach, the  eB-dependent vacuum part of the self energy is extracted by 
means of   dimensional regularization  where
the ultraviolet divergences corresponding to the pure vacuum self energy manifest as the pole singularities of
gamma as well as Hurwitz zeta functions. This improved regularization  procedure  
consistently reproduces the expected results in the vanishing magnetic field limit and can be used quite generally  in other  self energy calculations dealing with
arbitrary magnetic field strength.  In presence of the external magnetic field, the general 
Lorentz structure for the in-medium vector boson self energy 
is derived  which can also be implemented in case of the gauge bosons  such as  
photons and gluons. It has been shown that with vanishing perpendicular momentum of the external particle, essentially two  form factors are 
 sufficient to describe the self energy completely. Consequently, two distinct modes are observed in  the study of the effective mass,  dispersion relations and the 
spectral function  of $\rho^0$ where one of the modes possesses two fold degeneracy. 
  For large  baryonic chemical potential, it is observed that the  critical  magnetic field required
to block the   $\rho^0\rightarrow\pi^+\pi^-$ decay channel   increases significantly  
with temperature. However, in case of smaller values reaching down to vanishing chemical potential, 
the  critical field follows the opposite trend. 
\end{abstract}

\maketitle
%
\section{Introduction}\label{sec.intro}
 In  non-central heavy-ion collisions (HIC) at the LHC, the relative motion of the ions themselves can generate strong decaying magnetic pulse  of the order 
 $eB\sim$15$m_\pi^2$ ($B\sim 5\times10^{15}$Tesla) \cite{Skokov:2009qp}. While some of the studies support rapid decrease in the magnitude \cite{PhysRevLett.110.192301,MCLERRAN2014184}, an adiabatic decay is 
 expected~\cite{PhysRevC.88.024910,PhysRevC.87.024912,AdvHighEnergyPhys2013}
 due to the high conductivity of the produced medium. In spite of the ambiguities, the intensity of the produced magnetic field being much larger than the typical QCD scale,  the possibility of  magnetic modifications of different properties of the produced extreme state of matter  can not be refuted completely. In general, 
 high intensity magnetic fields can play a significant role in many astrophysical and cosmological phenomena \cite{VACHASPATI1991258,GRASSO2001163,S0218271804004530,0264-9381-35-8-084003}. Moreover, the magnetic influence on the properties of magnetars adds to the motivation of studying high density matter in presence of extreme magnetic fields \cite{annurev-astro, Ferrer:2005vd,Ferrer:2006vw,Ferrer:2007iw,Fukushima:2007fc,Feng:2009vt,Fayazbakhsh:2010gc,Fayazbakhsh:2010bh}.

 The study of  $\rho$ meson properties like the  effective mass and dispersion relations are important in the context of magnetic field induced vacuum 
superconductivity~\cite{Chernodub:2010qx,Chernodub:2012tf,VAFA1984173,PhysRevD.86.107703,LI2013141,PhysRevD.87.094502,PhysRevD.91.014017,PhysRevD.93.125027, PhysRevD.93.125027}.  Using Nambu Jona-Lasinio (NJL) model in presence of magnetic background, Liu. et.al.   have shown  
that the charged rho condensation in vacuum occurs at critical magnetic field $eB_c\sim0.2$GeV$^2$ \cite{PhysRevD.91.014017}. Generalization of  the study to finite temperature and density  shows that the condensation  survives even in presence of finite temperature and density \cite{1674-1137-40-2-023102}. At vanishing chemical potential, the corresponding critical magnetic field is observed to lie in the 
range 0.2 -0.6 GeV$^2$ for temperatures in between 0.2-0.5 GeV. However, the neutral $\rho$ meson in vacuum, having no trivial Landau shifts in the energy eigenvalue,  shows a slow decrease in the 
effective mass \cite{Ghosh:2016evc} in  weak magnetic field region. Thus, if neutral rho condensation is possible,  extremely large magnetic field values will be required to observe the condensation. It should be mentioned here that it has been shown using NJL model that the  effective mass of $\rho^0$ meson in fact increases at higher values of magnetic fields showing no possibility of condensation \cite{PhysRevD.91.014017}. In this scenario,  $\rho^0\rightarrow\pi^+\pi^-$ decay  may serve as an important probe to observe the influence of the magnetic field. 
As argued in Ref. \cite{Chernodub:2010qx}, even if point like $\rho^0$ meson is considered without any influence by magnetic field, there exists a critical value of the external magnetic field for which the $\rho^0$ to $\pi^+\pi^-$ decay stops due to  the trivial enhancement of the charged pion mass. Later the magnetic modification
arising from the loop corrections are taken into account at weak \cite{ aritra_weak,Ghosh:2016evc} as well as at strong field limits \cite{kawaguchi_strong} at zero temperature. An immediate  generalization of the previous works will be to incorporate the   medium effects of the $\rho^0$ meson which  may  reflect in the modification of the  decay rate and  the required  critical magnetic field.  
It should be noted here that apart from being important in the study of dense hadronic matter at extreme conditions usually expected to be present within compact stars, the incorporation of the medium effects is also essential for the  proper estimation  of   pion production in non-central heavy ion collisions.

 In this work  we  focus  on the  temperature and density modifications of neutral $\rho$ meson properties in presence of a static homogeneous magnetic background.
 The one loop self energy of $\rho$ meson is calculated for the effective $\rho\pi\pi$ and $\rhoNN$ interaction with magnetically modified pion and nucleon propagators corresponding to general field strength. After decomposing  the self energy 
 in terms of the form factors,  the decay width for  $\rho^0\rightarrow\pi^+\pi^-$ channel is obtained.
 It should be mentioned here that the spectral properties of rho meson in presence of finite temperature and magnetic field have been studied in our earlier work \cite{Ghosh:2017rjo}.
     However, unlike the previous case, dimensional regularization technique is used here to extract the ultraviolet divergence as pole singularities of gamma and Hurwitz zeta functions\cite{doi:10.1142/2065}. Also, instead of considering only the spin averaged thermal self energy contribution, the general Lorentz structure has been addressed in detail. 
     Apart from the technical differences,  the density dependence arising from the 
 charged nucleon loop serves as the most important extension of the previous study. Its importance can be understood as follows. It is well known that the general expression of decay width is related to  the  imaginary part of the self energy. Now, as far as the $\rho^0\rightarrow\pi^+\pi^-$ decay is  concerned, the invariant mass regime of interest does not allow the nucleon loop to directly contribute to the imaginary  part as the   unitary cut threshold of $NN$ loop  begins at much higher value. However, it should be noted that in the rest frame of the decaying particle, the  decay width  depends on its  effective mass.  The contribution from the nucleon loop incorporates significant modification in  the  effective mass of  $\rho^0$   which in turn  influences the decay. As we shall see, the critical field 
 required to stabilize the neutral $\rho$ against the $\pi^+\pi^-$ decay has a non-trivial dependence on the baryonic chemical potential.

 The article is organized as follows. In Sec.~\ref{sec.self.vac} the vacuum self energy of $\rho$ is discussed followed by evaluation of the in-medium $\rho$ self-energy
at zero magnetic field in Sec.~\ref{sec.self.med}. Next in Sec.~\ref{sec.self.med.eB}, the in-medium self energy at 
non-zero external magnetic field is presented. Sec.~\ref{sec.lorentz} is devoted to the discussion of the general Lorentz structure of the in-medium self energy function in presence of a constant background  magnetic field. After addressing the Lorentz structure of the interacting  $\rho$ propagator in Sec.\ref{sec.propagator}, the analytic structure of the self energy is discussed in Sec. \ref{sec.analytic}.
 Sec.~\ref{sec.numerical} contains the numerical results. Finally we summarize and conclude in Sec.~\ref{sec.summary}.
Some of the relevant calculational details are provided in the Appendix.


\section{$\rho^0$ Self Energy in the Vacuum} \label{sec.self.vac}

The effective Lagrangian for $\rho\pi\pi$ and $\rho NN$ interaction is~\cite{Krehl:1999km}
\begin{eqnarray}
\scrL_\text{int} = -g_\rhopipi \partial_\mu\vec{\rho}_\nu\cdot\FB{\partial^\mu\vec{\pi}\times\partial^\nu\vec{\pi}}
-g_\rhoNN \bar{\Psi}\TB{\gamma^\mu - \frac{\kappa_\rho}{2m_N}\sigma^\munu\partial_\nu}\vec{\tau}\cdot\vec{\rho}_\mu\Psi
\label{eq.Lagrangian}
\end{eqnarray}
where, $\Psi=\TB{\begin{array}{c} p \\ n \end{array}}$ is the nucleon isospin doublet, $\sigma^\munu = \frac{i}{2}\TB{\gamma^\mu,\gamma^\nu}$ and the components of $\vec{\tau}$  correspond to the  Pauli isospin matrices. It is understood that, the derivative within the square bracket in the above equation acts only 
on the $\rho$ field. The value of the coupling constants are given by $g_\rhopipi =20.72$ GeV$^{-2}$, $g_\rhoNN = 3.25$ and $\kappa_\rho = 6.1$ with $m_N = 939$ MeV as the mass of the nucleons. The metric tensor in this work is taken as $g^{\mu\nu}=diag(1,-1,-1,-1)$.
\begin{figure}[h]
	\begin{center}
		\includegraphics[angle=0, scale=0.35]{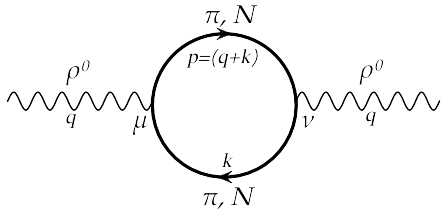}	
	\end{center}
\caption{Feynman diagram for the one-loop self energy of neutral $\rho$ meson.}
	\label{fig.feynman}
\end{figure}
Using Eq.~\eqref{eq.Lagrangian}, the one-loop vacuum self energy of $\rho^0$ is obtained as 
\begin{eqnarray}
\Pi^\munu_\text{pure-vac} = \FB{\Pi^\munu_\pi}_\text{pure-vac} + \FB{\Pi^\munu_\text{N}}_\text{pure-vac} 
\end{eqnarray} 
where, $\FB{\Pi^\munu_\pi}_\text{pure-vac}$ and $\FB{\Pi^\munu_\text{N}}_\text{pure-vac}$ are respectively the contributions from the $\pi\pi$-loop and $NN$-loop which are given by (applying Feynman rules to Fig.~\ref{fig.feynman})
\begin{eqnarray}
\FB{\Pi^\munu_\pi}_\text{pure-vac}(q) &=& i\int\frac{d^4k}{(2\pi)^4}\mathcal{N}_\pi^\munu(q,k)\Delta_F(k,m_\pi)\Delta_F(p=q+k,m_\pi) 
\label{eq.Pi.vac.pi.1}\\
\FB{\Pi^\munu_\text{N}}_\text{pure-vac}(q) &=& -i\int\frac{d^4k}{(2\pi)^4}\Tr\TB{\Gamma^\nu(q)S_\text{p}(p=q+k,m_N)\Gamma^\mu(-q)S_\text{p}(k,m_N)\frac{}{} \right. \nn \\ 
&& \left.+ \Gamma^\nu(q)S_\text{n}(p=q+k,m_N)\Gamma^\mu(-q)S_\text{n}(k,m_N)\frac{}{}}
\label{eq.Pi.vac.N.1}
\end{eqnarray}
where, 
\begin{eqnarray}
\Delta_F(k,m_\pi) = \frac{-1}{k^2-m_\pi^2+i\epsilon}
\end{eqnarray}
is the vacuum Feynman propagator for the charged pion. $S_\text{p}$ and $S_\text{n}$ are respectively the vacuum Feynman propagators for proton and neutron and are given by 
\begin{eqnarray}
S_\text{p}(k,m_N) = S_\text{n}(k,m_N) = (\cancel{k}+m_N)\Delta_F(k,m_N).
\end{eqnarray}
The second rank tensor $\mathcal{N}_\pi^\munu(q,k)$ and the vector $\Gamma^\mu(q)$ in  Eqs.~\eqref{eq.Pi.vac.pi.1} and \eqref{eq.Pi.vac.N.1} contain the factors coming from the  interaction vertices: 
\begin{eqnarray}
\mathcal{N}_\pi^\munu(q,k) &=& g_\rhopipi^2 \TB{q^4k^\mu k^\nu + (q\cdot k)^2q^\mu q^\nu - q^2(q\cdot k)(q^\mu k^\nu+q^\nu k^\mu) \frac{}{}} \label{eq.N.pi.1}\\
\Gamma^\mu(q) &=& g_\rhoNN\TB{\gamma^\mu - i\frac{\kappa_\rho}{2m_N}\sigma^\munu q_\nu}~.
\end{eqnarray}
The evaluations of the momentum integrals in Eqs.~\eqref{eq.Pi.vac.pi.1} and \eqref{eq.Pi.vac.N.1} are briefly sketched in Appendix~\ref{app.vac.self} and the final results can be read off from Eqs.~\eqref{eq.Pi.vac.pi.5} and \eqref{eq.Pi.vac.N.5}
\begin{eqnarray}
\FB{\Pi^\munu_\pi}_\text{pure-vac}(q) &=& (q^2g^\munu-q^\mu q^\nu)\FB{\frac{-g_\rhopipi^2q^2}{32\pi^2}}\int_{0}^{1}dx\Delta_\pi
\TB{\frac{1}{\varepsilon}-\gamma_\text{E}+1-\ln\FB{\frac{\Delta_\pi}{4\pi\Lambda_\pi}}}\Bigg|_{\varepsilon\rightarrow 0} 
\label{eq.Pi.vac.pi.6} \\
\FB{\Pi^\munu_\text{N}}_\text{pure-vac}(q) &=& (q^2g^\munu-q^\mu q^\nu)\FB{\frac{g_\rhoNN^2}{2\pi^2}}\int_{0}^{1}dx
\TB{ \SB{2x(1-x)+\kappa_\rho+\frac{\kappa_\rho^2}{2} -\frac{\kappa_\rho^2}{4m_N^2}\Delta_\text{N} } \times  \right. \nn \\ 
	&& \left. \SB{\frac{1}{\varepsilon}-\gamma_\text{E}-\ln\FB{\frac{\Delta_\tN}{4\pi\Lambda_\tN}}}
-\frac{\kappa_\rho^2}{4m_N^2}\Delta_\text{N}} \Bigg|_{\varepsilon\rightarrow 0}
\label{eq.Pi.vac.N.6}
\end{eqnarray}
where $\Delta_\pi$ and $\Delta_\tN$ are defined in Eqs.~\eqref{eq.Delta.pi} and \eqref{eq.Delta.N}. As can be seen from the above equations, the vacuum self energy is divergent and scale dependent which renormalizes the bare $\rho^0$ mass to its physical mass after adding proper vacuum counter terms in the Lagrangian. The particular Lorentz structure in the above equations renders the self energy transverse to the $\rho^0$ momentum i.e. $q_\mu\Pi^\munu_\text{pure-vac} = 0$. 


\section{$\rho^0$ Self Energy in the Medium} \label{sec.self.med}

In order to calculate the $\rho^0$ self energy at finite temperature and density, we employ the real time formalism (RTF) of finite temperature field theory where all the two point correlation functions such as the propagator and the self energy become $2\times2$ matrices in the thermal space~\cite{Bellac:2011kqa,Mallik:2016anp}. However, they can be put in a diagonal form where the diagonal elements   can be obtained from any one component (say the $11$-component) of the said $2\times2$ matrix. The $11$-components of real time thermal pion and nucleon propagators are
\begin{eqnarray}
D^{11}(k) &=& \Delta_F(k,m_\pi) + \eta(k\cdot u)\TB{\Delta_F(k,m_\pi)-\Delta^*_F(k,m_\pi)} \label{eq.D11}\\
S^{11}_\tpn(k) &=& S_\tpn(k,m_N) - \etatilde(k\cdot u)\TB{S_\tpn(k,m_N)-\gamma^0S^\dagger_\tpn(k,m_N)\gamma^0}
\label{eq.S11}
\end{eqnarray}
where $\eta(x)=\Theta(x)f(x)+\Theta(-x)f(-x)$ and $\etatilde(x)=\Theta(x)f^+(x)+\Theta(-x)f^-(-x)$ in which $f(x)$ and $f^\pm(x)$ are respectively the Bose-Einstein and Fermi-Dirac distribution functions corresponding to pions and nucleons: 
\begin{eqnarray}
f(x) = \TB{e^{x/T}-1}^{-1} ~~,~~ f^\pm(x) = \TB{e^{(x\mp\mu_B)/T}+1}^{-1}~.
\end{eqnarray} 
Here, $\Theta(x)$ is the unit step function, $u^\mu$ is the medium four-velocity; $T$ and $\mu_B$ are respectively the temperature and baryon chemical potential of the medium. In the local rest frame (LRF) of the medium,  $u_\text{LRF}^\mu\equiv(1,\vec{0})$.

For the evaluation of the $11$-component of the thermal self energy matrix, the vacuum pion and nucleon propagators in Eqs.~\eqref{eq.Pi.vac.pi.1} and \eqref{eq.Pi.vac.N.1} are replaced by the respective $11$-components of the thermal propagators given in Eqs.~\eqref{eq.D11} and \eqref{eq.S11} as~\cite{Mallik:2016anp}
\begin{eqnarray}
\FB{\Pi^\munu_\pi}_{11}(q) &=& i\int\frac{d^4k}{(2\pi)^4}\mathcal{N}_\pi^\munu(q,k)D^{11}(k,m_\pi)D^{11}(p=q+k,m_\pi) 
\label{eq.Pi.T.pi.1}\\
\FB{\Pi^\munu_\text{N}}_{11}(q) &=& -i\int\frac{d^4k}{(2\pi)^4}\Tr\TB{\Gamma^\nu(q)S_\text{p}^{11}(k,m_N)\Gamma^\mu(-q)S_\text{p}^{11}(p=q+k,m_N)\frac{}{} \right. \nn \\ 
	&& \left.+ \Gamma^\nu(q)S_\text{n}^{11}(k,m_N)\Gamma^\mu(-q)S_\text{n}^{11}(p=q+k,m_N)\frac{}{}}~.
\label{eq.Pi.T.N.1}
\end{eqnarray}
The analytic thermal self energy function of $\rho^0$ denoted by a bar 
$\RE\Pibar^\munu(q^0,\vec{q}) = \RE\Pibar^\munu_\pi(q^0,\vec{q}) + \RE\Pibar^\munu_\tN(q^0,\vec{q})$ is related to the above quantities by the relations~\cite{Mallik:2016anp} 
\begin{eqnarray}
\RE\Pibar^\munu_{\pi,N}(q^0,\vec{q}) &=& \FB{\RE\Pi^\munu_{\pi,N}}_{11}(q^0,\vec{q}) \label{eq.analytic.relation.re}\\
\IM\Pibar^\munu_{\pi,N}(q^0,\vec{q}) &=& \sign{q^0}\tanh\FB{\frac{q^0}{2T}}\FB{\IM\Pi^\munu_{\pi,N}}_{11}(q^0,\vec{q})
\label{eq.analytic.relation.im}
\end{eqnarray}
where, $\sign{x}=\Theta(x)-\Theta(-x)$. Rewriting Eqs.~\eqref{eq.D11} and \eqref{eq.S11} as
\begin{eqnarray}
D^{11}(k) &=& \Delta_F(k,m_\pi) + 2\pi i\eta(k\cdot u)\delta\FB{k^2-m_\pi^2} \label{eq.D11.2}\\
S^{11}_\tpn(k) &=& \FB{\cancel{k}+m_N}\TB{\Delta_F(k,m_N) - 2\pi i\etatilde(k\cdot u)\delta\FB{k^2-m_N^2}}
\label{eq.S11.2}
\end{eqnarray}
and substituting into Eqs.~\eqref{eq.Pi.T.pi.1} and \eqref{eq.Pi.T.N.1} and performing the $dk^0$ integration (using the Dirac delta functions) followed by using Eqs.~\eqref{eq.analytic.relation.re} and \eqref{eq.analytic.relation.im} we get the real parts as, 
\begin{eqnarray}
\RE\Pibar^\munu_\pi(q^0,\vec{q}) &=& \RE\FB{\Pi^\munu_\pi}_\text{pure-vac}(q) + \bigintssss\frac{d^3k}{(2\pi)^3}\mathcal{P}
\TB{\frac{f(\omega_k)}{2\omega_k}\SB{\frac{\mathcal{N}_\pi^\munu(k^0=-\omega_k)}{(q^0-\omega_k)^2-(\omega_p)^2}
		+ \frac{\mathcal{N}_\pi^\munu(k^0=\omega_k)}{(q^0+\omega_k)^2-(\omega_p)^2}} \right. \nn \\
	&& \hspace{3.0cm} \left.
	+ \frac{f(\omega_p)}{2\omega_p}\SB{\frac{\mathcal{N}_\pi^\munu(k^0=-q^0-\omega_p)}{(q^0+\omega_p)^2-(\omega_k)^2}
		+ \frac{\mathcal{N}_\pi^\munu(k^0=-q^0+\omega_p)}{(q^0-\omega_p)^2-(\omega_k)^2}}} \label{eq.repi.pi.1}\\
\RE\Pibar^\munu_\tN(q^0,\vec{q}) &=& \RE\FB{\Pi^\munu_\tN}_\text{pure-vac}(q) - \bigintssss\frac{d^3k}{(2\pi)^3}\mathcal{P}
\TB{\frac{1}{2\Omega_k}\SB{\frac{f^-(\Omega_k)\mathcal{N}_\tN^\munu(k^0=-\Omega_k)}{(q^0-\Omega_k)^2-(\Omega_p)^2}
		+ \frac{f^+(\Omega_k)\mathcal{N}_\tN^\munu(k^0=\Omega_k)}{(q^0+\Omega_k)^2-(\Omega_p)^2}} \right. \nn \\
	&& \hspace{3.0cm} \left.
	+ \frac{1}{2\Omega_p}\SB{\frac{f^-(\Omega_p)\mathcal{N}_\tN^\munu(k^0=-q^0-\Omega_p)}{(q^0+\Omega_p)^2-(\Omega_k)^2}
		+ \frac{f^+(\Omega_p)\mathcal{N}_\tN^\munu(k^0=-q^0+\Omega_p)}{(q^0-\Omega_p)^2-(\Omega_k)^2}}} \label{eq.repi.N.1}
\end{eqnarray}
and the imaginary parts as
\begin{eqnarray}
\IM\Pibar^\munu_\pi(q^0,\vec{q}) &=& -\sign{q^0}\tanh\FB{\frac{\beta q^0}{2}}
\pi\bigintssss\frac{d^3k}{(2\pi)^3}\frac{1}{4\omega_k\omega_p}
\TB{ \SB{1+f(\omega_k)+f(\omega_p)+2f(\omega_k)f(\omega_p)\frac{}{}} \right. \nn \\
	&& \hspace{-1.5cm}\left. \SB{\frac{}{}\mathcal{N}_\pi^\munu(k^0=-\omega_k)\delta(q^0-\omega_k-\omega_p)
		+\mathcal{N}_\pi^\munu(k^0=\omega_k)\delta(q^0+\omega_k+\omega_p) } 
	+ \SB{\frac{}{}f(\omega_k)+f(\omega_p)+2f(\omega_k)f(\omega_p)} \right. \nn \\
	&& \left. \SB{\mathcal{N}_\pi^\munu(k^0=-\omega_k)\delta(q^0-\omega_k+\omega_p)
		+\mathcal{N}_\pi^\munu(k^0=\omega_k)\delta(q^0+\omega_k-\omega_p) \frac{}{}}} \label{eq.impi.pi.1} \\
\IM\Pibar^\munu_\tN(q^0,\vec{q}) &=& -\sign{q^0}\tanh\FB{\frac{\beta q^0}{2}}
\pi\bigintssss\frac{d^3k}{(2\pi)^3}\frac{1}{4\Omega_k\Omega_p} \times \nn \\
&& \TB{ \SB{\frac{}{}1-f^-(\Omega_k)-f^+(\Omega_p)+2f^-(\Omega_k)f^+(\Omega_p)} \mathcal{N}_\tN^\munu(k^0=-\Omega_k)\delta(q^0-\Omega_k-\Omega_p)
	\right. \nn \\ && \left.
	+ \SB{\frac{}{}1-f^+(\Omega_k)-f^-(\Omega_p)+2f^+(\Omega_k)f^-(\Omega_p)} \mathcal{N}_\tN^\munu(k^0=\Omega_k)\delta(q^0+\Omega_k+\Omega_p)  
	\right. \nn \\ && \left.
	+ \SB{\frac{}{}-f^-(\Omega_k)-f^-(\Omega_p)+2f^-(\Omega_k)f^-(\Omega_p)} \mathcal{N}_\tN^\munu(k^0=-\omega_k)\delta(q^0-\Omega_k+\Omega_p)
	\right. \nn \\ && \left.
		+ \SB{\frac{}{}-f^+(\Omega_k)-f^+(\Omega_p)+2f^+(\Omega_k)f^+(\Omega_p)} \mathcal{N}_\tN^\munu(k^0=\Omega_k)\delta(q^0+\Omega_k-\Omega_p) } \label{eq.impi.N.1}
\end{eqnarray}
where, $\mathcal{P}$ denotes the Cauchy Principal value integration, $\omega_k=\sqrt{m_\pi^2+\vec{k}^2}$,  $\Omega_k=\sqrt{m_N^2+\vec{k}^2}$ and $\mathcal{N}_\tN(q,k)$ is defined in Eq.~\eqref{eq.Nmunu_N}. 


\section{$\rho^0$ Self Energy in the Magnetized Medium} \label{sec.self.med.eB}

In  presence of the external magnetic field $\vec{B}=B\hat{z}$, the propagations of the charged pion and proton are modified. One of the possible ways to incorporate the effect of external magnetic field is the Schwinger proper time formalism in which the $11$-components of charged pion and proton propagators respectively become~\cite{Schwinger:1951nm,Ayala:2004dx}
\begin{eqnarray}
D_B^{11}(k) &=& \Delta_B(k,m_\pi) + \eta(k\cdot u)\TB{\Delta_B(k,m_\pi)-\Delta^*_B(k,m_\pi)} ~~\text{and}\label{eq.DB11}\\
S_B^{11}(k) &=& S_B(k,m_N) - \etatilde(k\cdot u)\TB{S_B(k,m_N)-\gamma^0S^\dagger_B(k,m_N)\gamma^0}
\label{eq.SB11}
\end{eqnarray}
where, $\Delta_B(k,m_\pi)$ and $S_B(k,m_N)$ denote the momentum space vacuum (zero temperature) Schwinger proper time propagators for charged pion and proton respectively \cite{Schwinger:1951nm}:
\begin{eqnarray}
\Delta_B(k) &=& i\int_{0}^{\infty} ds \exp\TB{is\SB{k_\parallel^2+\frac{\tan(eBs)}{eBs}k_\perp^2-m_N^2}}  
\label{eq.schwinger.pi.1} \\
S_B(k) &=& i\int_{0}^{\infty} ds \exp\TB{is\SB{k_\parallel^2+\frac{\tan(eBs)}{eBs}k_\perp^2-m_N^2}}
\TB{\frac{}{}\FB{\cancel{k}_\parallel+m_N}\SB{1-\gamma^1\gamma^2\tan(eBs)}+\cancel{k}_\perp\sec^2(eBs)}~.
\label{eq.schwinger.N.1}
\end{eqnarray}
In the above equations, $e=\MB{e}$ is the charge of the proton; the four-vector $k$ is decomposed into $k=(k_\parallel+k_\perp)$ where $k_\parallel^\mu = g_\parallel^\munu k_\nu$ and $k_\perp^\mu = g_\perp^\munu k_\nu$ corresponding to the decomposition of the metric tensor 
$g^\munu=(g_\parallel^\munu+g_\perp^\munu)$ with $g_\parallel^\munu=\text{diag}(1,0,0,-1)$ and $g_\perp^\munu=\text{diag}(0,-1,-1,0)$. The above decomposition can be done in a Lorentz covariant way by introducing another four-vector 
\begin{eqnarray}
b^\mu=\frac{1}{B}G^\munu u_\nu
\end{eqnarray}
where $G^\munu=\frac{1}{2}\epsilon^{\mu\nu\alpha\beta}F_\alphabeta$ is the dual of the electromagnetic field tensor $F^\munu$. In the local rest frame of the medium, $b_\text{LRF}^\mu\equiv\FB{0,0,0,1}$, which is the direction of the external magnetic field. Using $b^\mu$, we can write 
\begin{eqnarray}
g_\parallel^\munu = \FB{u^\mu u^\nu - b^\mu b^\nu} ~~\text{and}~~ 
g_\perp^\munu = \FB{g^\munu - u^\mu u^\nu + b^\mu b^\nu}~.
\end{eqnarray}

It is important to note that, the coordinate space Schwinger propagator contains a gauge dependent translationally non-invariant phase factor. However, for the one-loop graphs containing equally charged particle in the loop, the phase factor gets canceled and the momentum space propagator is sufficient for the calculation of the self energy. The proper time integral in Eqs.~\eqref{eq.schwinger.pi.1} and \eqref{eq.schwinger.N.1} can be performed in order to express the propagators as a sum over discrete Landau levels as
\begin{eqnarray}
\Delta_B(k) &=& -\sum\limits_{l=0}^{\infty}\frac{2(-1)^le^{-\alpha_k}L_l(2\alpha_k)}{k_\parallel^2-m_\pi^2-(2l+1)eB+i\epsilon} \label{eq.DB}\\
S_B(k) &=& -\sum_{l=0}^{\infty}\TB{ \frac{(-1)^le^{-\alpha_k}\mathcal{D}_l(k)}{k_\parallel^2-m_N^2-2leB+i\epsilon}}
\label{eq.SB}
\end{eqnarray}
where, 
\begin{eqnarray}
\mathcal{D}_l(k) = \FB{\cancel{k}_\parallel+m_N}\TB{\FB{1+i\gamma^1\gamma^2}L_l(2\alpha_k)  
	-\FB{1-i\gamma^1\gamma^2}L_{l-1}(2\alpha_k)} - 4\cancel{k}_\perp L^1_{l-1}(2\alpha_k)
\end{eqnarray}
with $\alpha_k=-k_\perp^2/eB$. Here, $L^a_l(z)$ denotes the generalized Laguerre polynomial with $L^a_{-1}(z)=0$ and 
$L_l(z)=L_l^0(z)$. We now rewrite Eqs.~\eqref{eq.DB11} and \eqref{eq.SB11} using Eqs.~\eqref{eq.DB} and \eqref{eq.SB} as
\begin{eqnarray}
D_B^{11}(k) &=& \sum_{l=0}^{\infty}2(-1)^le^{-\alpha_k}L_l(2\alpha_k) \TB{\frac{-1}{k_\parallel^2-m_l^2+i\epsilon}
	+2\pi i\eta(k\cdot u)\delta\FB{k_\parallel^2-m_l^2}} \label{eq.DB11.2}\\
S^{11}_B(k) &=& \sum_{l=0}^{\infty}(-1)^le^{-\alpha_k}\mathcal{D}_l(k)
\TB{\frac{-1}{k_\parallel^2-M_l^2+i\epsilon}
	-2\pi i\etatilde(k\cdot u)\delta\FB{k_\parallel^2-M_l^2}} \label{eq.SB11.2}
\end{eqnarray}
where we have defined the Landau level dependent ``dimensionally reduced effective masses" (as a consequence of dimensional reduction) of pion and proton as
\begin{eqnarray}
m_l = \sqrt{m_\pi^2+(2l+1)eB} ~~\text{and}~~ M_l = \sqrt{m_N^2+2leB}~.
\label{eq.dim.mass}
\end{eqnarray}

We now replace the $11$-component of the charged pion and proton propagators in Eqs.~\eqref{eq.Pi.T.pi.1} and \eqref{eq.Pi.T.N.1} as $D^{11}\rightarrow D_B^{11}, S_\tp^{11}\rightarrow S_B^{11}$ i.e by the respective magnetized ones given in Eqs.~\eqref{eq.DB11.2} and \eqref{eq.SB11.2} and then perform the $dk^0$ integrations (using the Dirac delta functions). Following Eqs.~\eqref{eq.analytic.relation.re} and \eqref{eq.analytic.relation.im} we get the thermal self energy functions under external magnetic field which we will denote by a \textit{double bar} to distinguish them from the thermal self energy functions in the absence of magnetic field. 
Their explicit expressions are given by
\begin{eqnarray}
\RE\Pibarbar^\munu_\pi(q^0,\vec{q}) &=& \sum_{l=0}^{\infty}\sum_{n=0}^{\infty}\bigintssss\frac{d^3k}{(2\pi)^3}\mathcal{P}
\TB{\frac{f(\omega_k^l)}{2\omega_k^l}\SB{\frac{\mathcal{N}_{\pi,nl}^\munu(k^0=-\omega_k^l)}{(q^0-\omega_k^l)^2-(\omega_p^n)^2}
		+ \frac{\mathcal{N}_{\pi,nl}^\munu(k^0=\omega_k^l)}{(q^0+\omega_k^l)^2-(\omega_p^n)^2}} \right. \nn \\
	&& \hspace{1.0cm} \left.
	+ \frac{f(\omega_p^n)}{2\omega_p^n}\SB{\frac{\mathcal{N}_{\pi,nl}^\munu(k^0=-q^0-\omega_p^n)}{(q^0+\omega_p^n)^2-(\omega_k^l)^2}
		+ \frac{\mathcal{N}_{\pi,nl}^\munu(k^0=-q^0+\omega_p^n)}{(q^0-\omega_p^n)^2-(\omega_k^l)^2}}} 
	+ \RE\FB{\Pi^\munu_\pi}_\text{vac}(q,eB)\label{eq.repiB.pi.1}\\
\RE\Pibarbar^\munu_\tN(q^0,\vec{q}) &=& \frac{1}{2}\RE\Pibar^\munu_\tN(q^0,\vec{q}) - \sum_{l=0}^{\infty}\sum_{n=0}^{\infty}\bigintssss\frac{d^3k}{(2\pi)^3}\mathcal{P}
\TB{\frac{1}{2\Omega_k^l}\SB{\frac{f^-(\Omega_k^l)\mathcal{N}_{\tp,nl}^\munu(k^0=-\Omega_k^l)}{(q^0-\Omega_k^l)^2-(\Omega_p^n)^2}
		+ \frac{f^+(\Omega_k^l)\mathcal{N}_{\tp,nl}^\munu(k^0=\Omega_k^l)}{(q^0+\Omega_k^l)^2-(\Omega_p^n)^2}} \right. \nn \\
	&& \hspace{-0.5cm} \left.
	+ \frac{1}{2\Omega_p^n}\SB{\frac{f^-(\Omega_p^n)\mathcal{N}_{\tp,nl}^\munu(k^0=-q^0-\Omega_p^n)}{(q^0+\Omega_p^n)^2-(\Omega_k^l)^2}
		+ \frac{f^+(\Omega_p^n)\mathcal{N}_{\tp,nl}^\munu(k^0=-q^0+\Omega_p^n)}{(q^0-\Omega_p^n)^2-(\Omega_k^l)^2}}} 
	+ \RE\FB{\Pi^\munu_\tp}_\text{vac}(q,eB)  \label{eq.repiB.N.1}	
\end{eqnarray}
\begin{eqnarray}
\IM\Pibarbar^\munu_\pi(q^0,\vec{q}) &=& -\sign{q^0}\tanh\FB{\frac{\beta q^0}{2}}
\pi\sum_{l=0}^{\infty}\sum_{n=0}^{\infty}\bigintssss\frac{d^3k}{(2\pi)^3}\frac{1}{4\omega_k^l\omega_p^n}
\TB{ \SB{1+f(\omega_k^l)+f(\omega_p^n)+2f(\omega_k^l)f(\omega_p^n)\frac{}{}} \right. \nn \\
	&& \hspace{-1.5cm}\left. \SB{\frac{}{}\mathcal{N}_{\pi,nl}^\munu(k^0=-\omega_k^l)\delta(q^0-\omega_k^l-\omega_p^n)
		+\mathcal{N}_{\pi,nl}^\munu(k^0=\omega_k^l)\delta(q^0+\omega_k^l+\omega_p^n) } 
	+ \SB{\frac{}{}f(\omega_k^l)+f(\omega_p^n)+2f(\omega_k^l)f(\omega_p^n)} \right. \nn \\
	&& \left. \SB{\mathcal{N}_{\pi,nl}^\munu(k^0=-\omega_k^l)\delta(q^0-\omega_k^l+\omega_p^n)
		+\mathcal{N}_{\pi,nl}^\munu(k^0=\omega_k^l)\delta(q^0+\omega_k^l-\omega_p^n) \frac{}{}}} \label{eq.impiB.pi.1} \\
\IM\Pibarbar^\munu_\tN(q^0,\vec{q}) &=& \frac{1}{2}\IM\Pibar^\munu_\tN(q^0,\vec{q}) -\sign{q^0}\tanh\FB{\frac{\beta q^0}{2}}
\pi\sum_{l=0}^{\infty}\sum_{n=0}^{\infty}\bigintssss\frac{d^3k}{(2\pi)^3}\frac{1}{4\Omega_k^l\Omega_p^n} \times \nn \\
&& \TB{ \SB{\frac{}{}1-f^-(\Omega_k^l)-f^+(\Omega_p^n)+2f^-(\Omega_k^l)f^+(\Omega_p^n)} \mathcal{N}_{\tp,nl}^\munu(k^0=-\Omega_k^l)\delta(q^0-\Omega_k^l-\Omega_p^n)
	\right. \nn \\ && \left.
	+ \SB{\frac{}{}1-f^+(\Omega_k^l)-f^-(\Omega_p^n)+2f^+(\Omega_k^l)f^-(\Omega_p^n)} \mathcal{N}_{\tp,nl}^\munu(k^0=\Omega_k^l)\delta(q^0+\Omega_k^l+\Omega_p^n)  
	\right. \nn \\ && \left.
	+ \SB{\frac{}{}-f^-(\Omega_k^l)-f^-(\Omega_p^n)+2f^-(\Omega_k^l)f^-(\Omega_p^n)} \mathcal{N}_{\tp,nl}^\munu(k^0=-\omega_k^l)\delta(q^0-\Omega_k^l+\Omega_p^n)
	\right. \nn \\ && \left.
	+ \SB{\frac{}{}-f^+(\Omega_k^l)-f^+(\Omega_p^n)+2f^+(\Omega_k^l)f^+(\Omega_p^n)} \mathcal{N}_{\tp,nl}^\munu(k^0=\Omega_k^l)\delta(q^0+\Omega_k^l-\Omega_p^n) } \label{eq.impiB.N.1}
\end{eqnarray}
where, 
\begin{eqnarray}
\mathcal{N}^\munu_{\pi,nl}(q,k) &=& 4(-1)^{n+l}e^{-\alpha_k-\alpha_p}L_l(2\alpha_k)L_n(2\alpha_p) \mathcal{N}^\munu_\pi(q,k) 
\label{eq.N.pi.2}\\
\mathcal{N}^\munu_{\tp,nl}(q,k) &=& -g_\rhoNN^2(-1)^{n+l}e^{-\alpha_k-\alpha_p}
\Tr\TB{\Gamma^\nu(q)\mathcal{D}_n(q+k)\Gamma^\mu(-q)\mathcal{D}_l(k)} \label{eq.N.p.2} \\
\omega_k^l &=& \sqrt{k_z^2+m_l^2} = \sqrt{k_z^2+m_\pi^2+(2l+1)eB}  \\
\Omega_k^l &=& \sqrt{k_z^2+M_l^2} = \sqrt{k_z^2+m_N^2+2leB}~. 
\end{eqnarray} 
The first terms on the RHS of Eqs.~\eqref{eq.repiB.N.1} and \eqref{eq.impiB.N.1} are the contributions from the neutron-neutron loop which are not affected by the external magnetic field. The last terms on the RHS of Eqs.~\eqref{eq.repiB.pi.1} and \eqref{eq.repiB.N.1} are the contributions from $\pi\pi$ and proton-proton loop which depend on the external magnetic field but independent of temperature. Their explicit forms are given by
\begin{eqnarray}
\RE\FB{\Pi^\munu_\pi}_\text{vac}(q,eB) = \RE \sum_{l=0}^{\infty}\sum_{n=0}^{\infty}i\int\frac{d^4k}{(2\pi)^4}
\mathcal{N}^\munu_{\pi,nl}\Delta_F(k_\parallel,m_l)\Delta_F(q_\parallel+k_\parallel,m_n) \label{eq.ebvac.pi.1}\\
\RE\FB{\Pi^\munu_\tp}_\text{vac}(q,eB) = \RE \sum_{l=0}^{\infty}\sum_{n=0}^{\infty}i\int\frac{d^4k}{(2\pi)^4}
\mathcal{N}^\munu_{\tp,nl}\Delta_F(k_\parallel,M_l)\Delta_F(q_\parallel+k_\parallel,M_n)~. \label{eq.ebvac.p.1}
\end{eqnarray}
It is important to note that, the above quantities respectively contain the divergent pure vacuum contributions $\FB{\Pi^\munu_\pi}_\text{pure-vac}(q)$ and $\frac{1}{2}\FB{\Pi^\munu_\tN}_\text{pure-vac}(q)$ in a nontrivial way (as the above equations seem to appear non-perturbative in eB). In contrast, for the case of weak magnetic field expansion of the Schwinger propagator, the pure vacuum contribution to the self energy trivially decouples from the magnetic field dependent terms. Since we are working with the full propagator including all the Landau levels, we have to properly regularize the above expressions in order to extract the pure vacuum contributions from these quantities. We use dimensional regularization in which the ultraviolet divergence appear as the pole of Gamma and Hurwitz zeta function the details of which are provided in the Appendices~\ref{app.eb.vac.pi} and \ref{app.eb.vac.p}. Here, we take the transverse momentum of $\rho^0$ to be zero i.e. $q_\perp=0$ which makes substantial simplifications of the analytic calculations. The final result can be read off from Eqs.~\eqref{eq.ebvac.pi.8} and \eqref{eq.ebvac.p.8} as
\begin{eqnarray}
\FB{\Pi^\munu_\pi}_\text{vac}(q_\parallel,eB) &=& \FB{\Pi^\munu_\pi}_\text{pure-vac}(q_\parallel) + 
\FB{\Pi^\munu_\pi}_\text{eB-vac}(q_\parallel,eB) \label{eq.ebvac.pi.10} \\
\FB{\Pi^\munu_\tp}_\text{vac}(q_\parallel,eB) &=& \frac{1}{2}\FB{\Pi^\munu_\tN}_\text{pure-vac}(q_\parallel) + 
\FB{\Pi^\munu_\tp}_\text{eB-vac}(q_\parallel,eB)
\label{eq.ebvac.p.10}
\end{eqnarray}
where, the scale dependent divergent pure-vacuum parts are completely decoupled as the first term on the RHS of the above equation;  
the scale independent and finite ``eB-dependent vacuum contributions" to the real part of the self energy functions are
\begin{eqnarray}
\FB{\Pi^\munu_\pi}_\text{eB-vac}(q_\parallel,eB) &=& \frac{-g_\rhopipi^2q_\parallel^2}{32\pi^2}\int_{0}^{1}dx
\TB{ \SB{\ln\FB{\frac{\Delta_\pi(q_\perp=0)}{2eB}}-1 } \Delta_\pi(q_\perp=0)
	(q_\parallel^2g^\munu-q_\parallel^\mu q_\parallel^\nu) \right. \nn \\ && \left.
	- (q_\parallel^2g_\parallel^\munu-q_\parallel^\mu q_\parallel^\nu)2eB
	\SB{\ln\Gamma\FB{z_\pi+\frac{1}{2}}-\ln\sqrt{2\pi}} \right. \nn \\ && \left.
	+ q_\parallel^2g_\perp^\munu\SB{ \Delta_\pi(q_\perp=0)+\frac{eB}{2}-\frac{1}{2}\Delta_\pi(q_\perp=0)
		\SB{\psi\FB{z_\pi+\frac{1}{2}} + \psi\FB{z_\pi+x+\frac{1}{2}  } } } } \label{eq.ebvac.pi.11} 
\end{eqnarray}
\begin{eqnarray}
\FB{\Pi^\munu_\tp}_\text{eB-vac}(q_\parallel,eB) &=& \frac{g_\rhoNN^2}{4\pi^2}\int_{0}^{1}dx \Bigg[\Bigg.
\ln\FB{\frac{\Delta_N(\qper=0)}{2eB}} \SB{2x(1-x)+\kappa_\rho+\frac{\kappa_\rho^2}{2}-\frac{\kappa_\rho^2}{4m_N^2}\Delta_N(\qper=0)} 
(q_\parallel^2g^\munu-q_\parallel^\mu q_\parallel^\nu) \nn \\
&& -2x(1-x)\FB{ \psi(z_N)+\frac{1}{2z_N}}(q_\parallel^2\gpll^\munu-q_\parallel^\mu q_\parallel^\nu) 
+ 2eB \gper^\munu \SB{ \FB{z_N-\frac{m_N^2}{eB}}\psi(z_N+x)+z_N \right. \nn \\ && \left.
	+ \ln\Gamma(z+x)-\ln\sqrt{2\pi}\frac{}{}} - \kappa_\rho \SB{ (q_\parallel^2\gpll^\munu-q_\parallel^\mu q_\parallel^\nu) 
	\FB{\psi(z_N)+\frac{1}{2z_N}}  + \qpll^2\gper^\munu\psi(z+x) } \nn \\ &&
+ \frac{\kappa_\rho^2}{4m_N^2}2eB\TB{ (q_\parallel^2\gpll^\munu-q_\parallel^\mu q_\parallel^\nu)
	\SB{ -\frac{m_N^2}{eB}\FB{\psi(z_N)+\frac{1}{2z_N}} + \frac{1}{2}\ln(z_N)+\ln\Gamma(z_N)-\ln\sqrt{2\pi} } \right. \nn \\  && \left.
	-\qpll^2\gper^\munu\SB{ \FB{\frac{m_N^2}{eB}-z_N}\psi(z_N+x)+\Delta_N(\qper=0) } 
	+ \frac{\kappa_\rho^2}{4m_N^2}(\qpll^2g^\munu-\qpll^\mu\qpll^\nu)\Delta_N(\qper=0) }~.
\label{eq.ebvac.p.11}
\end{eqnarray}
Eqs.~\eqref{eq.ebvac.pi.10} and \eqref{eq.ebvac.p.10} imply that the vacuum counter terms are sufficient to renormalize the theory and thus the external magnetic field does not create additional divergences. For $q_\perp=0$, the $d^2\kper$ integrals in Eqs.~\eqref{eq.repiB.pi.1}-\eqref{eq.impiB.N.1} can be analytically performed (see Appendix~\ref{app.d2k.integral}) and we finally get, 
\begin{eqnarray}
\RE\Pibarbar^\munu_\pi(q^0,q_z) &=& \RE\FB{\Pi^\munu_\pi}_\text{pure-vac}(\qpll) + \sum_{n=0}^{\infty}~\sum_{l=(n-1)}^{(n+1)}\bigintssss_{-\infty}^{\infty}\frac{dk_z}{2\pi}\mathcal{P}
\TB{\frac{f(\omega_k^l)}{2\omega_k^l}\SB{\frac{\tilde{\mathcal{N}}_{\pi,nl}^\munu(k^0=-\omega_k^l)}{(q^0-\omega_k^l)^2-(\omega_p^n)^2}
		+ \frac{\tilde{\mathcal{N}}_{\pi,nl}^\munu(k^0=\omega_k^l)}{(q^0+\omega_k^l)^2-(\omega_p^n)^2}} \right. \nn \\
	&& \hspace{0.5cm} \left.
	+ \frac{f(\omega_p^n)}{2\omega_p^n}\SB{\frac{\tilde{\mathcal{N}}_{\pi,nl}^\munu(k^0=-q^0-\omega_p^n)}{(q^0+\omega_p^n)^2-(\omega_k^l)^2}
		+ \frac{\tilde{\mathcal{N}}_{\pi,nl}^\munu(k^0=-q^0+\omega_p^n)}{(q^0-\omega_p^n)^2-(\omega_k^l)^2}}} 
+ \RE\FB{\Pi^\munu_\pi}_\text{eB-vac}(\qpll,eB)\label{eq.repiB.pi.2}\\
\RE\Pibarbar^\munu_\tN(q^0,q_z) &=& \RE\Pibar^\munu_\tN(q^0,q_z) - \sum_{n=0}^{\infty}~\sum_{l=(n-1)}^{(n+1)}\bigintssss_{-\infty}^{\infty}\frac{dk_z}{2\pi}\mathcal{P}
\TB{\frac{1}{2\Omega_k^l}\SB{\frac{f^-(\Omega_k^l)\tilde{\mathcal{N}}_{\tp,nl}^\munu(k^0=-\Omega_k^l)}{(q^0-\Omega_k^l)^2-(\Omega_p^n)^2}
		+ \frac{f^+(\Omega_k^l)\tilde{\mathcal{N}}_{\tp,nl}^\munu(k^0=\Omega_k^l)}{(q^0+\Omega_k^l)^2-(\Omega_p^n)^2}} \right. \nn \\
	&& \hspace{-1.5cm} \left.
	+\frac{1}{2\Omega_p^n}\SB{\frac{f^-(\Omega_p^n)\tilde{\mathcal{N}}_{\tp,nl}^\munu(k^0=-q^0-\Omega_p^n)}{(q^0+\Omega_p^n)^2-(\Omega_k^l)^2}
		+ \frac{f^+(\Omega_p^n)\tilde{\mathcal{N}}_{\tp,nl}^\munu(k^0=-q^0+\Omega_p^n)}{(q^0-\Omega_p^n)^2-(\Omega_k^l)^2}}} 
+ \RE\FB{\Pi^\munu_\tp}_\text{eB-vac}(\qpll,eB)  \label{eq.repiB.N.2}
\end{eqnarray}
\begin{eqnarray}
\IM\Pibarbar^\munu_\pi(q^0,q_z) &=& -\sign{q^0}\tanh\FB{\frac{\beta q^0}{2}}
\pi\sum_{n=0}^{\infty}~\sum_{l=(n-1)}^{(n+1)}\bigintssss_{-\infty}^{\infty}\frac{dk_z}{2\pi}\frac{1}{4\omega_k^l\omega_p^n}
\TB{ \SB{1+f(\omega_k^l)+f(\omega_p^n)+2f(\omega_k^l)f(\omega_p^n)\frac{}{}} \right. \nn \\
	&& \hspace{-1.5cm}\left. \SB{\frac{}{}\tilde{\mathcal{N}}_{\pi,nl}^\munu(k^0=-\omega_k^l)\delta(q^0-\omega_k^l-\omega_p^n)
		+\tilde{\mathcal{N}}_{\pi,nl}^\munu(k^0=\omega_k^l)\delta(q^0+\omega_k^l+\omega_p^n) } 
	+ \SB{\frac{}{}f(\omega_k^l)+f(\omega_p^n)+2f(\omega_k^l)f(\omega_p^n)} \right. \nn \\
	&& \left. \SB{\tilde{\mathcal{N}}_{\pi,nl}^\munu(k^0=-\omega_k^l)\delta(q^0-\omega_k^l+\omega_p^n)
		+\tilde{\mathcal{N}}_{\pi,nl}^\munu(k^0=\omega_k^l)\delta(q^0+\omega_k^l-\omega_p^n) \frac{}{}}} \label{eq.impiB.pi.2} \\
\IM\Pibarbar^\munu_\tN(q^0,q_z) &=& \frac{1}{2}\IM\Pibar^\munu_\tN(q^0,q_z) -\sign{q^0}\tanh\FB{\frac{\beta q^0}{2}}
\pi\sum_{n=0}^{\infty}~\sum_{l=(n-1)}^{(n+1)}\bigintssss_{-\infty}^{\infty}\frac{dk_z}{2\pi}\frac{1}{4\Omega_k^l\Omega_p^n} \times \nn \\
&& \TB{ \SB{\frac{}{}1-f^-(\Omega_k^l)-f^+(\Omega_p^n)+2f^-(\Omega_k^l)f^+(\Omega_p^n)} \tilde{\mathcal{N}}_{\tp,nl}^\munu(k^0=-\Omega_k^l)\delta(q^0-\Omega_k^l-\Omega_p^n)
	\right. \nn \\ && \left.
	+ \SB{\frac{}{}1-f^+(\Omega_k^l)-f^-(\Omega_p^n)+2f^+(\Omega_k^l)f^-(\Omega_p^n)} \tilde{\mathcal{N}}_{\tp,nl}^\munu(k^0=\Omega_k^l)\delta(q^0+\Omega_k^l+\Omega_p^n)  
	\right. \nn \\ && \left.
	+ \SB{\frac{}{}-f^-(\Omega_k^l)-f^-(\Omega_p^n)+2f^-(\Omega_k^l)f^-(\Omega_p^n)} \tilde{\mathcal{N}}_{\tp,nl}^\munu(k^0=-\omega_k^l)\delta(q^0-\Omega_k^l+\Omega_p^n)
	\right. \nn \\ && \left.
	+ \SB{\frac{}{}-f^+(\Omega_k^l)-f^+(\Omega_p^n)+2f^+(\Omega_k^l)f^+(\Omega_p^n)} \tilde{\mathcal{N}}_{\tp,nl}^\munu(k^0=\Omega_k^l)\delta(q^0+\Omega_k^l-\Omega_p^n) } \label{eq.impiB.N.2}
\end{eqnarray}
where, $\tilde{\mathcal{N}}^\munu_{\pi,nl}(q_\parallel,k_\parallel)$ and $\tilde{\mathcal{N}}^\munu_{\tp,nl}(q_\parallel,k_\parallel)$ can be read off from Eq.~\eqref{eq.N.pi.5} and \eqref{eq.N.p.5}. The presence of Kronecker delta functions in the expressions of  $\tilde{\mathcal{N}}^\munu_{\pi,nl}(q_\parallel,k_\parallel)$ and $\tilde{\mathcal{N}}^\munu_{\tp,nl}(q_\parallel,k_\parallel)$ 
 has eliminated one of the double sums or in other words, the sum over index $l$ now runs from $(n-1)$ to $(n+1)$.

\section{Lorentz Structure of the vector boson self energy in magnetized medium} \label{sec.lorentz}

In this section, we will derive the tensorial decomposition of the massive vector boson self energy. We note that, the self energy $\Pi^\munu(q)$ being a second rank tensor, has sixteen components which will mix among themselves with the change of frame. It is useful to use linearly independent basis tensors (constructed with the available vectors and tensors) to express $\Pi^\munu(q)$ so that the form factors (corresponding to each basis) remain Lorentz invariant. This will also enable one to solve the Dyson-Schwinger equation in order to obtain the complete interacting vector boson propagator. In order to proceed, we first note that the vector boson self energy satisfies the following constrain
\begin{eqnarray}
\Pi^\munu(q) = \Pi^\numu(q) ~~ \text{and} ~~ q_\mu\Pi^\munu(q) = 0~. \label{eq.Pimunu.constraint}
\end{eqnarray}

Let us first consider the pure vacuum case i.e. for zero temperature and zero external magnetic field. In this case, the only available vector is the momentum $q^\mu$ along with the metric tensor $g^\munu$ so that $\Pi^\munu(q)$ is a linear combination of $q^\mu q^\nu$ and $g^\munu$ i.e $\Pi^\munu(q) = (\alpha_1 g^\munu + \alpha_2 q^\mu q^\nu)$. Imposing the constrains of Eq.~\eqref{eq.Pimunu.constraint}, 
we get $\alpha_1+\alpha_2q^2 = 0$ which makes the only possible Lorentz structure of the self energy as
\begin{eqnarray}
\Pi^\munu = \alpha_1\FB{g^\munu-\frac{q^\mu q^\nu}{q^2}} \label{eq.pidecom.v}
\end{eqnarray}
where the Lorentz invariant form factor $\alpha_1 = \alpha_1(q^2) = \frac{1}{3}\Pi^\mu_{~\mu}$. Note that, with $q^\mu$ and $g^\munu$, the only possible 
Lorentz scalar that can be formed by contracting with $\Pi^\munu(q)$ is the quantity $g_\munu\Pi^\munu = \Pi^\mu_{~\mu}$ implying the existence of only one form factor.

We now consider the case with finite temperature but zero magnetic field. In this case we have an additional four vector $u^\mu$ (medium four-velocity) along with $q^\mu$ and $g^\munu$. This makes $\Pi^\munu$ to be a linear combination of $g^\munu$, $q^\mu q^\nu$, $u^\mu u^\nu$, $q^\mu u^\nu$ and $q^\nu u^\mu$ i.e.
\begin{eqnarray}
\Pi^\munu(q) = (\alpha_1 g^\munu + \alpha_2q^\mu q^\nu + \alpha_3u^\mu u^\nu + \alpha_4q^\mu u^\nu + \alpha_5 q^\nu q^\mu)
\end{eqnarray}
However, imposing the constrains in Eq.~\eqref{eq.Pimunu.constraint}, we find the following relationship among the coefficients
\begin{eqnarray}
\alpha_5 &=& \alpha_4 \\
\alpha_1 + \alpha_2q^2+\alpha_4(q\cdot u) &=& 0 \\
\alpha_3 (q\cdot u) + \alpha_4q^2 &=& 0 
\end{eqnarray}
which makes only two of the coefficients independent. Choosing $\alpha_1$ and $\alpha_2$ as  independent, we get,

\begin{eqnarray}
\Pi^\munu(q) = \alpha_1 \TB{g^\munu+\frac{q^2}{(q\cdot u)}u^\mu u^\nu - \frac{1}{(q\cdot u)}(q^\mu u^\nu + q^\nu u^\mu) } 
+ \alpha_2 \TB{q^\mu q^\nu +\frac{q^4}{(q\cdot u)^2}u^\mu u^\nu - \frac{q^2}{(q\cdot u)}(q^\mu u^\nu + q^\nu u^\mu) } ~.
\label{eq.pidecom.t.1}
\end{eqnarray}
where the Lorentz invariant form factors $\alpha_1 = \alpha_1(q^2,q\cdot u)$ and $\alpha_2 = \alpha_2(q^2,q\cdot u)$ can be obtained by contracting both side of the above equations with $g_\munu$ and $u_\mu u_\nu$ so that the form factors will become functions of the Lorentz scalars 
$g_\munu\Pi^\munu = \Pi^\mu_{~\mu}$ and $u_\mu u_\nu\Pi^\munu$. Note that, with $q^\mu$, $u^\mu$ and $g^\munu$, only two possible 
Lorentz scalars that can be formed by contracting with $\Pi^\munu(q)$ are the quantities $\Pi^\mu_{~\mu}$ and $u_\mu u_\nu\Pi^\munu$ implying the existence of only two form factors. Unlike the pure vacuum case given in Eq.~\eqref{eq.pidecom.v}, here the  decomposition of $\Pi^\munu$ in Eq.~\eqref{eq.pidecom.t.1} is not unique. As already mentioned, it is useful to construct linearly independent (and mutually orthogonal) basis tensors (note that the basis tensors within square brackets in Eq.~\eqref{eq.pidecom.t.1} are not  mutually orthogonal). One such choice of orthogonal tensor basis could be 
\begin{eqnarray}
P_1^\munu = \FB{g^\munu - \frac{q^\mu q^\nu}{q^2} -\frac{\utilde^\mu\utilde^\nu}{\utilde^2}} ~~ \text{and}~~ 
P_2^\munu = \FB{\frac{\utilde^\mu\utilde^\nu}{\utilde^2}} \label{eq.proj.tensor.t1}
\end{eqnarray}
where 
\begin{eqnarray}
\utilde^\mu = u^\mu - \frac{(q\cdot u)}{q^2}q^\mu,  \label{eq.utilde}
\end{eqnarray}
which is constructed from $u^\mu$ by subtracting out its projection along $q^\mu$. It is easy to check that $P_1^\munu$ and $P_2^\munu$ satisfy all the properties of projection tensors i.e. 
\begin{eqnarray}
g_\alphabeta P_i^{\mu\alpha} P_j^{\beta\nu} = \delta_{ij} P_i^\munu ~~\text{and}~~
g_\alphabeta g_\munu P_i^{\mu\alpha} P_j^{\beta\nu} = \delta_{ij}~.
\label{eq.form.factor.t.orthogonality}
\end{eqnarray}
Therefore, $\Pi^\munu$ can be written as
\begin{eqnarray}
\Pi^\munu(q) = \Pi_1(q^2,\qdotu) P_1^\munu + \Pi_2(q^2,\qdotu) P_2^\munu  \label{eq.Pi.t}
\end{eqnarray}
where the form factors are 
\begin{eqnarray}
\Pi_1(q^2,\qdotu) = \frac{1}{2}\FB{\Pimumu - \frac{1}{\utilde^2} u_\mu u_\nu\Pi^\munu} ~~ \text{and}~~ 
\Pi_2(q^2,\qdotu) = \FB{ \frac{1}{\utilde^2} u_\mu u_\nu\Pi^\munu} ~. \label{eq.Pi1.Pi2}
\end{eqnarray}
Care should be taken when considering the special case like $\vec{q}=\vec{0}$~\cite{Mallik:2016anp}. To see this, let us consider $q^i=|\vec{q}|n^i$ so that the spatial components of the projectors at $\vec{q}=\vec{0}$ become (in the LRF)
\begin{eqnarray}
P_1^{ij} = g^{ij}+n^in^j ~~ \text{and}~~ P_2^{ij} = -n^in^j~.
\end{eqnarray} 
This implies that the spatial components of self energy at vanishing three momentum
\begin{eqnarray}
\Pi^{ij}(q^0,\vec{q}=\vec{0}) = \Pi_1g^{ij}+n^in^j\FB{\Pi_1-\Pi_2}
\end{eqnarray}
depend on the direction of $\vec{q}$ even at $|\vec{q}|=0$. This ambiguity is eliminated by setting additional constraint on the form factors as $\Pi_1(q^0,\vec{q}=\vec{0})=\Pi_2(q^0,\vec{q}=\vec{0})$.

Following the same strategy, we now construct suitable orthogonal tensor basis for the  vector bososn self energy at finite temperature under external magnetic field. In this case we have an additional four vector $b^\mu$ (corresponding to the magnetic field direction) along with $q^\mu$, $u^\mu$ and $g^\munu$. This makes the symmetric $\Pi^\munu$ to be a linear combination of seven tensors as
\begin{eqnarray}
\Pi^\munu(q) = \alpha_1g^\munu + \alpha_2 q^\mu q^\nu + \alpha_3 u^\mu u^\nu + \alpha_4b^\mu b^\nu 
+\alpha_5(q^\mu u^\nu+q^\nu u^\mu) + \alpha_6(q^\mu b^\nu+q^\nu b^\mu)+ \alpha_7(u^\mu b^\nu+u^\nu b^\mu)
\end{eqnarray}
However, imposing the constrains in Eq.~\eqref{eq.Pimunu.constraint}, we find the following relationship among the coefficients
\begin{eqnarray}
\alpha_1 + \alpha_2 q^2  + \alpha_5(\qdotu) + \alpha_6 (\qdotb) &=& 0 \\
\alpha_3 + \alpha_5 q^2  + \alpha_7(\qdotb) &=& 0 \\
\alpha_4 (\qdotb) + \alpha_6q^2 + \alpha_7(\qdotu)   &=& 0 
\end{eqnarray}
which makes only (7-3=4) four of the coefficients independent. The Lorentz invariant form factors $\alpha_i = \alpha_i(q^2,q\cdot u,\qdotb)$ with $i=1,2,..., 7$ can be obtained by contracting both side of the above equations separately with $g_\munu$, $u_\mu u_\nu$, $b_\mu b_\nu$ and $u_\mu b_\nu$ 
so that the form factors will become functions of the Lorentz scalars $\Pi^\mu_{~\mu}$, $u_\mu u_\nu\Pi^\munu$, $b_\mu b_\nu\Pi^\munu$ and $u_\mu b_\nu\Pi^\munu$. Note that, with $q^\mu$, $u^\mu$, $b^\mu$ and $g^\munu$, only four possible 
Lorentz scalars that can be formed by contracting with $\Pi^\munu(q)$ are the quantities $\Pi^\mu_{~\mu}$, $u_\mu u_\nu\Pi^\munu$, $b_\mu b_\nu\Pi^\munu$ and $u_\mu b_\nu\Pi^\munu$ implying the existence of only four form factors. Like the finite temperature case, here the the decomposition of $\Pi^\munu$ is also not unique. One convenient choice of tensor basis could be 
\begin{eqnarray}
P_1^\munu &=& \FB{g^\munu - \frac{q^\mu q^\nu}{q^2} -\frac{\utilde^\mu\utilde^\nu}{\utilde^2} -\frac{\btilde^\mu\btilde^\nu}{\btilde^2} } 
\label{eq.proj.tensor.tb1} \\
P_2^\munu &=& \FB{\frac{\utilde^\mu\utilde^\nu}{\utilde^2}} \label{eq.proj.tensor.tb2} \\
P_3^\munu &=& \FB{\frac{\btilde^\mu\btilde^\nu}{\btilde^2}}  \label{eq.proj.tensor.tb3} \\
Q^\munu &=& \frac{1}{\sqrt{\utilde^2\btilde^2}}\FB{\utilde^\mu\btilde^\nu+\utilde^\nu\btilde^\mu}
\label{eq.proj.tensor.tb4}
\end{eqnarray}
where $\utilde^\mu$ is defined in Eq.~\eqref{eq.utilde} and $\btilde^\mu$ is defined as
\begin{eqnarray}
\btilde^\mu = b^\mu - \frac{(\qdotb)}{q^2}q^\mu - \frac{\bdotut}{\utilde^2}\utilde^\mu~. \label{eq.btilde}
\end{eqnarray}
The basis tensors in Eqs.~\eqref{eq.proj.tensor.tb1}-\eqref{eq.proj.tensor.tb4} satisfy the following relations:
\begin{eqnarray}
&& g_\alphabeta g_\munu P_i^{\mu\alpha} P_j^{\beta\nu} = \delta_{ij} \label{eq.proj.tensor.relation.1}\\
&& g_\alphabeta g_\munu P_i^{\mu\alpha} Q^{\beta\nu} = 0 \\
&& g_\alphabeta g_\munu Q^{\mu\alpha} Q^{\beta\nu} = 2 \\
&& g_\alphabeta P_i^{\mu\alpha} P_j^{\beta\nu} = \delta_{ij} P_i^\munu \\
&& g_\alphabeta Q^{\mu\alpha} Q^{\beta\nu} = P_2^\munu + P_3^\munu \\
&& g_\alphabeta P_1^{\mu\alpha} Q^{\beta\nu} = g_\alphabeta Q^{\mu\alpha} P_1^{\beta\nu} = 0 \\
&& g_\alphabeta P_2^{\mu\alpha} Q^{\beta\nu} = g_\alphabeta Q^{\mu\alpha} P_3^{\beta\nu}  
= \frac{\utilde^\mu\btilde^\nu}{\sqrt{\utilde^2\btilde^2}} \\
&& g_\alphabeta P_3^{\mu\alpha} Q^{\beta\nu} = g_\alphabeta Q^{\mu\alpha} P_2^{\beta\nu}  
= \frac{\utilde^\nu\btilde^\mu}{\sqrt{\utilde^2\btilde^2}} \label{eq.proj.tensor.relation.8}
\end{eqnarray}
Using the basis given in Eqs.~\eqref{eq.proj.tensor.tb1}-\eqref{eq.proj.tensor.tb4}, the self energy at finite temperature under external magnetic field can be written as
\begin{eqnarray}
\Pi^\munu(q) = \Pi_\alpha P_1^\munu + \Pi_\beta P_2^\munu + \Pi_\gamma P_3^\munu + \Pi_\delta Q^\munu
\label{eq.Pi.BT}
\end{eqnarray}
where the form factors are obtained as
\begin{eqnarray}
\Pi_\beta &=& \frac{1}{\utilde^2}u_\mu u_\nu\Pi^\munu \label{eq.Pi.beta}\\
\Pi_\gamma &=& \frac{1}{\btilde^2}\TB{b_\mu b_\nu\Pi^\munu + \frac{(\bdotut)^2}{\utilde^4}u_\mu u_\nu\Pi^\munu 
	-2\frac{(\bdotut)}{\utilde^2}u_\mu b_\nu\Pi^\munu} \label{eq.Pi.gamma} \\
\Pi_\delta &=& \frac{1}{\sqrt{\utilde^2\btilde^2}}\TB{ u_\mu b_\nu\Pi^\munu - \frac{(\bdotut)}{\utilde^2}u_\mu u_\nu\Pi^\munu} 
\label{eq.Pi.delta}\\
\Pi_\alpha &=& \FB{\Pi^\mu_{~\mu}-\Pi_\beta-\Pi_\gamma} \label{eq.Pi.alpha}
\end{eqnarray}
Analogous to the case of only finite temperature, care should be taken while considering the special case $\qper=0$. To see this, let us consider $q_\perp^i=|\vec{q}_\perp|n^i$ with $i=1,2$ so that the following components of self energy at vanishing $\qper$ become (in the LRF)
\begin{eqnarray}
\Pi_{ij}(q^0,\qper=0,q_z) &=& \Pi_\alpha g_{ij}+n_in_j\FB{\Pi_\alpha-\Pi_\gamma} \\
\Pi_{i3}(q^0,\qper=0,q_z) &=& \frac{q^0}{\sqrt{\qpll^2}}n_i\Pi_\delta
\end{eqnarray}
which depend on the direction of $\vec{q}_\perp$ even at $\qper=0$. This ambiguity is eliminated by setting additional constraints on the form factors as 
\begin{eqnarray}
\Pi_\alpha(q^0,\qper=0,q_z)=\Pi_\gamma(q^0,\qper=0,q_z) ~~\text{and}~~ \Pi_\delta(q^0,\qper=0,q_z)= 0~.
\label{eq.additional.constraint}
\end{eqnarray}


\section{The Interacting $\rho$ meson Propagator and its Lorentz Structure} \label{sec.propagator}

Let us first consider the zero temperature and zero magnetic field case for which the complete interacting $\rho$ propagator $D^\munu$ is obtained by solving the Dyson-Schwinger equation
\begin{eqnarray}
D^\munu = \Delta^\munu - \Delta^{\mu\alpha}\Pi_{\alpha\beta}D^{\beta\nu} \label{eq.ds.vac}
\end{eqnarray}
where 
\begin{eqnarray}
\Delta^\munu = \FB{-g^\munu + \frac{q^\mu q^\nu}{m_\rho^2}}\Delta_F(q,m_\rho)
\end{eqnarray}
is the free vacuum Feynman propagator and $\Pi^\munu$ is the one-loop self energy of $\rho$ meson which has the Lorentz structure given  in Eq.~\eqref{eq.pidecom.v} as
\begin{eqnarray}
\Pi^\munu = \FB{g^\munu-\frac{q^\mu q^\nu}{q^2}}\Pi \label{eq.pi.vac}
\end{eqnarray}
with the form factor $\Pi = \frac{1}{3}\Pi^\mu_{~\mu}$. In order to solve Eq.~\eqref{eq.ds.vac}, we rewrite it as 
\begin{eqnarray}
\FB{D^\munu}^{-1} = \FB{\Delta^\munu}^{-1} +\Pi^\munu
\end{eqnarray}
where $\FB{\Delta^\munu}^{-1} = (q^2-m_\rho^2)g^\munu-q^\mu q^\nu$ which satisfies $\Delta^{\mu\alpha}\FB{\Delta_{\alpha\nu}}^{-1} = g^\mu_{~\nu}$. Substituting $\Pi^\munu$ from Eq.~\eqref{eq.pi.vac} in the above equation, we get the inverse of the complete propagator which can be inverted using the relation $D^{\mu\alpha}\FB{D_{\alpha\nu}}^{-1}= g^\mu_{~\nu}$ to obtain the complete propagator as
\begin{eqnarray}
D^\munu(q) = \FB{-g^\munu+\frac{q^\mu q^\nu}{q^2}}\FB{\frac{-1}{q^2-m_\rho^2+\Pi}} - \frac{q^\mu q^\nu }{q^2m_\rho^2}
\end{eqnarray}

We now consider the case of finite temperature and zero magnetic field. As already mentioned in Sec.~\ref{sec.self.med}, in RTF of finite
temperature field theory all the two point correlation functions become $2\times2$ matrices in thermal space. In this case the Dyson-Schwinger equation also becomes a matrix equation~\cite{Mallik:2016anp}
\begin{eqnarray}
\bf{D}^\munu = \bf{\Delta}^\munu - \bf{\Delta}^{\mu\alpha}\bf{\Pi}_{\alpha\beta}\bf{D}^{\beta\nu}~. \label{eq.ds.t}
\end{eqnarray}
Each term of the above equation can be diagonalized in terms of the respective analytic functions (denoted by a bar) so that the above equation becomes an algebric one
\begin{eqnarray}
\overline{D}^\munu = \overline{\Delta}^\munu - \overline{\Delta}^{\mu\alpha}\overline{\Pi}_{\alpha\beta}\overline{D}^{\beta\nu} \label{eq.ds.t2}
\end{eqnarray}
where $\overline{\Delta}^\munu = \Delta^\munu$. The above equation can be rewritten as
\begin{eqnarray}
\FB{\overline{D}^\munu}^{-1} = \FB{\overline{\Delta}^\munu}^{-1} +\overline{\Pi}^\munu~.
\label{eq.ds.t3}
\end{eqnarray}
In this case, the Lorentz structure of the thermal self energy function is given in Eq.~\eqref{eq.Pi.t} as 
\begin{eqnarray}
\overline{\Pi}^\munu(q) = \Pi_1(q^2,\qdotu) P_1^\munu + \Pi_2(q^2,\qdotu) P_2^\munu  \label{eq.Pi.t.2}
\end{eqnarray}
where the projection tensors and form factors are respectively defined in Eqs.~\eqref{eq.proj.tensor.t1} and \eqref{eq.Pi1.Pi2}. 
Substituting the above equation in Eq.~\eqref{eq.ds.t3}, we get the inverse of the complete propagator. In order to obtain the complete propagator, we write 
\begin{eqnarray}
\overline{D}^\munu = A_1 P_1^\munu + A_2 P_2^\munu + \xi q^\mu q^\nu
\end{eqnarray}
and use the relation $\overline{D}^{\mu\alpha}\FB{\overline{D}_{\alpha\nu}}^{-1}= g^\mu_{~\nu}$ to extract $A_1$, $A_2$ and $\xi$. The final form of the complete interacting thermal propagator is obtained as
\begin{eqnarray}
\overline{D}^\munu = \frac{P_1^\munu}{q^2-m_\rho^2+\Pi_1} + \frac{P_2^\munu}{q^2-m_\rho^2+\Pi_2} - \frac{q^\mu q^\nu }{q^2m_\rho^2}
\end{eqnarray}

Finally we consider the case with both finite temperature and external magnetic field. In this case we need to solve the Dyson-Schwinger equation 
\begin{eqnarray}
\FB{\overline{\overline{D}}^\munu}^{-1} = \FB{\overline{\Delta}^\munu}^{-1} +\Pibarbar^\munu~.
\label{eq.ds.tb1}
\end{eqnarray}
where a double bar is used to denote thermal self energy function and complete propagator under external magnetic field as discussed in Sec.~\ref{sec.self.med.eB}. In this case, the Lorentz structure of the thermal self energy function is given in Eq.~\eqref{eq.Pi.BT} as 
\begin{eqnarray}
\Pibarbar^\munu(q) = \Pi_\alpha P_1^\munu + \Pi_\beta P_2^\munu + \Pi_\gamma P_3^\munu + \Pi_\delta Q^\munu
\end{eqnarray}
where the basis tensors and form factors are given in Eqs.~\eqref{eq.proj.tensor.tb1}-\eqref{eq.proj.tensor.tb4} and \eqref{eq.Pi.beta}-\eqref{eq.Pi.alpha}. Substituting the above equation in Eq.~\eqref{eq.ds.tb1}, we get the inverse of the complete propagator. In order to obtain the complete propagator, we write 
\begin{eqnarray}
\overline{\overline{D}}^\munu = A_\alpha P_1^\munu + A_\beta P_2^\munu + A_\gamma P_3^\munu + A_\delta Q^\munu + \xi q^\mu q^\nu
\label{eq.complete.prop}
\end{eqnarray}
and use the relation $\overline{\overline{D}}^{\mu\alpha}\FB{\overline{\overline{D}}_{\alpha\nu}}^{-1}= g^\mu_{~\nu}$ to extract the 
coefficients as
\begin{eqnarray}
A_\alpha &=& \frac{1}{q^2-m_\rho^2+\Pi_\alpha} \label{eq.A_alpha}\\
A_\beta &=& \frac{q^2-m_\rho^2+\Pi_\gamma}{\FB{q^2-m_\rho^2+\Pi_\gamma}\FB{q^2-m_\rho^2+\Pi_\beta}-\Pi_\delta^2} \\
A_\gamma &=& \frac{q^2-m_\rho^2+\Pi_\beta}{\FB{q^2-m_\rho^2+\Pi_\beta}\FB{q^2-m_\rho^2+\Pi_\gamma}-\Pi_\delta^2} \\
A_\delta &=& \frac{-\Pi_\delta}{\FB{q^2-m_\rho^2+\Pi_\beta}\FB{q^2-m_\rho^2+\Pi_\gamma}-\Pi_\delta^2} \\
\xi &=& \frac{-1}{q^2m_\rho^2}~. \label{eq.xi}
\end{eqnarray}


\section{Analytic Structure of the Self Energy} \label{sec.analytic}

In this work, we have considered the transverse momentum of the rho meson to be zero i.e. $q_\perp=0$. As shown in Eq.~\eqref{eq.additional.constraint}, for the special case $q_\perp=0$, the additional constraints to be imposed on the form factors are
\begin{eqnarray}
\Pi_\alpha(q^0,\qper=0,q_z)=\Pi_\gamma(q^0,\qper=0,q_z) ~~\text{and}~~ \Pi_\delta(q^0,\qper=0,q_z)= 0~.
\label{eq.constraints}
\end{eqnarray}
Using the above constraints, we get from Eqs.~\eqref{eq.Pi.beta}-\eqref{eq.Pi.alpha}
\begin{eqnarray}
\Pi_\alpha &=& \Pi_\gamma = \frac{1}{2}\FB{\Pibarbar^\mu_{~\mu}-\frac{1}{\utilde^2}u_\mu u_\nu\Pibarbar^\munu} \label{eq.Pi.alpha.2} \\
\Pi_\beta &=& \frac{1}{\utilde^2}u_\mu u_\nu\Pibarbar^\munu \label{eq.Pi.beta.2}\\
\Pi_\delta &=& 0 \label{eq.Pi.delta.2}
\end{eqnarray}
which imply that we need to calculate only the two quantities quantities $\Pibarbar^\mu_{~\mu}$ and $u_\mu u_\nu\Pibarbar^\munu = \Pibarbar^{00}$. These are obtained from Eqs.~\eqref{eq.repiB.pi.2}-\eqref{eq.impiB.N.2} by contracting them with $g_\munu$ 
and $u_\mu u_\nu$. This essentially means replacing  $\mathcal{N}^\munu$ for all the loops with $\mathcal{N}^\mu_{~\mu}$ or 
$\mathcal{N}^{00}$, an explicit list for which has been provided in Appendix~\ref{app.Nmumu00}.

Let us now discuss the analytic structure of the self energy functions. We first consider the zero magnetic field case. The imaginary part of the self energy function for $\pi\pi$ and NN loops as given in Eqs.~\eqref{eq.impi.pi.1} and \eqref{eq.impi.N.1} each contains four Dirac delta functions. These delta functions represent energy-momentum conservation and they are non vanishing in certain kinematic domain. They are termed as the Unitary-I, Unitary-II, Landau-II and Landau-I cuts as they appear in those equations. The kinematic regions for the Unitary-I and Unitary-II cuts are given by~\cite{Mallik:2016anp} $\sqrt{\vec{q}^2+4m_L^2}<q^0<\infty$ and $-\infty < q^0 < -\sqrt{\vec{q}^2+4m_L^2}$ whereas the same for the two Landau cuts are $|q^0|<|\vec{q}|$ where $m_L$ is the mass of the loop particle i.e. $m_L=m_\pi$ or $m_N$. These cuts correspond to different physical processes such as decay or scattering. For example, Unitary cuts correspond to the decay of $\rho^0$ into a $\pi^+\pi^-$ or $N\bar{N}$ pair and the Landau cuts correspond to the scattering of a $\rho^0$ with a pion or nucleon producing the same in the final state along with their time reversed processes. If we restrict ourselves to the physical timelike kinematic regions defined in terms of $q^0>0$ and $q^2>0$, then only the Unitary-I cut contributes. It is important to note that, a non-trivial Landau cut appears in the physical timelike region only if the loop particles have different masses and lie in the kinematic domain $|\vec{q}|<q^0 < \sqrt{\vec{q}^2+\Delta m^2}$ where $\Delta m$ is the mass difference of the loop particles.

Let us now consider the case of both finite temperature and non zero external magnetic field. In this
case the imaginary parts of the self energy as given in Eqs.~\eqref{eq.impiB.pi.2} and \eqref{eq.impiB.N.2} 
also contain  four Dirac delta functions corresponding to the Unitary and Landau cuts. It is important to 
note that the arguments of the delta functions contain only the longitudinal dynamics (because of dimensional 
reduction) which implies that the analytic structure of the self energy functions will only depend on the 
longitudinal momentum of $\rho$. On the other hand, the transverse dynamics has appeared as Landau level dependent 
``dimensionally reduced effective mass" to the loop particles as given in Eq.~\eqref{eq.dim.mass}. Therefore, even 
if the loop particles have the same masses, a non-trivial Landau cut may appear in the physical timelike kinematic 
domain if the  two loop particles reside in different Landau levels. Physically, this means that $\rho^0$ can get 
absorbed in a scattering with a pion or a proton in a lower Landau level producing another pion or proton in a higher Landau level as the final state.
A detailed discussions on the analytic structure in presence of external magnetic field can be found in Refs.~\cite{Ghosh:2017rjo,Ghosh:2018xhh}. 
The Unitary-I and Unitary-II terms for the $\pi\pi$ loop are non-vanishing in the kinematic domains $\sqrt{q_z^2+4(m_\pi^2+eB)} < q^0 < \infty$ and $-\infty < q^0 < -\sqrt{q_z^2+4(m_\pi^2+eB)}$ whereas the kinematic domain for both the Landau cuts is
\begin{eqnarray}
|q^0|<\sqrt{q_z^2+(\sqrt{m_\pi^2+eB}-\sqrt{m_\pi^2+3eB})^2}~. \label{eq.LC.threshold.pipi}
\end{eqnarray}
The corresponding kinematic domains for the NN loop are $\sqrt{q_z^2+4m_N^2} < q^0 < \infty$ and $-\infty < q^0 < -\sqrt{q_z^2+4m_N^2}$ for the Unitary-I and Unitary-II cuts respectively and  
\begin{eqnarray}
|q^0|<\sqrt{q_z^2+(m_N-\sqrt{m_N^2+2eB})^2} \label{eq.LC.threshold.pp}
\end{eqnarray}
for the Landau cuts. Note that, the threshold of the Landau cuts appears when the ``dimensionally reduced effective mass" difference between the loop particles is the maximum. As can be seen from Eqs.~\eqref{eq.impiB.pi.2} and \eqref{eq.impiB.N.2}, for a particular value of the index $n$, the sum over the index $l$ runs only for three values $(n-1)$, $n$ and $(n+1)$ which implies that, the Landau level difference between the loop particles can be at most one. Thus the maximum difference in their ``dimensionally reduced effective mass" appears when one of them is at the lowest Landau level and the other one is at the first Landau level which in turn defines the Landau cut threshold in Eqs.~\eqref{eq.LC.threshold.pipi} and \eqref{eq.LC.threshold.pp}.

We now simplify the expressions of the imaginary parts given in Eqs.~\eqref{eq.impi.pi.1}, \eqref{eq.impi.N.1}, \eqref{eq.impiB.pi.2} and \eqref{eq.impiB.N.2} by evaluating one of the integrals using the Dirac delta functions. For the imaginary parts at zero magnetic field, 
we evaluate the $d(\cos\theta)$ integrals and get (after imposing the kinematic restrictions discussed above), 
\begin{eqnarray}
\IM\Pibar_{\pi,\tN}^\munu(q^0,\vec{q}) &=& -\sign{q^0}\tanh\FB{\frac{q^0}{2T}}
\frac{1}{16\pi|\vec{q}|}\TB{\int_{\omega_-}^{\omega_+}d\FB{\omega_k,\Omega_k} \FB{U_1^{\pi,\tN}}^\munu(\cos\theta=\cos\theta_0^{\pi,\tN})
	\Theta\FB{q^0-\sqrt{\vec{q}^2+4m_{\pi,\tN}^2}} \right. \nn \\ && \left.
	+\int_{-\omega_+}^{-\omega_-}d\omega_k \FB{U_2^{\pi,\tN}}^\munu(\cos\theta=\cos\theta_0'^{\pi,\tN})
	\Theta\FB{-q^0-\sqrt{\vec{q}^2+4m_{\pi,\tN}^2}} \right. \nn \\ && \left. +
	\int_{-\omega_+}^{\infty}d\omega_k \FB{L_1^{\pi,\tN}}^\munu(\cos\theta=\cos\theta_0'^{\pi,\tN})\Theta\FB{-|q^0|+|\vec{q}|}
	\right. \nn \\ && \left.
	+ \int_{\omega_-}^{\infty}d\omega_k \FB{L_2^{\pi,\tN}}^\munu(\cos\theta=\cos\theta_0^{\pi,\tN})\Theta\FB{-|q^0|+|\vec{q}|}}
\label{eq.impi.t.simple.1}
\end{eqnarray}
where, 
\begin{eqnarray}
\omega_\pm = \left\{\begin{array}{c} \frac{1}{2q^2}\TB{q^0q^2\pm|\vec{q}|\lambda^{1/2}\FB{q^2,m_\pi^2,m_\pi^2}} ~~\text{for}~\pi\pi~\text{loop} \\ \frac{1}{2q^2}\TB{q^0q^2\pm|\vec{q}|\lambda^{1/2}\FB{q^2,m_N^2,m_N^2}} ~~\text{for NN loop} \end{array} \right.~,
\end{eqnarray}
\begin{eqnarray}
\FB{U_1^\pi}^\munu &=& \SB{1+f(\omega_k)+f(\omega_p)+2f(\omega_k)f(\omega_p)} N_\pi^\munu (k^0=-\omega_k)~, \label{eq.U1.t.pi}\\
\FB{U_2^\pi}^\munu &=& \SB{1+f(\omega_k)+f(\omega_p)+2f(\omega_k)f(\omega_p)} N_\pi^\munu (k^0=\omega_k)~, \\
\FB{L_1^\pi}^\munu &=& \SB{f(\omega_k)+f(\omega_p)+2f(\omega_k)f(\omega_p)} N_\pi^\munu (k^0=\omega_k)~, \\
\FB{L_2^\pi}^\munu &=& \SB{f(\omega_k)+f(\omega_p)+2f(\omega_k)f(\omega_p)} N_\pi^\munu (k^0=-\omega_k)~,
\end{eqnarray}
\begin{eqnarray}
\FB{U_1^\tN}^\munu &=& \SB{1-f^-(\Omega_k)-f^+(\Omega_p)+2f^-(\Omega_k)f^+(\Omega_p)} N_\tN^\munu (k^0=-\Omega_k)~, \label{eq.U1.t.N} \\
\FB{U_2^\tN}^\munu &=& \SB{1-f^+(\Omega_k)-f^-(\Omega_p)+2f^+(\Omega_k)f^-(\Omega_p)} N_\tN^\munu (k^0=\Omega_k)~, \\
\FB{L_1^\tN}^\munu &=& \SB{-f^+(\Omega_k)-f^+(\Omega_p)+2f^+(\Omega_k)f^+(\Omega_p)} N_\tN^\munu (k^0=\Omega_k)~, \\
\FB{L_2^\tN}^\munu &=& \SB{-f^-(\Omega_k)-f^-(\Omega_p)+2f^-(\Omega_k)f^-(\Omega_p)} N_\tN^\munu (k^0=-\Omega_k)~,
\end{eqnarray}
\begin{eqnarray}
\cos\theta_0^{\pi} &=& \FB{\frac{-2q^0\omega_k+q^2}{2|\vec{q}||\vec{k}|}}~, \\
\cos\theta_0'^{\pi} &=& \FB{\frac{2q^0\omega_k+q^2}{2|\vec{q}||\vec{k}|}}~, \\
\cos\theta_0^{\tN} &=& \FB{\frac{-2q^0\Omega_k+q^2}{2|\vec{q}||\vec{k}|}}~~\text{and} \\
\cos\theta_0'^{\tN} &=& \FB{\frac{2q^0\Omega_k+q^2}{2|\vec{q}||\vec{k}|}}~,
\end{eqnarray}
with $\lambda(x,y,z)=x^2+y^2+z^2-2xy-2yz-2zx$ being the K\"all\'en function.

For the imaginary parts at finite magnetic field, we evaluate the $dk_z$ integrals in Eqs.~\eqref{eq.impiB.pi.2} and \eqref{eq.impiB.N.2}  using the Dirac delta functions. The imaginary part due to $\pi\pi$ loop simplifies to 
\begin{eqnarray}
\IM\Pibarbar^\munu_\pi(q^0,q_z) &=& -\sign{q^0}\tanh\FB{\frac{q^0}{2T}}
\sum_{n=0}^{\infty} ~\sum_{l=(n-1)}^{(n+1)}\frac{1}{4\lambda^{1/2}(q_\parallel^2,m_{l}^2,m_{n}^2)} \nn \\ && \sum_{\tilde{k}_z\in\tilde{k}_z^\pm}^{}
\TB{\frac{}{}\FB{\tilde{U}^\pi_{1,nl}}^\munu(k_z=\tilde{k}_z)	\Theta\FB{q^0-\sqrt{q_z^2+(m_{l}+m_{n})^2}}
		\right. \nn \\ && \hspace{0cm}\left.
	+ \FB{\tilde{U}^\pi_{2,nl}}^\munu(k_z=\tilde{k}_z) \Theta\FB{-q^0-\sqrt{q_z^2+(m_{l}+m_{n})^2}}
	\right. \nn \\ && \hspace{0cm}\left.
	+ \FB{\tilde{L}^\pi_{1,nl}}^\munu(k_z=\tilde{k}_z) \Theta\FB{q^0-\min\FB{q_z,E_\pm}}\Theta\FB{-q^0+\max\FB{q_z,E_\pm}}
	\right. \nn \\ && \hspace{0cm}\left.	 
	+ \FB{\tilde{L}^\pi_{2,nl}}^\munu(k_z=\tilde{k}_z) \Theta\FB{-q^0-\min\FB{q_z,E_\pm}}\Theta\FB{q^0+\max\FB{q_z,E_\pm}\frac{}{}}
} \label{eq.impi.simple.tb.1}
\end{eqnarray}
where, 
\begin{eqnarray}
\FB{\tilde{U}^\pi_{1,nl}}^\munu &=& \SB{1+f(\omegatilde_k^l)+f(\omegatilde_p^n)+2f(\omegatilde_k^l)f(\omegatilde_p^n)}\tilde{N}^\munu_{\pi,nl} (k^0=-\omegatilde_k^l)~, 
\label{eq.U1.tb.pi} \\
\FB{\tilde{U}^\pi_{1,nl}}^\munu &=& \SB{1+f(\omegatilde_k^l)+f(\omegatilde_p^n)+2f(\omegatilde_k^l)f(\omegatilde_p^n)} \tilde{N}^\munu_{\pi,nl} (k^0=\omegatilde_k^l)~, \\
\FB{\tilde{U}^\pi_{1,nl}}^\munu &=& \SB{f(\omegatilde_k^l)+f(\omegatilde_p^n)+2f(\omegatilde_k^l)f(\omegatilde_p^n)} \tilde{N}^\munu_{\pi,nl} (k^0=\omegatilde_k^l)~, \\
\FB{\tilde{U}^\pi_{1,nl}}^\munu &=& \SB{f(\omegatilde_k^l)+f(\omegatilde_p^n)+2f(\omegatilde_k^l)f(\omegatilde_p^n)} \tilde{N}^\munu_{\pi,nl} (k^0=-\omegatilde_k^l)~,
\end{eqnarray}
with, $\tilde{k}_z^\pm = \frac{1}{2q_\parallel^2}\TB{-yq_z\pm|q^0|\lambda^{1/2}\FB{q_\parallel^2,m_{l}^2,m_{n}^2}}$, 
$y=(q_\parallel^2+m_{l}^2-m_{n}^2)$, $\tilde{\omega}_k^l = \sqrt{\tilde{k}_z^2+m_{l}^2}$, 
and $E_\pm = \frac{m_{l}-m_{n}}{\MB{m_{l}\pm m_{n}}}\sqrt{q_z^2+(m_{l}\pm m_{n})^2}$.

The corresponding expression of the imaginary part due to NN loop reads
\begin{eqnarray}
\IM\Pibarbar^\munu_\tN(q^0,q_z) &=& \frac{1}{2}\IM\Pibar^\munu_\tN(q^0,q_z) -\sign{q^0}\tanh\FB{\frac{q^0}{2T}}
\sum_{n=0}^{\infty} ~\sum_{l=(n-1)}^{(n+1)}\frac{1}{4\lambda^{1/2}(q_\parallel^2,M_{l}^2,M_{n}^2)} \nn \\ && \sum_{\tilde{k}_z\in\tilde{K}_z^\pm}^{}
\TB{\frac{}{}\FB{\tilde{U}^\tp_{1,nl}}^\munu(k_z=\tilde{k}_z)	\Theta\FB{q^0-\sqrt{q_z^2+(M_{l}+M_{n})^2}}
	\right. \nn \\ && \hspace{0cm}\left.
	+ \FB{\tilde{U}^\tp_{2,nl}}^\munu(k_z=\tilde{k}_z) \Theta\FB{-q^0-\sqrt{q_z^2+(M_{l}+M_{n})^2}}
	\right. \nn \\ && \hspace{0cm}\left.
	+ \FB{\tilde{L}^\tp_{1,nl}}^\munu(k_z=\tilde{k}_z) \Theta\FB{q^0-\min\FB{q_z,E'_\pm}}\Theta\FB{-q^0+\max\FB{q_z,E'_\pm}}
	\right. \nn \\ && \hspace{0cm}\left.	 
	+ \FB{\tilde{L}^\tp_{2,nl}}^\munu(k_z=\tilde{k}_z) \Theta\FB{-q^0-\min\FB{q_z,E'_\pm}}\Theta\FB{q^0+\max\FB{q_z,E'_\pm}\frac{}{}}
} \label{eq.impi.simple.tb.2}
\end{eqnarray}
where, 
\begin{eqnarray}
\FB{\tilde{U}^\tp_{1,nl}}^\munu &=& \SB{1-f^-(\Omegatilde_k^l)-f^+(\Omegatilde_p^n)+2f^-(\Omegatilde_k^l)f^+(\Omegatilde_p^n)}\tilde{N}^\munu_{\tp,nl} (k^0=-\Omegatilde_k^l)~, \\
\FB{\tilde{U}^\tp_{1,nl}}^\munu &=& \SB{1-f^+(\Omegatilde_k^l)-f^-(\Omegatilde_p^n)+2f^+(\Omegatilde_k^l)f^-(\Omegatilde_p^n)} \tilde{N}^\munu_{\tp,nl} (k^0=\Omegatilde_k^l)~, \\
\FB{\tilde{U}^\tp_{1,nl}}^\munu &=& \SB{-f^+(\Omegatilde_k^l)-f^+(\Omegatilde_p^n)+2f(\Omegatilde_k^l)f(\Omegatilde_p^n)} \tilde{N}^\munu_{\tp,nl} (k^0=\Omegatilde_k^l)~, \\
\FB{\tilde{U}^\tp_{1,nl}}^\munu &=& \SB{-f^-(\Omegatilde_k^l)-f^-(\Omegatilde_p^n)+2f(\Omegatilde_k^l)f(\Omegatilde_p^n)} \tilde{N}^\munu_{\tp,nl} (k^0=-\Omegatilde_k^l)~,
\end{eqnarray}
with, $\tilde{K}_z^\pm = \frac{1}{2q_\parallel^2}\TB{-Yq_z\pm|q^0|\lambda^{1/2}\FB{q_\parallel^2,M_{l}^2,M_{n}^2}}$, 
$Y=(q_\parallel^2+M_{l}^2-M_{n}^2)$, $\tilde{\Omega}_k^l = \sqrt{\tilde{K}_z^2+M_{l}^2}$, 
and ~~ $E'_\pm = \frac{M_{l}-M_{n}}{\MB{M_{l}\pm M_{n}}}\sqrt{q_z^2+(M_{l}\pm M_{n})^2}$. The first term on the RHS of Eq.~\eqref{eq.impi.simple.tb.2} is the contribution from the neutron-neutron loop (which is not affected by the external magnetic field )
whose simplified form is given in Eq.~\eqref{eq.impi.t.simple.1}.

\section{Numerical Results} \label{sec.numerical}

\begin{figure}[h]
	\begin{center}
		\includegraphics[angle=-90,scale=0.35]{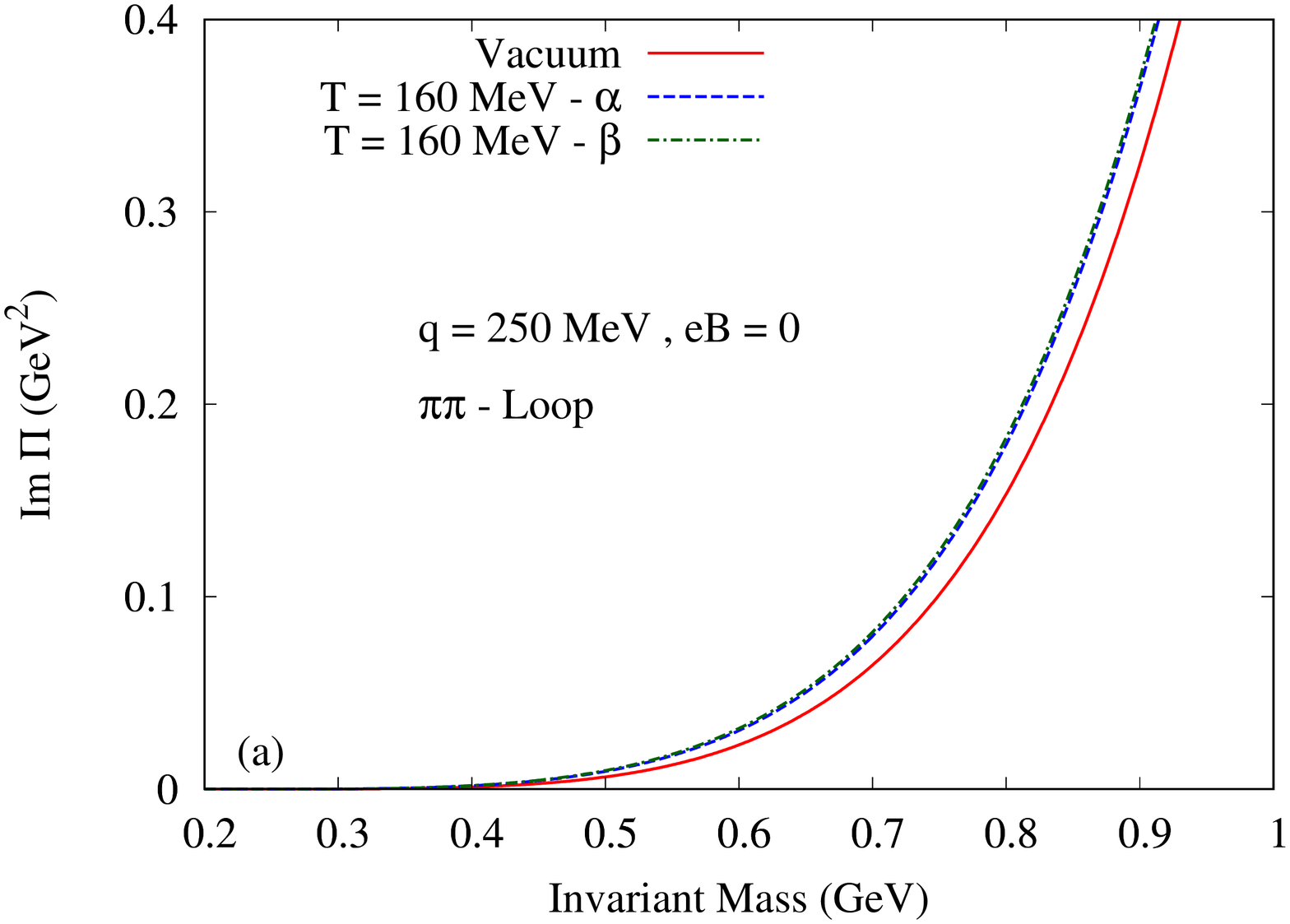} 	
		\includegraphics[angle=-90,scale=0.35]{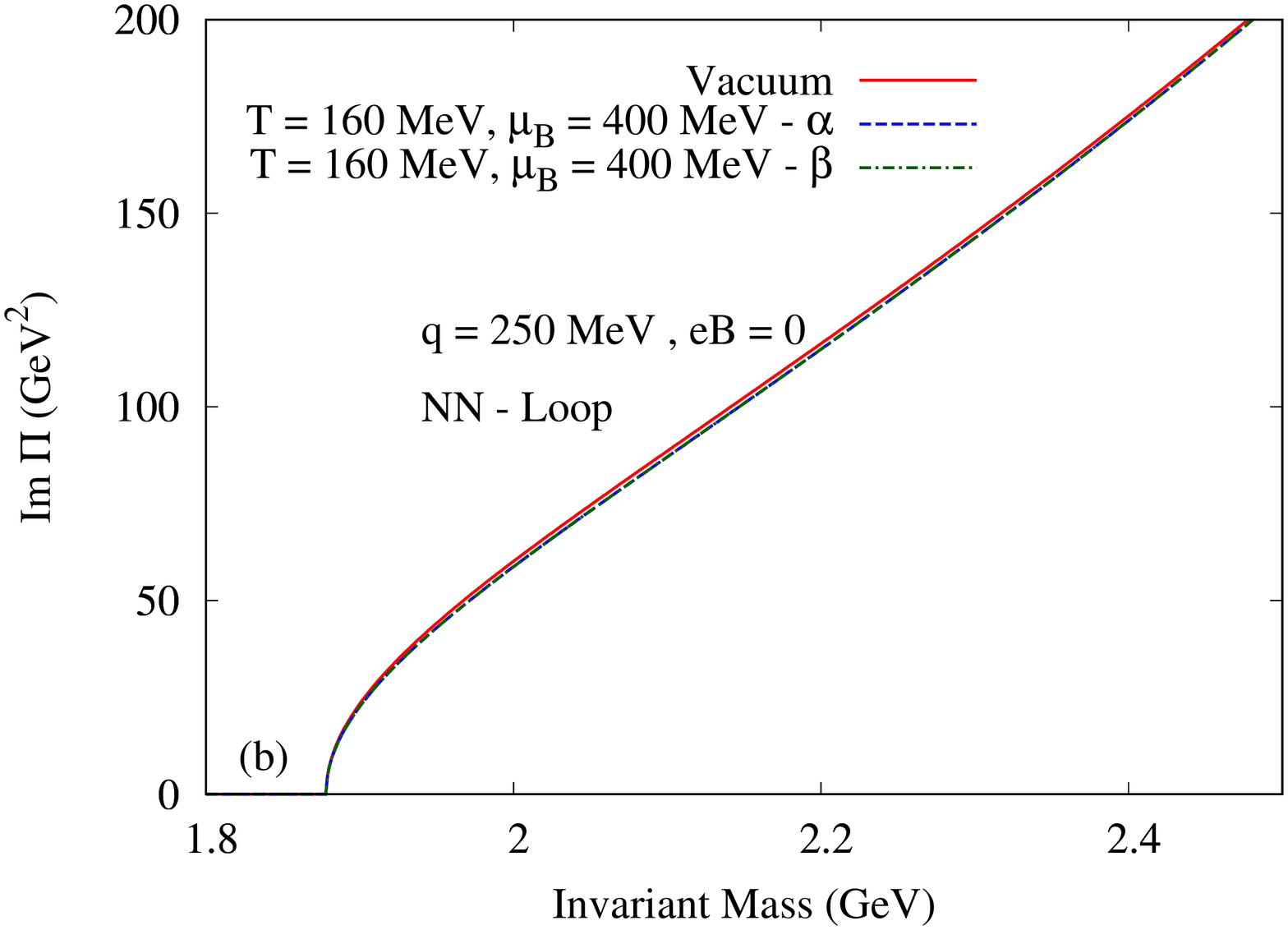}
	\end{center}
	\caption{Imaginary part of the self energy of $\rho^0$ as a function of invariant mass at zero magnetic field and at $\rho^0$ three momentum 
	$|\hspace{-0.05cm}\vec{\hspace{0.05cm}q}|=250$ MeV. The vacuum self energy for $T=\mu_B=0$ is compared with the in-medium one obtained at temperature $T=160$ MeV and baryon chemical potential $\mu_B = 400$ MeV for the (a) $\pi\pi$ loop and (b) NN Loop.}
	\label{fig.impi.eB0}
\end{figure}

\begin{figure}[h]
	\begin{center}
		\includegraphics[angle=-90,scale=0.35]{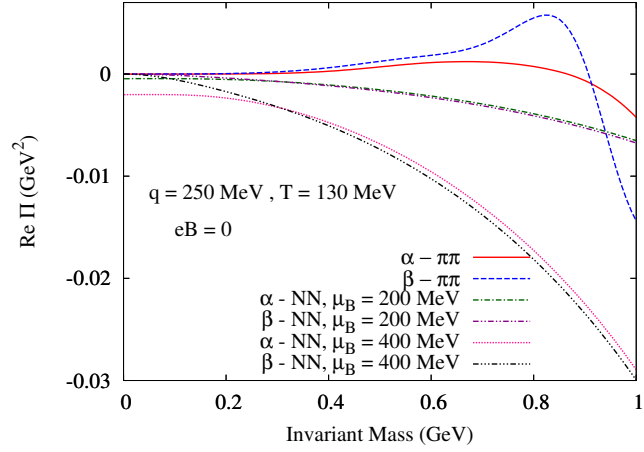}
	\end{center}
	\caption{Real part of the self energy of $\rho^0$  as a function of invariant mass  at zero magnetic field and at temperature $T=130$ MeV with $\rho^0$ three momentum $|\hspace{-0.05cm}\vec{\hspace{0.05cm}q}|=250$ MeV. The contributions from NN loop is shown for two different values of baryon chemical potential ($\mu_B=$ 200 and 400 MeV respectively).}
	\label{fig.repi.eB0}
\end{figure}

We begin this section by presenting the real and imaginary parts of the in-medium self energy functions of $\rho^0$. As can be seen from Eqs.~\eqref{eq.Pi.alpha}-\eqref{eq.Pi.delta.2}, we have only two non-zero form factors for the self energy which are $\Pi_\alpha$ and $\Pi_\beta$ for $q_\perp=0$. Let us first consider the zero magnetic field case for which the imaginary and real parts of $\Pi_\alpha$ and $\Pi_\beta$ are depicted in Figs.~\ref{fig.impi.eB0} and \ref{fig.repi.eB0} respectively.
In Fig.~\ref{fig.impi.eB0}(a), $\IM\Pi_\alpha$ and $\IM\Pi_\beta$ due to $\pi\pi$ loop are plotted as a function of invariant mass ($\sqrt{q^2}$) of $\rho^0$ for vacuum as well as for medium ($T=160$ MeV and $\mu_B=400$ MeV) with $q_z=250$ MeV. It is to be understood that in the case of vacuum the two form factors are equal. In this case, the only contribution comes from the Unitary-I cut which starts at $2m_\pi$ in the invariant mass axis. With the increase in temperature , the degeneracy between the form factor get lifted as well as they are enhanced with respect to the vacuum. This is due to the enhancement of the thermal factor in Eq.~\eqref{eq.U1.t.pi} which increases the available phase space with the increase in temperature. The corresponding results for the NN loop is shown in Fig.~\ref{fig.impi.eB0}(b) for which the threshold of the Unitary-I cut is $2m_N$. In this case, with the increase in temperature and density, the imaginary part decreases slightly with respect to the vacuum which can be understood from Eq.~\eqref{eq.U1.t.N} where, because of the negative signs in front of the thermal distribution functions of the nucleons, the thermal factor reduces with the increase in temperature thus showing opposite behaviour as compared to the $\pi\pi$ loop.

In Fig.~\ref{fig.repi.eB0}, $\RE\Pi_\alpha$ and $\RE\Pi_\beta$ are shown as a function of $\rho^0$ invariant mass at zero external magnetic field with $\rho^0$ longitudinal momentum $q_z=250$ MeV at temperature $T=130$ MeV. For the $\pi\pi$ loop, the real part is positive at low invariant mass and becomes negative in the high invariant mass region in contrast to the NN loop for which the contribution to the real part is always negative. The real part due to NN loop is shown for two different values of baryon chemical potential $\mu_B=$ 200 and 400 MeV respectively. For low values of $\mu_B$, the contribution of the NN loop is almost of the  same order as $\pi\pi$ loop, however at high $\mu_B$, the contribution from NN loop dominates over the $\pi\pi$ loop. 
\begin{figure}[h]
	\begin{center}
		\includegraphics[angle=-90,scale=0.35]{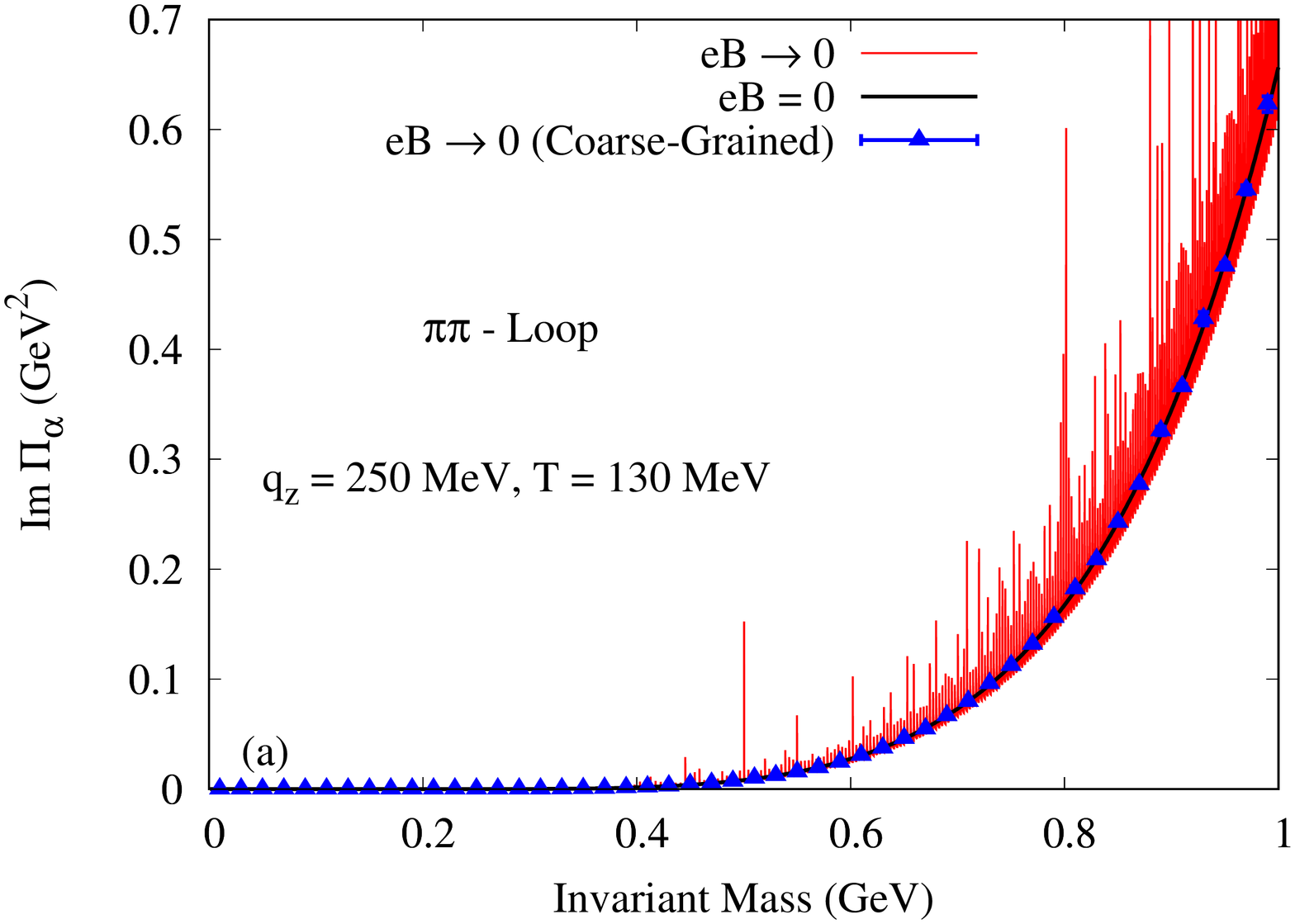} 	
		\includegraphics[angle=-90,scale=0.35]{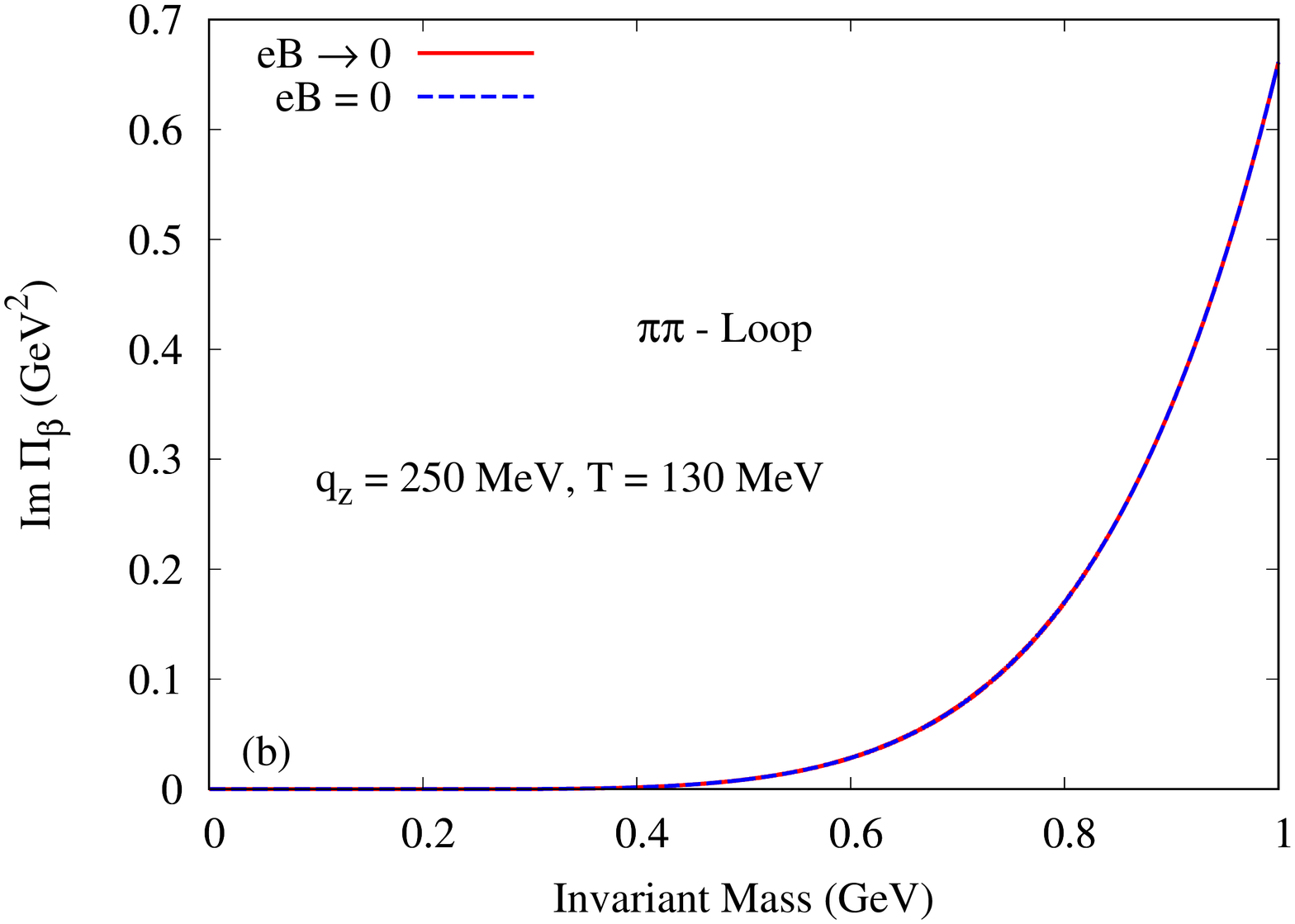} \\
		\includegraphics[angle=-90,scale=0.35]{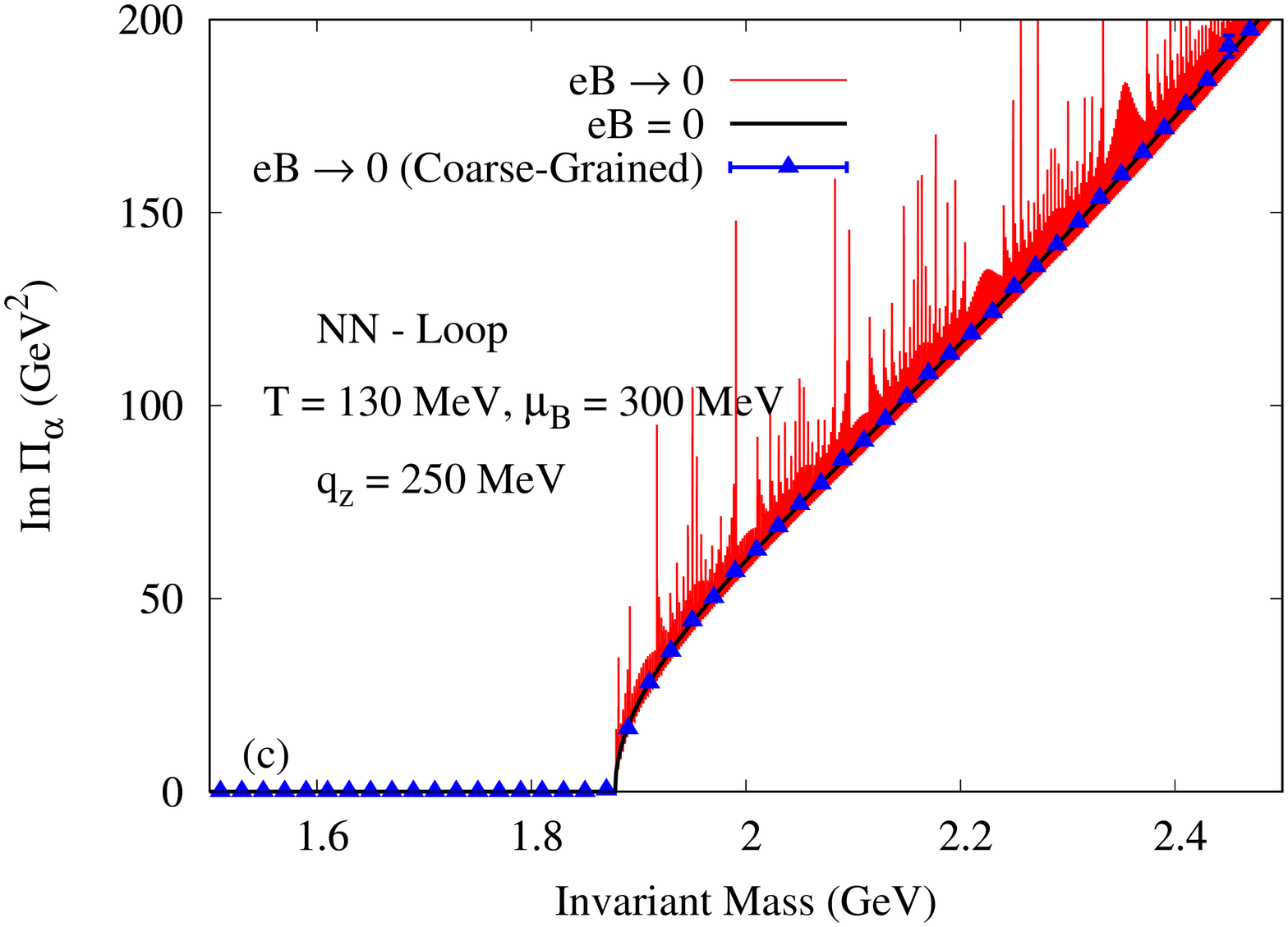} 	
		\includegraphics[angle=-90,scale=0.35]{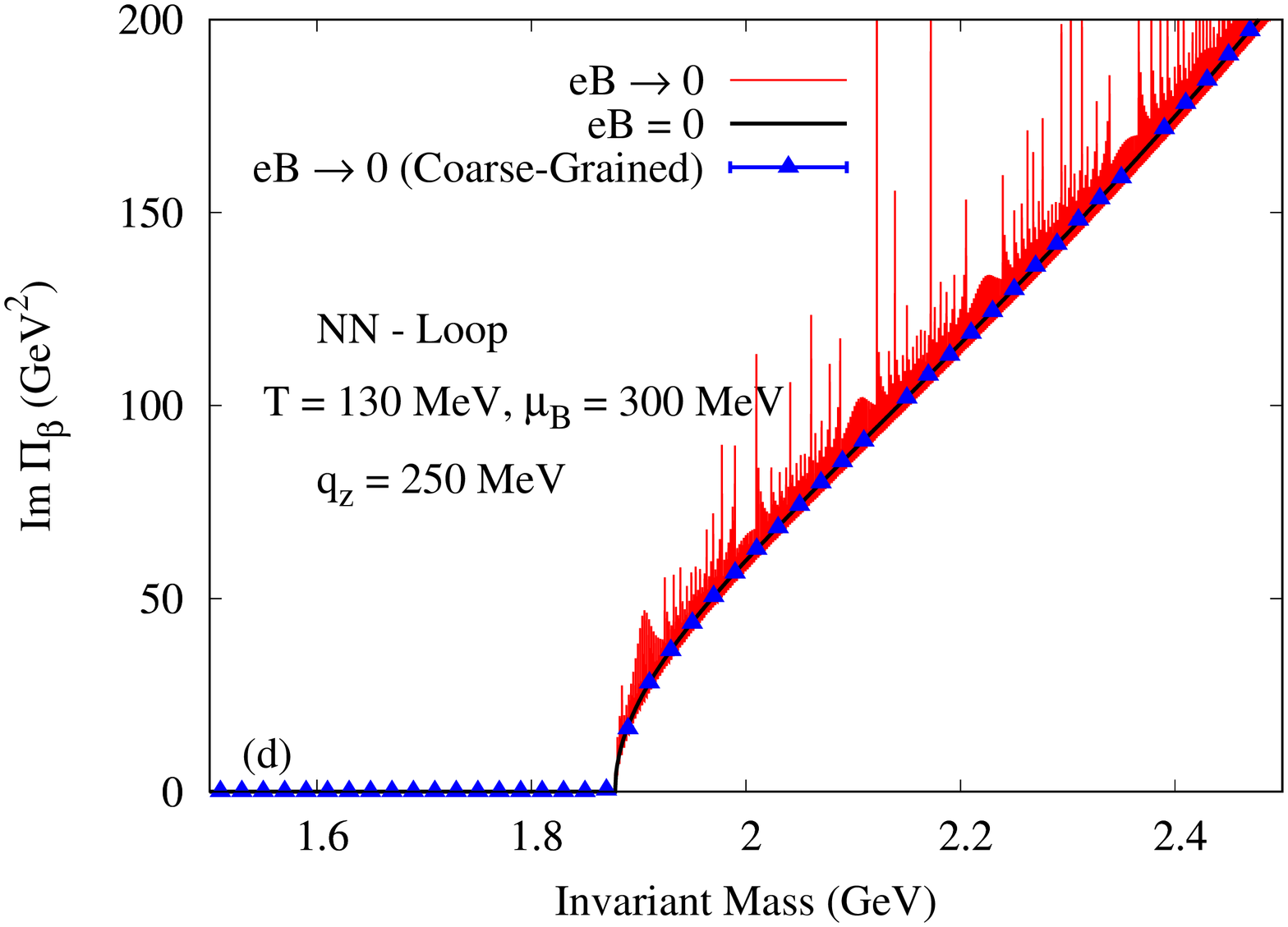}
	\end{center}
	\caption{The imaginary part of the form factors  as a function of the invariant mass at $eB=0$ have been compared with the imaginary part at non zero magnetic field in the  numerical limit $eB\rightarrow 0$ at temperature $T=130$ MeV and at baryon chemical potential $\mu_B=300$ MeV with $\rho^0$ longitudinal momentum $q_z=250$ MeV. The contribution due the $\pi\pi$ loop from the form factors $\Pi_\alpha$ and  $\Pi_\beta$ are shown in panels (a) and (b) respectively.  The corresponding contributions due the NN loop are shown in panels (c) and (d). The respective coarse-grained (CG) quantities from the $eB\rightarrow 0$ results are also shown in (a), (c) and (d).}
	\label{fig.impi.eB0.limit}
\end{figure}
\begin{figure}[h]
	\begin{center}
		\includegraphics[angle=-90,scale=0.35]{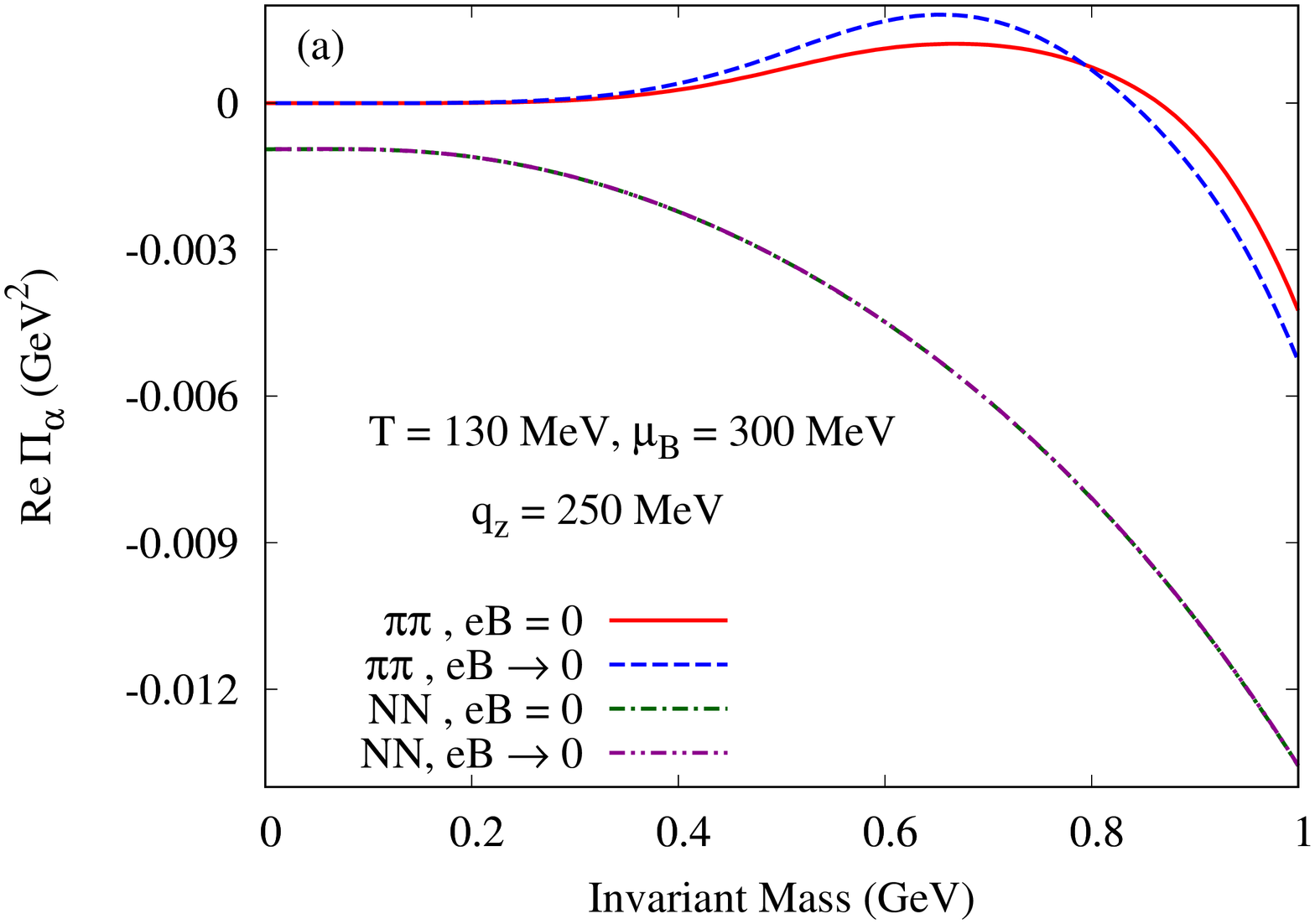}
		\includegraphics[angle=-90,scale=0.35]{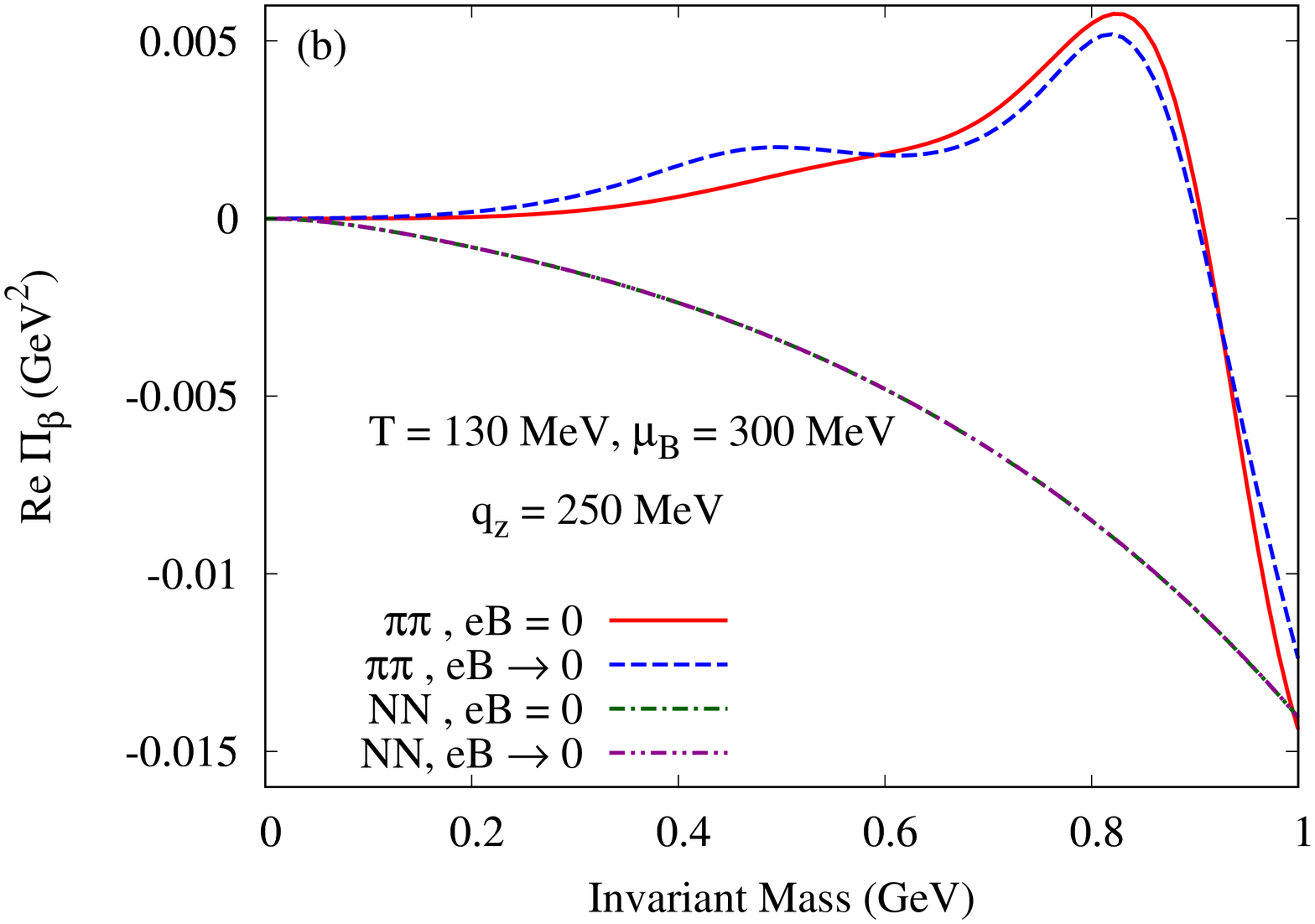}
	\end{center}
	\caption{The real part of the form factors  as a function of the invariant mass at $eB=0$ have been compared with the real part at non zero magnetic field in the  numerical limit $eB\rightarrow 0$ at temperature $T=130$ MeV and at baryon chemical potential $\mu_B=300$ with $\rho^0$ longitudinal momentum $q_z=250$ MeV. The contribution from the form factors (a) $\Pi_\alpha$ and (b) $\Pi_\beta$ are shown separately due to $\pi\pi$ and NN loop.}
	\label{fig.repi.eB0.limit}
\end{figure}

We now turn on the external magnetic field. For the check of consistency of the calculation at non-zero magnetic field, it is essential that  $eB\rightarrow0$ limit of non-zero magnetic field results reproduces the $eB=0$ one. In order to take the $eB\rightarrow0$ limit numerically, we have considered upto 500 Landau levels for a convergent result. We have shown the imaginary part of the self energy as a function of invariant mass of $\rho^0$ with longitudinal momentum $q_z=250$ MeV at temperature $T=130$ MeV and at baryon chemical potential $\mu_B=300$ MeV for the two cases: $eB=0$ and $eB\rightarrow0$ in Fig.~\ref{fig.impi.eB0.limit} separately for the $\pi\pi$ and NN loops. Fig.~\ref{fig.impi.eB0.limit}(a) shows $\IM\Pi_\alpha$ for the $\pi\pi$ loop in which the $eB\rightarrow0$ graph has a series of spikes infinitesimally separated from each other all over the whole invariant mass region whereas the $eB=0$ graph is finite and well behaved. Interestingly, the $eB\rightarrow0$ graph does not miss the $eB=0$ curve which implies that when average is done, the $eB =0$ line will be exactly reproduced. The appearance of these spikes are due to the ``threshold singularities"~\cite{Chakraborty:2017vvg,Ghosh:2017rjo,Ghosh:2018xhh} at each Landau level as can be understood from Eq.~\eqref{eq.impi.simple.tb.1} where the K\"all\'en function goes to zero at each threshold of the Unitary and Landau cuts defined in terms of the unit step functions therein, which is a consequence of the dimensional reduction. In order to extract physical and finite results out of these spikes, we have used Ehrenfest's coarse-graining (CG)~\cite{Gorban,Ehrenfest,Ghosh:2018xhh}. In this method, the whole invariant mass region has been discretized in small bins followed by  bin averages. In other words, the self energy at a given $\sqrt{q_\parallel^2}$ is approximated by its average over the neighbourhood around that point. This in turn smears out the spike like structures. As can be seen in the figure, after CG,  $\IM\Pi_\alpha$ exactly matches with the analytic $eB=0$ graph. The corresponding comparison of $eB\rightarrow0$ and $eB=0$ result for  $\IM\Pi_\beta$ due to $\pi\pi$ loop is shown in Fig.~\ref{fig.impi.eB0.limit}(b). In this case, $eB\rightarrow0$ graph is finite and free from the threshold singularities and it matches exactly with the $eB=0$ graph. The absence of the threshold singularities in this case is due to an overall factor of K\"all\'en functions coming from $\tilde{\mathcal{N}}^{00}_{\pi,nl}$ in Eq.~\eqref{eq.U1.tb.pi} which cancels the K\"all\'en functions in the denominator of Eq.~\eqref{eq.impi.simple.tb.1}. Thus the $\IM\Pi_\beta$ due to the $\pi\pi$ loop does not require to be coarse grained.

The corresponding results for the NN loop is depicted in Figs.~\ref{fig.impi.eB0.limit}(c) and \ref{fig.impi.eB0.limit}(d). In this case, both the $\IM\Pi_\alpha$ and $\IM\Pi_\beta$ suffer threshold singularities as there is no overall K\"all\'en functions coming from $\tilde{\mathcal{N}}^{\munu}_{\tp,nl}$. So both the form factors have to be coarse grained after which they exactly reproduce the $eB=0$ graphs.

We now turn our attention to the real part of the self energy at non-zero magnetic field and show how a numerical limit of $eB\rightarrow0$ agrees with the $eB=0$ results. This has been shown in Fig.~\ref{fig.repi.eB0.limit} where the real part of the form factors is shown as a function of $\rho^0$ invariant mass with longitudinal momentum $q_z=250$ MeV at temperature $T=130$ MeV and at baryon chemical potential $\mu_B=300$ MeV for the two cases $eB\rightarrow0$ and $eB=0$. The contributions from the $\pi\pi$ and NN loops are shown separately. Fig.~\ref{fig.repi.eB0.limit}(a) depicts $\RE\Pi_\alpha$ whereas Fig.~\ref{fig.repi.eB0.limit}(b) shows $\RE\Pi_\beta$. As can be seen from the figure, the $eB\rightarrow0$ graphs exactly reproduce the $eB=0$ for the case of NN loop. Whereas, for the $\pi\pi$ loop, $eB\rightarrow0$ is slightly deviated from the $eB=0$ graph but with an excellent qualitative agreement in their behaviour with respect to the variation of invariant mass of $\rho^0$. This small disagreement between the $eB\rightarrow0$ and $eB=0$ graph is due to the inaccuracy in the numerical principal value integration of Eqs.~\eqref{eq.repi.pi.1} and \eqref{eq.repiB.pi.2} for which the two particle bound state threshold $\sqrt{q^2_\parallel}>2m_\pi = 280$ MeV is less than the $\rho^0$ mass pole $m_\rho=0.770$ (in contrast, for the NN loop, the two particle bound state threshold is at $\sqrt{q^2_\parallel}>2m_N = 1.878$ GeV much higher than the range of the plot). 
\begin{figure}[h]
	\begin{center}
		\includegraphics[angle=-90,scale=0.35]{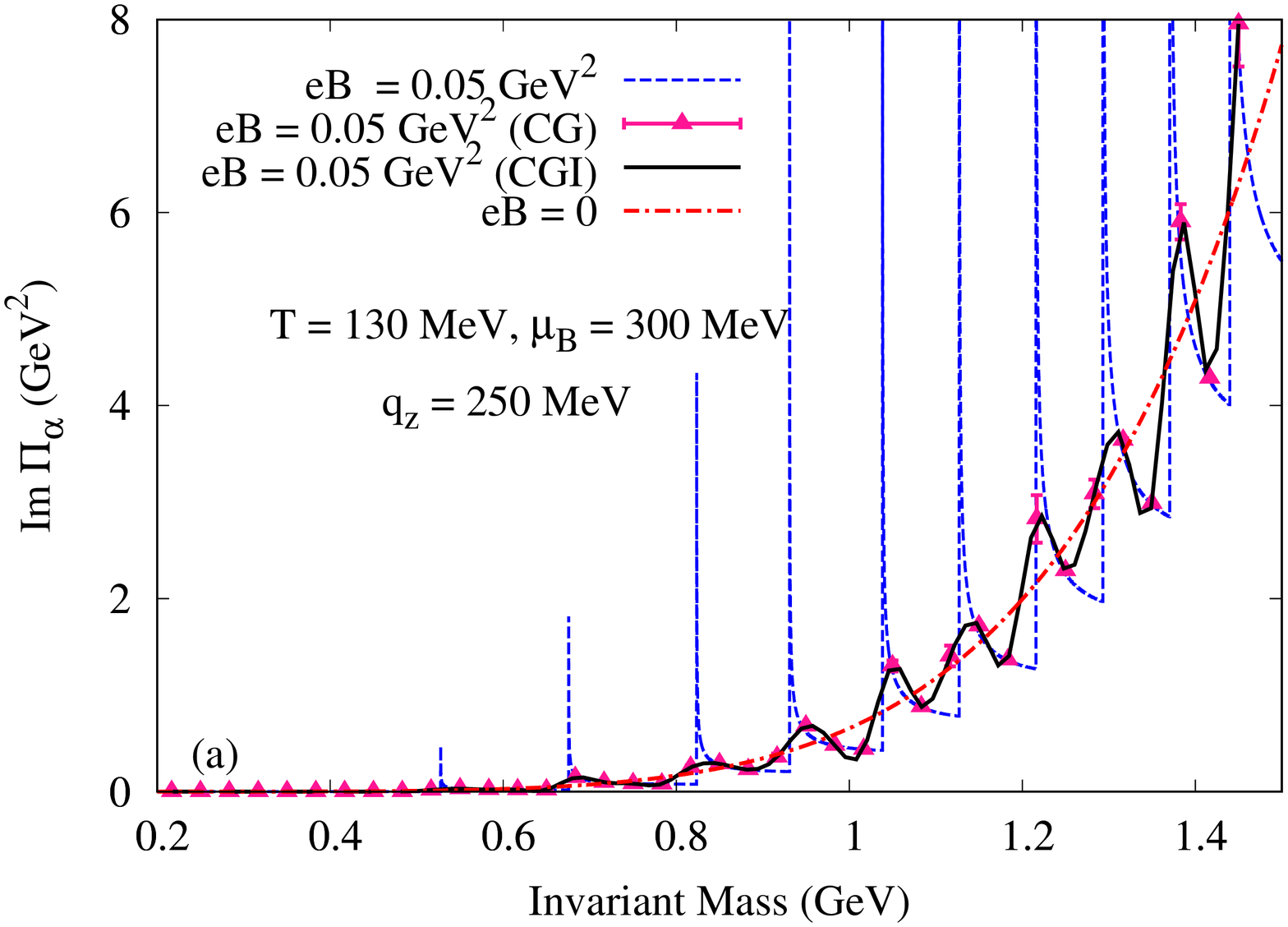} 	
		\includegraphics[angle=-90,scale=0.35]{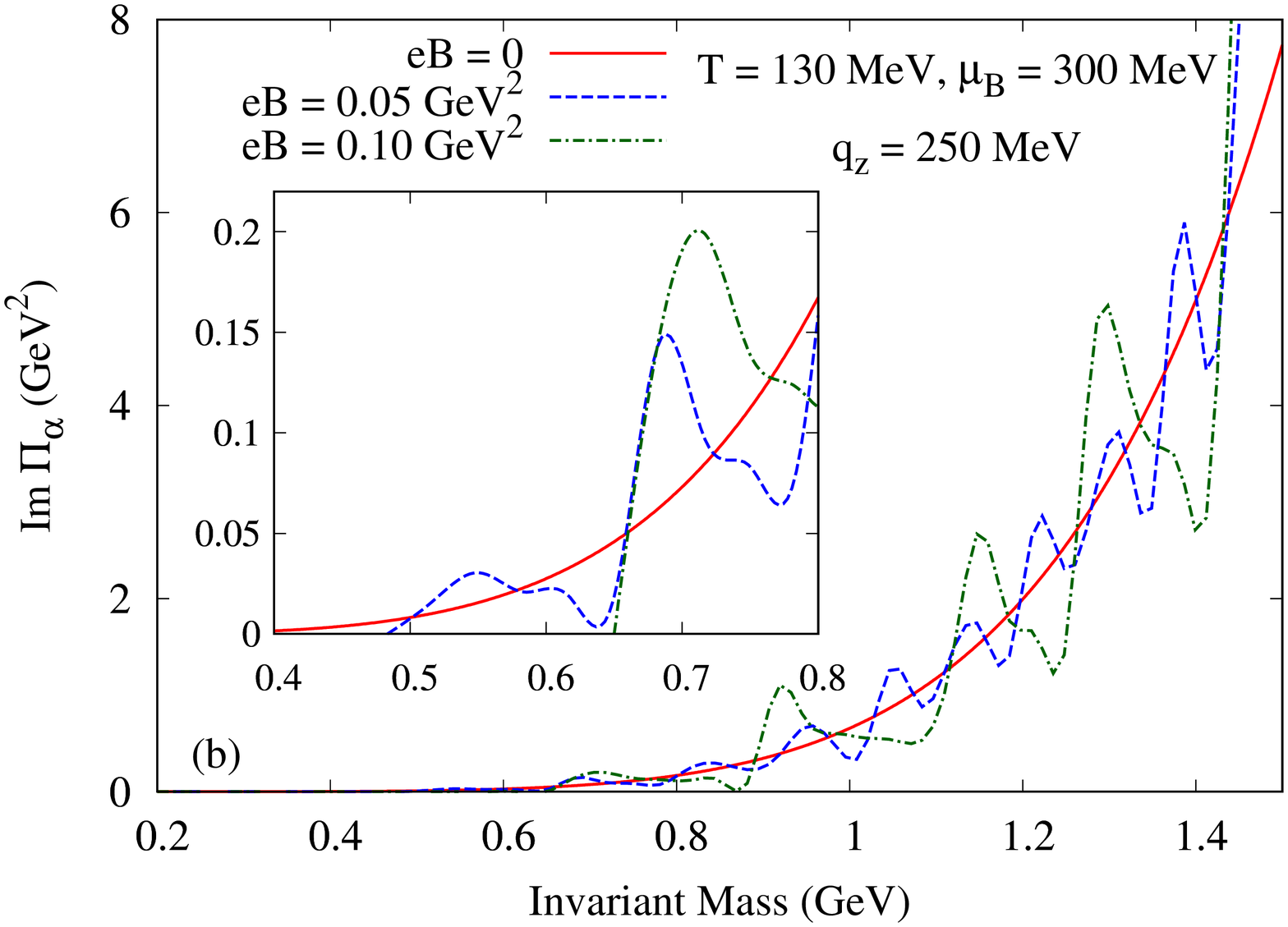}
	\end{center}
	\caption{The contribution from the form factor $\IM\Pi_\alpha$ to the imaginary part of the $\rho^0$ self energy is shown as a function of invariant mass at temperature $T=130$ MeV and at baryon chemical potential $\mu_B=300$ with $\rho^0$ longitudinal momentum $q_z=250$ MeV for (a) two different values of magnetic field ($eB=$ 0 and 0.05 GeV$^2$ respectively) and (b) three different values of magnetic field ($eB=$ 0, 0.05 and 0.10 GeV$^2$ respectively). The coarse-grained (CG) as well as coarse-grained interpolated (CGI) results are shown in (a) whereas (b) shows only the CGI results. The inset plot in (b) shows the movement of the Unitary cut threshold by focusing in smaller range of invariant mass.}
	\label{fig.impi.eB.1}
\end{figure}
\begin{figure}[h]
	\begin{center}
		\includegraphics[angle=-90,scale=0.35]{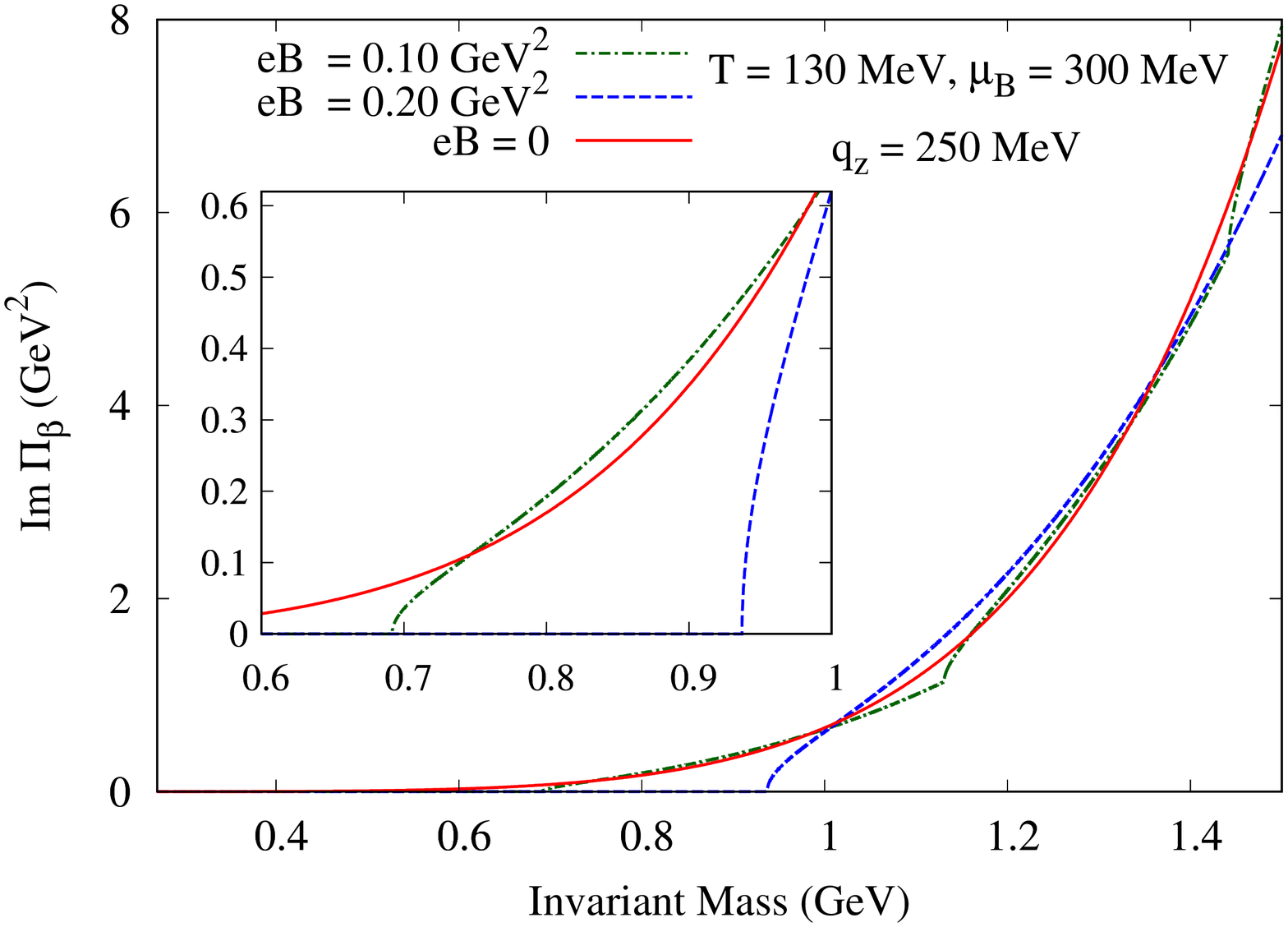} 	
	\end{center}
	\caption{The contribution from the form factor $\IM\Pi_\beta$ to the imaginary part of the $\rho^0$ self energy is shown as a function of invariant mass at temperature $T=130$ MeV and at baryon chemical potential $\mu_B=300$ with $\rho^0$ longitudinal momentum $q_z=250$ MeV for three different values of magnetic field ($eB=$ 0, 0.05 and 0.10 GeV$^2$ respectively). The inset plot shows the movement of the Unitary cut threshold by focusing in smaller range of invariant mass.}
	\label{fig.impi.eB.2}
\end{figure}

Having checked the consistency of the non-zero magnetic field calculations, we now proceed to present the imaginary part of the self energy for nonzero values of the magnetic field. In Fig.~\ref{fig.impi.eB.1}, the variation of $\IM\Pi_\alpha$ is shown as a function of $\rho^0$ invariant mass with longitudinal momentum $q_z=250$ MeV at temperature $T=130$ MeV and at baryon chemical potential $\mu_B=300$ MeV. 
We have plotted the self energy upto $\sqrt{\qpll^2}=1.5$ GeV for which the Unitary cut of the NN loop does not contribute. Fig.~\ref{fig.impi.eB.1}(a) depicts $\IM\Pi_\alpha$ at magnetic field $eB=0.05$ GeV$^2$ in which the spikes get separated from each other by a finite value and it oscillates about the $eB=0$ graph. This is more clearly visible in the CG points which are used to obtain a coarse-grained interpolated (CGI) graph. Fig.~\ref{fig.impi.eB.1}(b) shows the CGI imaginary part at two different values of the magnetic field ($eB=0.05$ and 0.10 GeV$^2$ respectively); both of them are found to oscillate about the $eB=0$ graph. Moreover, with the increase in magnetic field, the oscillation frequency decreases with an increase in the oscillation amplitude. This behavior of the imaginary part with increasing magnetic field is consistent with Fig.~\ref{fig.impi.eB0.limit}, where for the $eB\rightarrow0$ case, the oscillation frequency becomes infinite and amplitude becomes zero, thus reproducing the $eB=0$ graph. Also with the increase in magnetic field, the threshold of the unitary cut moves towards the higher invariant mass value as discussed in Sec.~\ref{sec.analytic}. This has been shown clearly in the inset plot.

The corresponding results for the $\IM\Pi_\beta$ due to $\pi\pi$ loop as a function of $\rho^0$ invariant mass with longitudinal momentum $q_z=250$ MeV at temperature $T=130$ MeV and at baryon chemical potential $\mu_B=300$ MeV are shown in Fig.~\ref{fig.impi.eB.2} for the two different values of the magnetic field $eB=0.10$ and 0.20 GeV$^2$. Analogous to $\IM\Pi_\alpha$, $\IM\Pi_\beta$ also oscillates about $eB=0$ curve, but in this case the oscillation frequency is much smaller as compared to $\IM\Pi_\alpha$. The threshold of the Unitary cut moves towards higher invariant mass with the increase in magnetic field as clearly depicted in the inset plot. 
\begin{figure}[h]
	\begin{center}
		\includegraphics[angle=-90,scale=0.35]{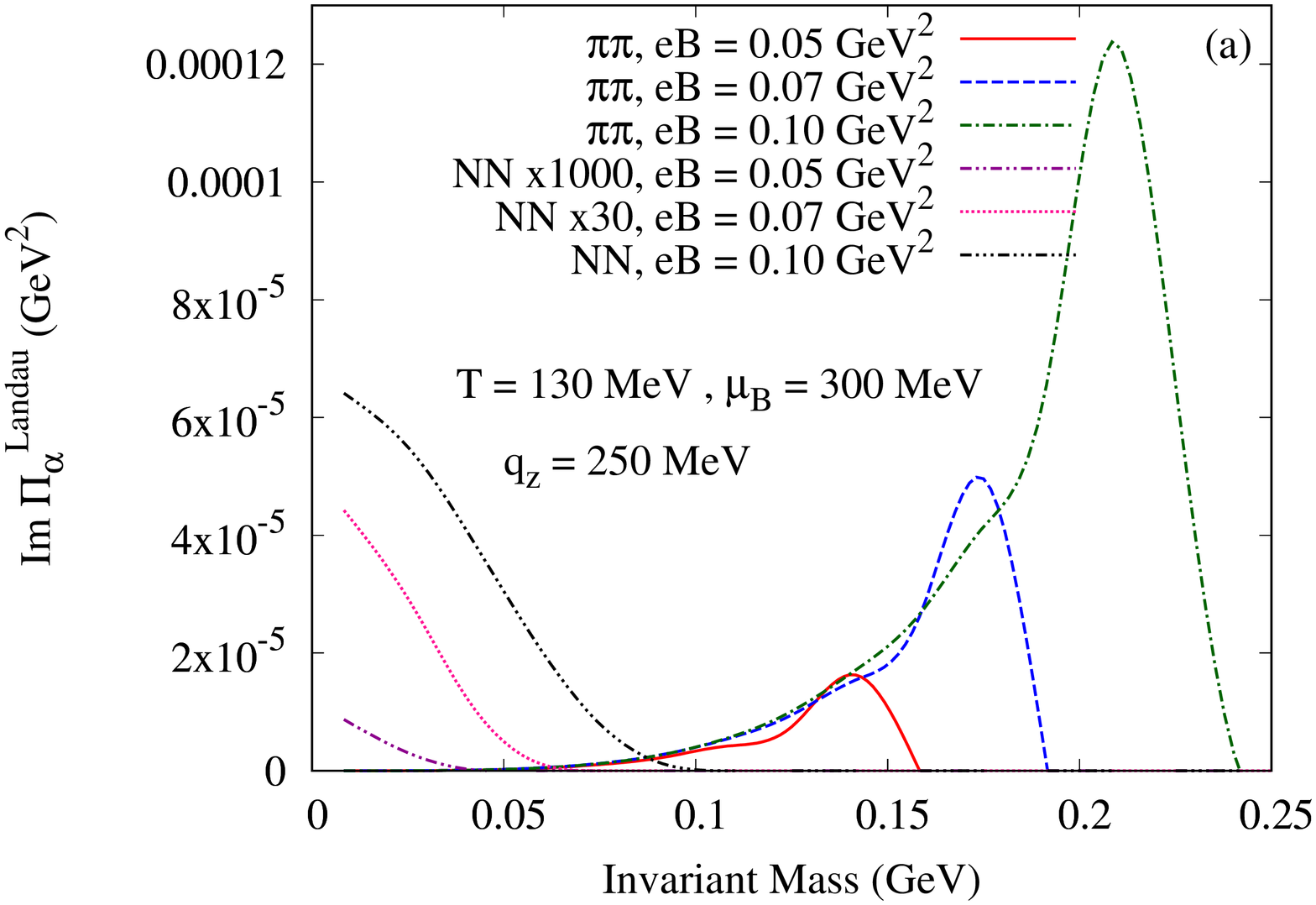} 	
		\includegraphics[angle=-90,scale=0.35]{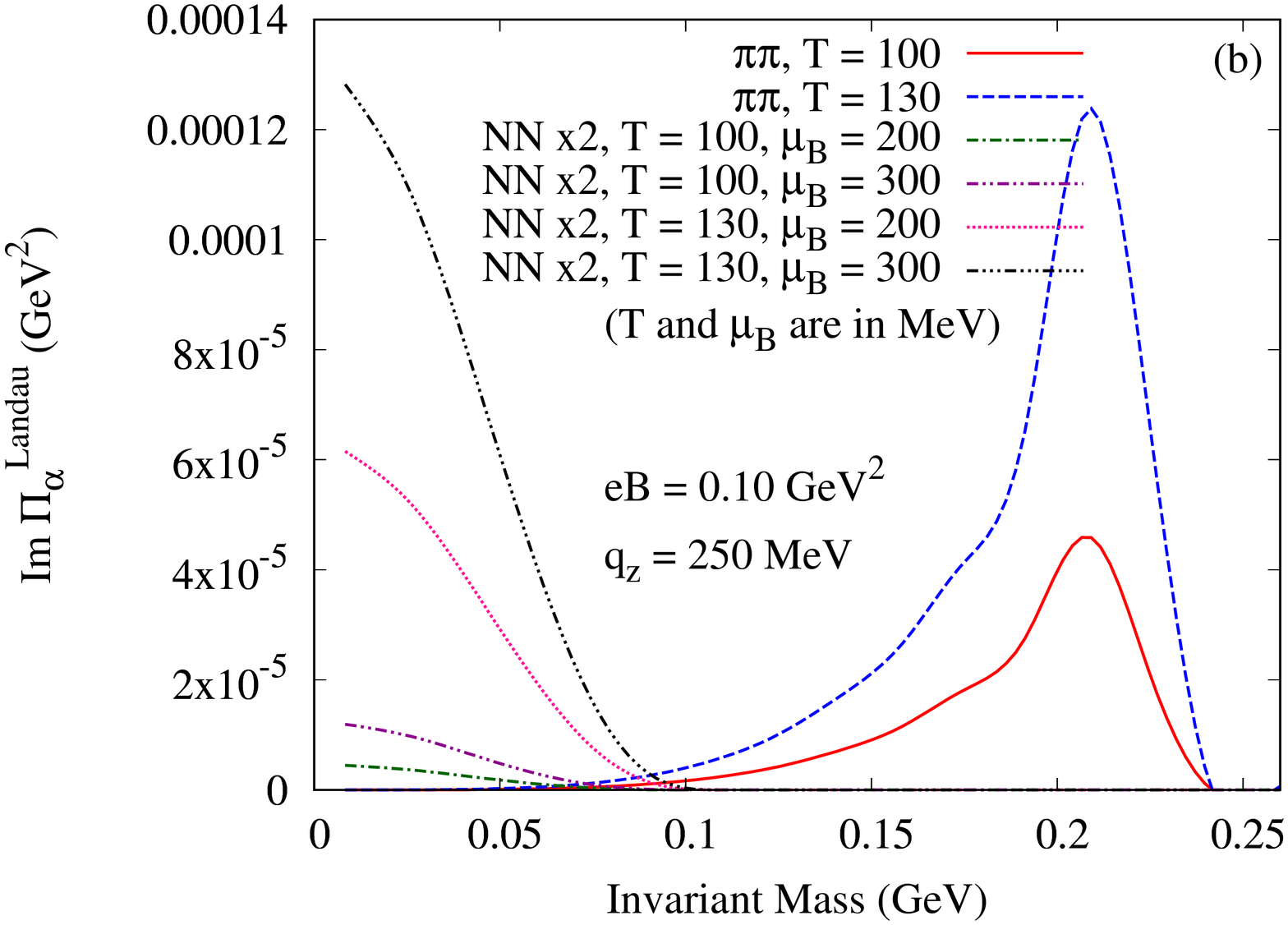}	
	\end{center}
	\caption{The contribution from the form factor $\Pi_\alpha$ to the Landau cut of the coarse grained (CG) imaginary part of the $\rho^0$ self energy is shown as a function of invariant mass with $\rho^0$ longitudinal momentum $q_z=250$ MeV (a) at temperature $T=130$ MeV and at baryon chemical potential $\mu_B=300$ for three different values of magnetic field ($eB$ = 0.05, 0.07 and 0.10 GeV$^2$ respectively) and (b) at magnetic field $eB=0.10$ GeV$^2$ for two different values of temperature ($T=$ 100 and 130 MeV respectively) and at baryon chemical potential ($\mu_B=$ 200 and 300 MeV respectively). The contribution from the $\pi\pi$ and NN loops are shown separately in which the later is scaled with different factors for the sake of presentation.}
	\label{fig.impi.eB.3}
\end{figure}

As discussed in Sec.~\ref{sec.analytic}, a non trivial Landau cut contribution in presence of external magnetic field may appear even if the loop particles have the same mass. In this case, we have observed Landau cut contribution only in $\IM\Pi_\alpha$, whereas the Landau cut does not appear in $\IM\Pi_\beta$. This can be understood from the expressions of trace and $00$ component of  $\tilde{\mathcal{N}}^\munu_{\pi,nl}$ and $\tilde{\mathcal{N}}^\munu_{\tp,nl}$ as given in Appendix~\ref{app.Nmumu00}. It can be noticed that, for both the $\pi\pi$ and proton-proton loops, the expression for the trace (i.e $\tilde{\mathcal{N}}^\mu_{~\mu}$) contains two additional Kronecker delta functions $\delta_l^{n\pm1}$ along with  $\delta_l^n$ which is absent in the expressions for the $00$ component (i.e $\tilde{\mathcal{N}}^{00}$) (see Eqs.\eqref{eq.N.pi.mumu}-\eqref{eq.N.p.00}). This implies that, for $\IM\Pi_\alpha$, the loop particles can be in different Landau levels whereas for $\IM\Pi_\beta$ the loop particles will always stay in the same Landau levels. Thus, as discussed in Se.~\ref{sec.analytic}, the non-trivial Landau cuts will appear only in $\IM\Pi_\alpha$ and not in $\IM\Pi_\beta$. The contribution of the CGI Landau cuts to  $\IM\Pi_\alpha$ as a function of $\rho^0$ invariant mass with longitudinal momentum $q_z=250$ MeV is shown in Fig.~\ref{fig.impi.eB.3}. It is to be noted that, the Landau cuts also contain the threshold singularities and thus have to be coarse grained. Fig.~\ref{fig.impi.eB.3}(a) shows the variation of $\IM\Pi_\alpha$ at a temperature $T=130$ MeV and at baryon chemical potential $\mu_B=300$ MeV for three different values of the magnetic field ($eB=0.05$, 0.07 and 0.10 GeV$^2$ respectively), whereas Fig.~\ref{fig.impi.eB.3}(b) shows the corresponding variation at magnetic field ($eB=0.10$ GeV$^2$) for two different values of temperature ($T=100$ and 130 MeV respectively). The contributions due to $\pi\pi$ loop and proton-proton loops are shown separately and in Fig.~\ref{fig.impi.eB.3}(b); the contribution due to proton-proton loop is shown for two different values of baryon chemical potential ($\mu_B=200$ and 300 MeV). As can be seen from the figures, the threshold of the Landau cuts due to $\pi\pi$ loop is different (greater) than that of proton-proton loop which can be understood from the discussions of Sec.~\ref{sec.analytic}. The threshold for $\pi\pi$ loop is $\sqrt{\qpll^2} < \FB{\sqrt{m_\pi^2+eB}-\sqrt{m_\pi^2+3eB}}$, whereas the same for proton-proton loop is $\sqrt{\qpll^2} < \FB{m_N-\sqrt{m_N^2+2eB}}$. The shift of the Landau cut threshold towards the higher invariant mass values with the increase in magnetic field can be clearly seen in Fig.~\ref{fig.impi.eB.3}(a). It is observed the the magnitude of the Landau cut contribution due to proton-proton loop is much less than that of $\pi\pi$ loop at lower values of the magnetic field and they become comparable to each other only at $eB\gtrsim 0.10$ GeV$^2$. In Fig.~\ref{fig.impi.eB.3}(a), we observe that with the increase in temperature and density, the Landau cut contribution increases without changing its threshold in the invariant mass axis.
\begin{figure}[h]
	\begin{center}
		\includegraphics[angle=-90,scale=0.35]{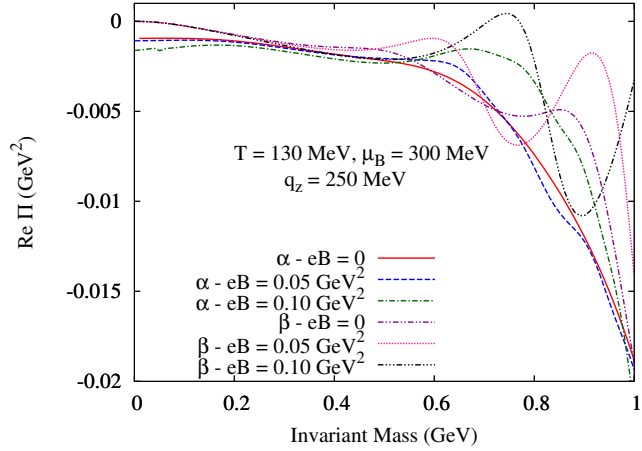}
	\end{center}
	\caption{The  real part of the thermal self energy of $\rho^0$  as a function of invariant mass at temperature $T=130$ MeV and at baryon chemical potential $\mu_B=$ 300 MeV with $\rho^0$ longitudinal momentum $q_z=250$ MeV is shown for three different values of magnetic field (0, 0.05 and 0.10 GeV$^2$ respectively).}
	\label{fig.repi.eB.1}
\end{figure}

\begin{figure}[h]
	\begin{center}
		\includegraphics[angle=-90,scale=0.35]{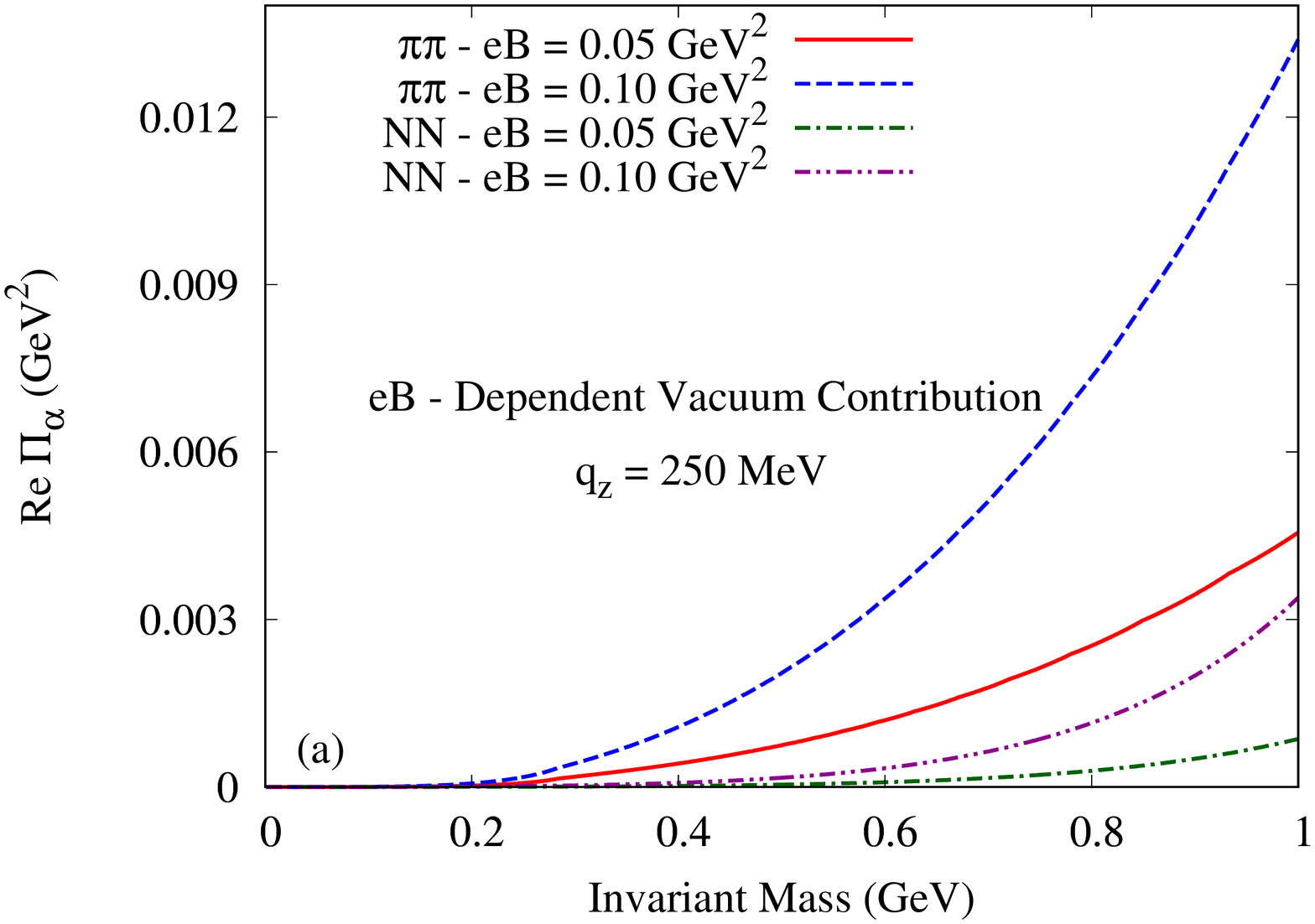}
		\includegraphics[angle=-90,scale=0.35]{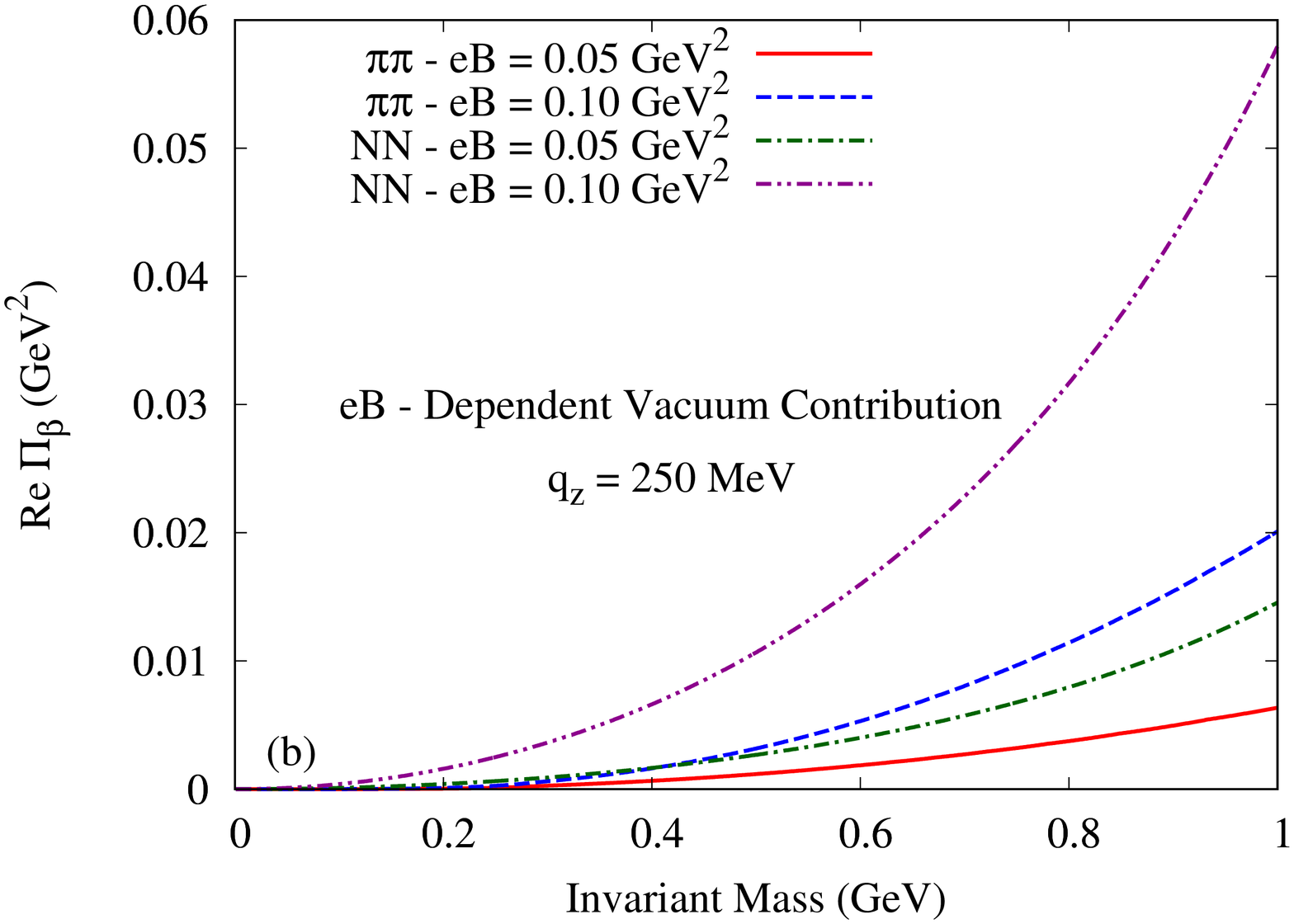}		
	\end{center}
	\caption{The eB-dependent vacuum contribution to the real part of the self energy of $\rho^0$  as a function of invariant mass with $\rho^0$ longitudinal momentum $q_z=250$ MeV is shown at two different values of magnetic field ($eB=$ 0.05 and 0.10 Gev$^2$ respectively ) for the form factors (a) $\Pi_\alpha$ and (b) $\Pi_\beta$. The contribution due to $\pi\pi$ and proton-proton loops are shown separately.}
	\label{fig.repi.eB.2}
\end{figure}

We now turn our attention to the real part of the self energy at finite temperature under external magnetic field. In Fig.~\ref{fig.repi.eB.1}, we have shown the thermal contribution to the real part of the self energy as a function of invariant mass with $\rho^0$ longitudinal momentum $q_z=250$ MeV at temperature $T=130$ MeV and at baryon chemical potential $\mu_B=300$ MeV for two different values of the magnetic field ($eB=0.05$ and 0.10 GeV$^2$ respectively). The contributions from the $\pi\pi$ and NN loops are summed up in this figure. We notice that, with the increase in magnetic field, the thermal contribution to the real part of the self energy oscillates about the $eB=0$ curve. The oscillation frequency and the oscillation amplitude respectively decreases and increases with the  magnetic field. 

Next in Fig.~\ref{fig.repi.eB.2}, the ``eB-dependent vacuum" contribution to the real part of the self energy is shown as a function of $\rho^0$ invariant mass with longitudinal momentum $q_z=250$ MeV for two different values of magnetic field ($eB=0.10$ and 0.20 GeV$^2$ respectively). Figs.~\ref{fig.repi.eB.2}(a) and \ref{fig.repi.eB.2}(b) show the contributions from $\Pi_\alpha$ and $\Pi_\beta$ respectively. The contributions due to $\pi\pi$ and proton-proton loops are shown separately. First of all, we note that at $eB=0$, these term will vanish. With the increase of the magnetic field, the eB-dependent vacuum term also increases and the contribution of $\Pi_\beta$
is more than $\Pi_\alpha$. 

Having obtained the real and imaginary parts of the self energy, we now proceed to evaluate the in-medium spectral functions of $\rho^0$ under external magnetic field. We have from Eq.~\eqref{eq.complete.prop}, the complete $\rho^0$ propagator as 
\begin{eqnarray}
\overline{\overline{D}}^\munu = A_\alpha P_1^\munu + A_\beta P_2^\munu + A_\gamma P_3^\munu + A_\delta Q^\munu + \xi q^\mu q^\nu
\label{eq.complete.prop.2}
\end{eqnarray}
where the coefficients are given in Eqs.~\eqref{eq.A_alpha}-\eqref{eq.xi} and the basis tensors are provided in Eqs.~\eqref{eq.proj.tensor.tb1}-\eqref{eq.proj.tensor.tb4}. Since we will be considering the special case $q_\perp=0$ for which $\Pi_\alpha = \Pi_\gamma$ and $\Pi_\delta=0$ as given in Eq.~\eqref{eq.constraints}, the coefficients in the above equation become
\begin{eqnarray}
A_\alpha &=& \FB{\frac{1}{q_\parallel^2-m_\rho^2+\Pi_\alpha}} \label{eq.A_alpha.1}\\
A_\beta &=& \FB{\frac{1}{q_\parallel^2-m_\rho^2+\Pi_\beta}} \\
A_\gamma &=& \FB{\frac{1}{q_\parallel^2-m_\rho^2+\Pi_\gamma}} \\
A_\delta &=& 0  \\
\xi &=& \frac{-1}{q_\parallel^2m_\rho^2} \label{eq.xi.1}
\end{eqnarray}
so that the complete in-medium interacting propagator is given by
\begin{eqnarray}
\overline{\overline{D}}^\munu(q^0,q_z) = \frac{P_1^\munu}{\FB{q_\parallel^2-m_\rho^2+\Pi_\alpha}}  + \frac{P_2^\munu}{\FB{q_\parallel^2-m_\rho^2+\Pi_\beta}} + \frac{P_3^\munu}{\FB{q_\parallel^2-m_\rho^2+\Pi_\alpha}}  - \frac{q_\parallel^\mu q_\parallel^\nu}{q_\parallel^2m_\rho^2}~.
\label{eq.complete.prop.3}
\end{eqnarray}
It is clear from the above equation, that there will be three modes for the propagation of $\rho^0$ meson in magnetized medium for vanishing transverse momentum of $\rho^0$. Of the three modes, two are found to be degenerate (the first and third term in the RHS of above equation) leaving two distinct modes for the propagation of $\rho^0$ which 
we denote as Mode-A and Mode-B respectively.
\begin{figure}[h]
	\begin{center}
		\includegraphics[angle=-90,scale=0.35]{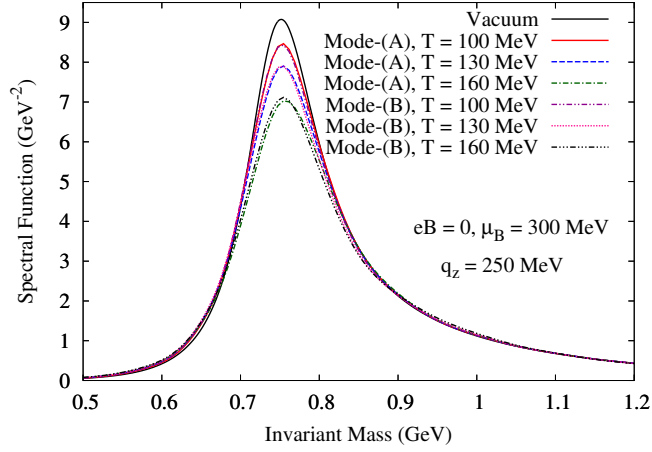}		
	\end{center}
	\caption{The in-medium spectral function of $\rho^0$ as a function of invariant mass at zero magnetic field and at baryon chemical potential $\mu_B$=300 MeV with $\rho^0$ longitudinal momentum $q_z=250$ MeV is shown for three different values of temperature ($T=$ 100, 130 and 160 MeV) and for different modes. The vacuum spectral function is also shown for comparison.}
	\label{fig.spectra.eB0}
\end{figure}

We now define the spectral function $S_\rho$ of $\rho^0$ for the two distinct modes as the the imaginary part of the complete propagator which is obtained from Eq.~\eqref{eq.complete.prop.3} as
\begin{eqnarray}
S_\rho^{(A)} &=& \IM\TB{\frac{-1}{q_\parallel^2-m_\rho^2+\Pi_\alpha}} 
= \frac{\IM\Pi_\alpha}{(q_\parallel^2-m_\rho^2+\RE\Pi_\alpha)^2 + (\IM\Pi_\alpha)^2 } \\ ~~&& \hspace{-5cm}\text{and}~~\nn \\
S_\rho^{(B)} &=& \IM\TB{\frac{-1}{q_\parallel^2-m_\rho^2+\Pi_\beta}} 
= \frac{\IM\Pi_\beta}{(q_\parallel^2-m_\rho^2+\RE\Pi_\beta)^2 + (\IM\Pi_\beta)^2 }. 
\end{eqnarray}
\begin{figure}[h]
	\begin{center}
		\includegraphics[angle=-90,scale=0.35]{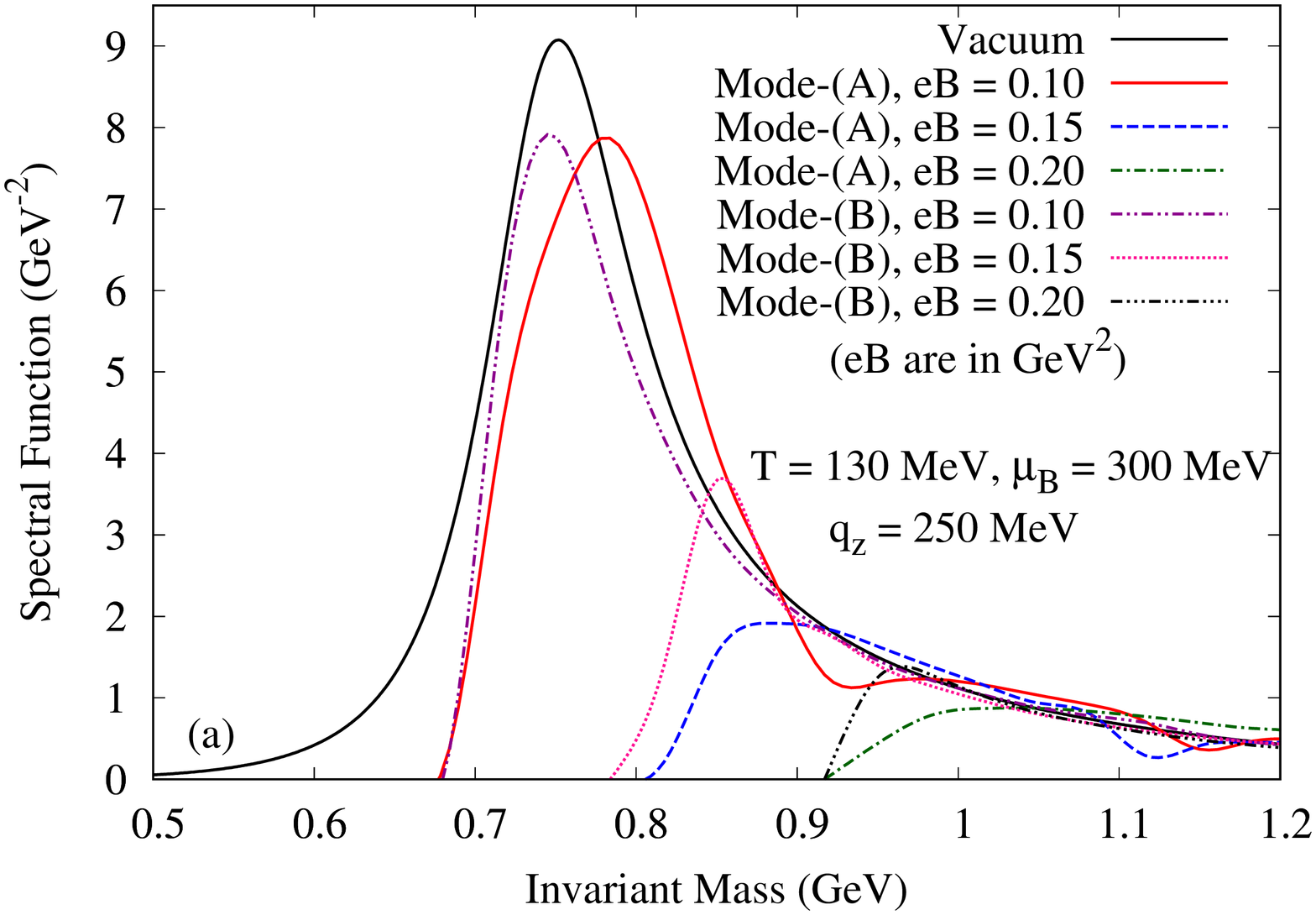}		
		\includegraphics[angle=-90,scale=0.35]{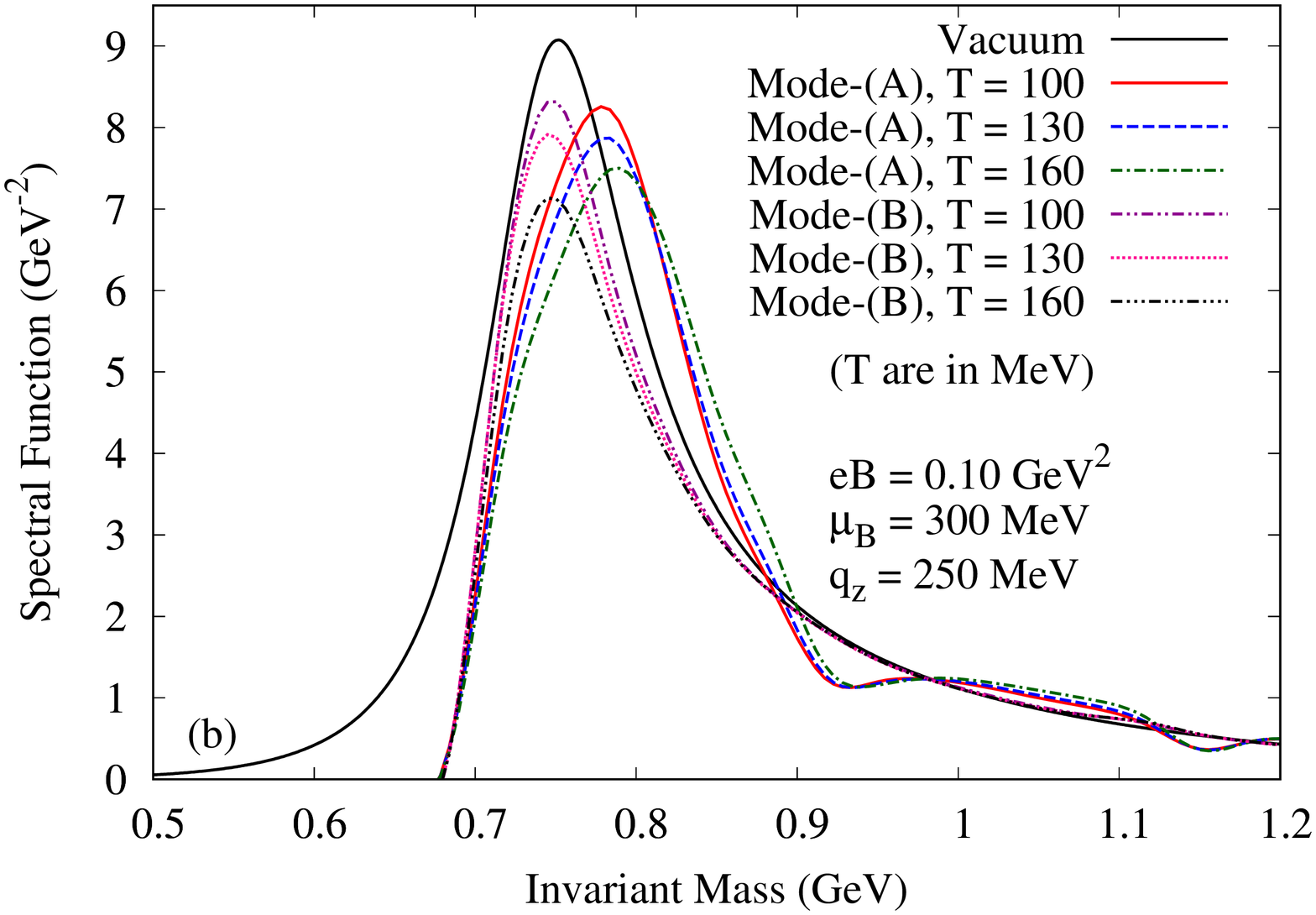}		
		\includegraphics[angle=-90,scale=0.35]{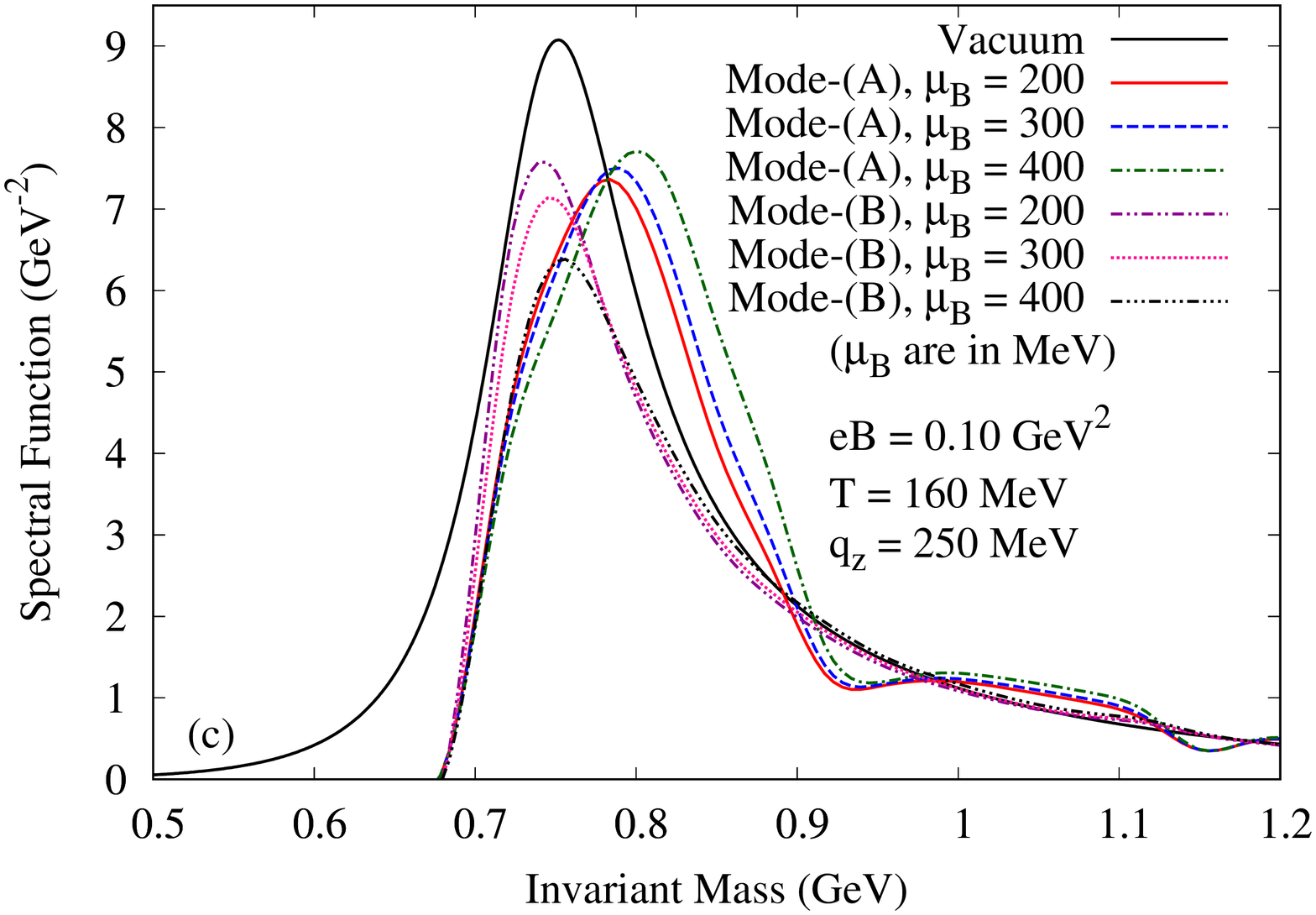}		
		\includegraphics[angle=-90,scale=0.35]{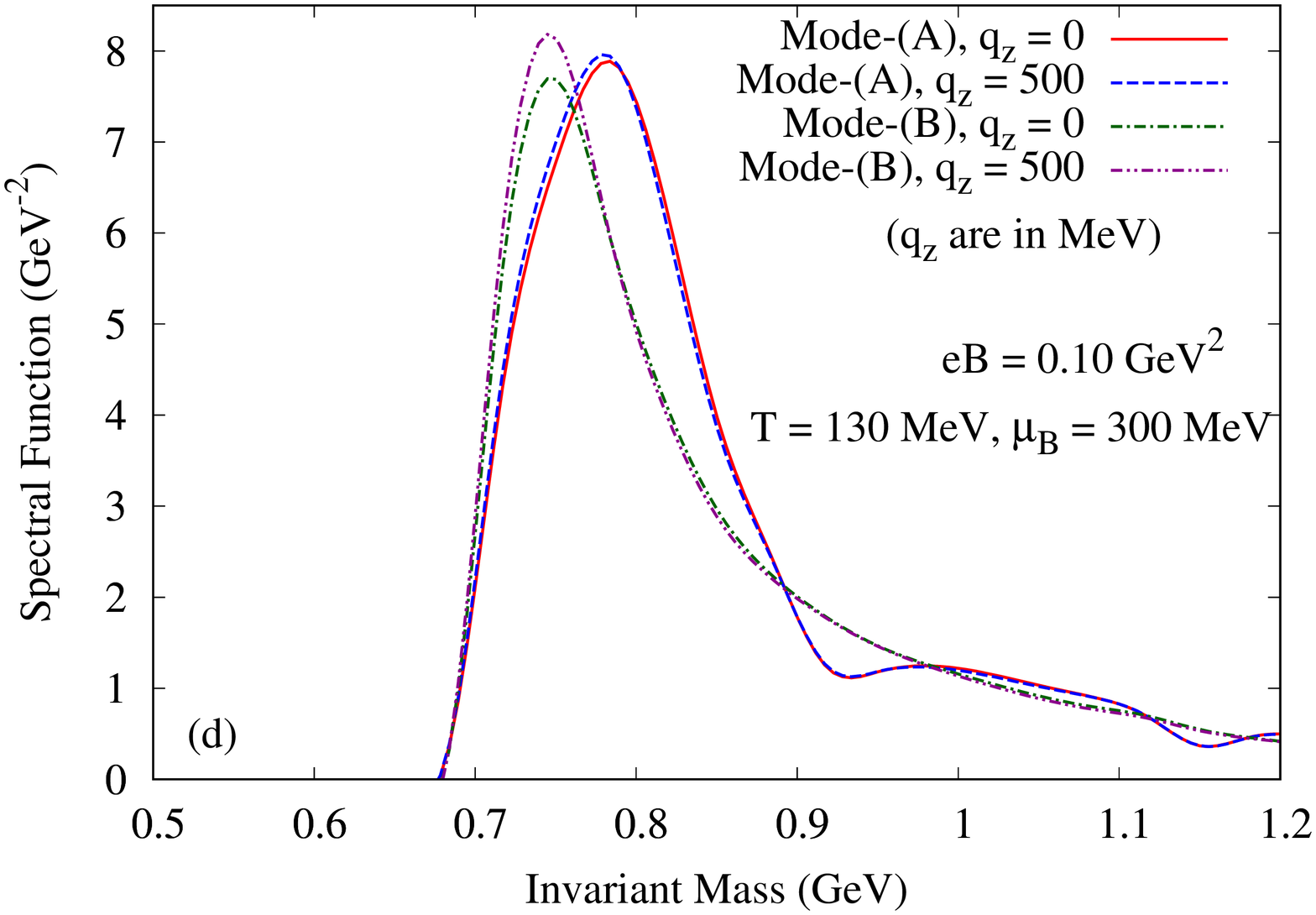}						
	\end{center}
	\caption{The in-medium spectral functions of $\rho^0$ as a function of invariant mass is shown for different modes (a) at temperature $T=$ 130 MeV and at baryon chemical potential $\mu_B=$ 300 MeV with $\rho^0$ longitudinal momentum $q_z=250$ MeV for three different values of magnetic field ($eB=$ 0.10, 0.15 and 0.20 GeV$^2$ respectively) (b) at magnetic field $eB=$ 0.10 GeV$^2$ and at baryon chemical potential $\mu_B=$ 300 MeV with $\rho^0$ longitudinal momentum $q_z=250$ MeV for three different values of temperature ($T=$ 100, 130 and 160 MeV respectively) (c) at magnetic field $eB=$ 0.10 GeV$^2$ and at temperature $T=$ 160 MeV with $\rho^0$ longitudinal momentum $q_z=250$ MeV for three different values of baryon chemical potential ($\mu_B=$ 200, 300 and 400 MeV respectively) and (d) at magnetic field $eB=$ 0.10 GeV$^2$ and at temperature $T=$ 130 MeV with baryon chemical potential $\mu_B=$ 300 MeV for two different values of $\rho^0$ longitudinal momentum ($q_z=$ 0 and 500 MeV respectively). The vacuum spectral function is also shown for comparison. }
	\label{fig.spectra.eB.1}
\end{figure}

In Fig.~\ref{fig.spectra.eB0}, the spectral function for the two modes at zero magnetic field is shown as a function of $\rho^0$ invariant mass with $\rho^0$ longitudinal momentum $q_z=250$ MeV at baryon chemical potential $\mu_B=300$ MeV for three different values of temperature  ($T=100$, 130 and 160 MeV respectively). The vacuum spectral function (which is  same for the two modes) is also shown for comparison. 
We find that, the spectral functions have a nice Breit-Wigner shape around the $\rho^0$ mass pole with a width $\mathcal{O}$(150 MeV) corresponding 
to the decay of $\rho^0\rightarrow\pi^+\pi^-$. With the increase in temperature, the width of the spectral function increases and the peak decreases. Physically, it corresponds to the enhancement of the decay process in the medium implying that the $\rho^0$ become more unstable at a high temperature. It is important to note that, for the invariant mass region shown in the plot, the imaginary part of the self energy that enters in the calculation of spectral function is completely due to the Unitary-I cut of $\pi\pi$ loop. On the other hand, the real part of the self energy that enters in the  spectral function calculation has contributions from both the $\pi\pi$ and NN loops.

 It can  be noticed that, even at a higher temperature ($T\sim 160$ MeV), the peak of the spectral functions have 
	marginal shifts over the invariant mass axis which correspond to a negligible mass shift of the $\rho$ meson with respect to 
	its vacuum mass. 
	This is in agreement with the fact that, based on consideration of chiral symmetry alone, the mass of the $\rho$ meson does not 
	change to $\mathcal{O}(T^2)$~\cite{Dey:1990ba}. At and above the critical temperature, chiral symmetry requires that the vector and 
	axial-vector spectral function are identical~\cite{Kapusta:1993hq} and is demonstrated in Ref.~\cite{Hohler:2013eba} 
	using sum rule approach. 
	However, scenarios of $\rho$ mass shift proposed by Brown and Rho~\cite{Brown:1991kk} are also not ruled out and the behaviour of the 
	$\rho$ meson mass can only indirectly be related to the chiral symmetry restoration. 
	Though significant shift of $\rho$ mass has also been reported in Ref~\cite{Alam:1999sc} using Walecka model, yet the 
	underlying phenomena behind this effect can not be related to the partial restoration of chiral symmetry of QCD. 
	Moreover, majority of experiments does not find evidence for mass shift of the rho meson in the medium but 
	rather a broadening of the spectral function is reported~\cite{Leupold:2009kz}.

We now turn on the external magnetic field and show the spectral function of $\rho^0$ as a function of its invariant mass for the two modes in Fig.~\ref{fig.spectra.eB.1}. The range of the invariant mass axis is taken as 0.5-1.2 GeV which is dominated by the Unitary cut contributions from the $\pi\pi$ loop. In 
Fig.~\ref{fig.spectra.eB.1}(a), the spectral function with $\rho^0$ longitudinal momentum $q_z=250$ MeV at temperature $T=130$ MeV and at baryon chemical potential $\mu_B=300$ MeV is shown for three different values of the magnetic field ($eB=0.10$, 0.15 and 0.20 GeV$^2$ respectively). 
It is observed that, with the increase in  the magnetic field, the two modes get well separated from each other and the threshold of the spectral function moves towards higher values of invariant mass corresponding to the magnetic field dependent Unitary cut threshold of the imaginary part of the self energy. At sufficiently high values of the magnetic field, the spectral function misses the $\rho^0$ mass pole (770 MeV) so that it looses its Breit-Wigner shape which may be termed as $\rho^0$ ``melting" in presence of magnetic field. The critical value of the magnetic field for a given temperature and baryon chemical potential for which the $\rho^0$ will melt is discussed later.


In Fig.~\ref{fig.spectra.eB.1}(b), the spectral function with $\rho^0$ longitudinal momentum $q_z=250$ MeV at a magnetic field $eB=0.10$ GeV$^2$ and at  a baryon chemical potential $\mu_B=300$ MeV is shown for three different values of temperature ($T=100$, 130 and 160 MeV respectively). In this case, the threshold of the spectral function remains fixed and for both the modes, the spectral function becomes shorter and wider with the increase in temperature with a marginal shift of its peak. The shift of the peak is due to the modification in the real part of the self energy with the change in temperature.

Fig.~\ref{fig.spectra.eB.1}(c) depicts the spectral function with $\rho^0$ longitudinal momentum $q_z=250$ MeV at a magnetic field $eB=0.10$ GeV$^2$ and at a temperature $T=160$ MeV for three different values of the baryon chemical potential ($\mu_B=200$, 300 and 400 MeV respectively). Analogous to the previous case, the threshold of the spectral function remains fixed for both the modes. Since the baryon chemical potential only affects the real part of the self energy in the given kinematic region, the peak of the spectral function changes its position (keeping the width almost same) with the change in baryon chemical potential. It can be noticed, that in contrast to Fig.~\ref{fig.spectra.eB.1}(b), the peak position of the spectral function is more sensitive to $\mu_B$ as compared to the temperature which is due to the dominant contribution coming from NN loop.

In Fig.~\ref{fig.spectra.eB.1}(d), the spectral function at a magnetic field $eB=0.10$ GeV$^2$ and at a temperature $T=130$ MeV with  baryon chemical potential $\mu_B=300$ MeV is shown for two different values of $\rho^0$ longitudinal momentum ($q_z=0$ and 500 MeV). In this case, the threshold of the spectral function remains same and the height of the spectral function increases with the increase of the longitudinal momentum.

\begin{figure}[h]
	\begin{center}
		\includegraphics[angle=-90,scale=0.35]{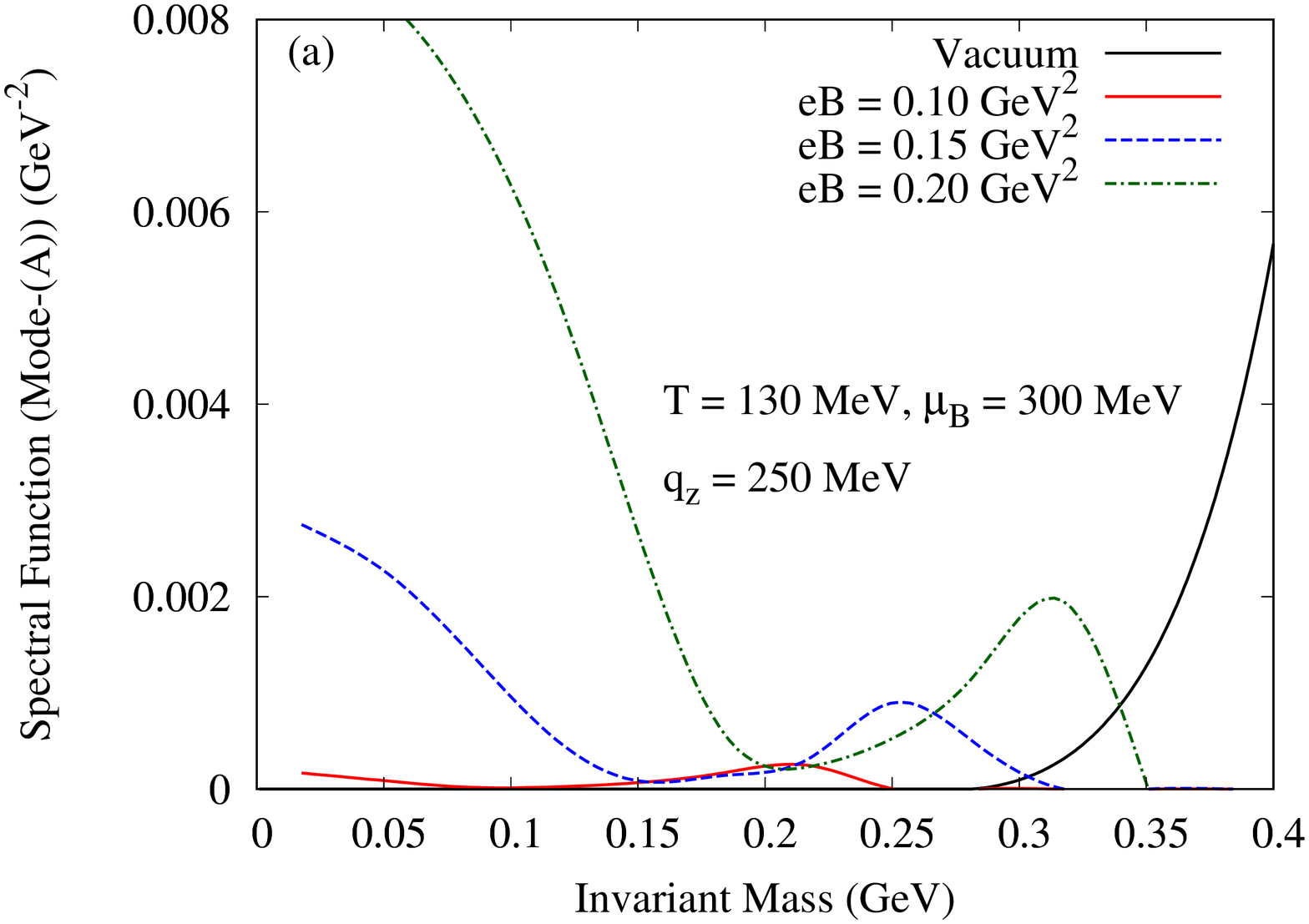}		
		\includegraphics[angle=-90,scale=0.35]{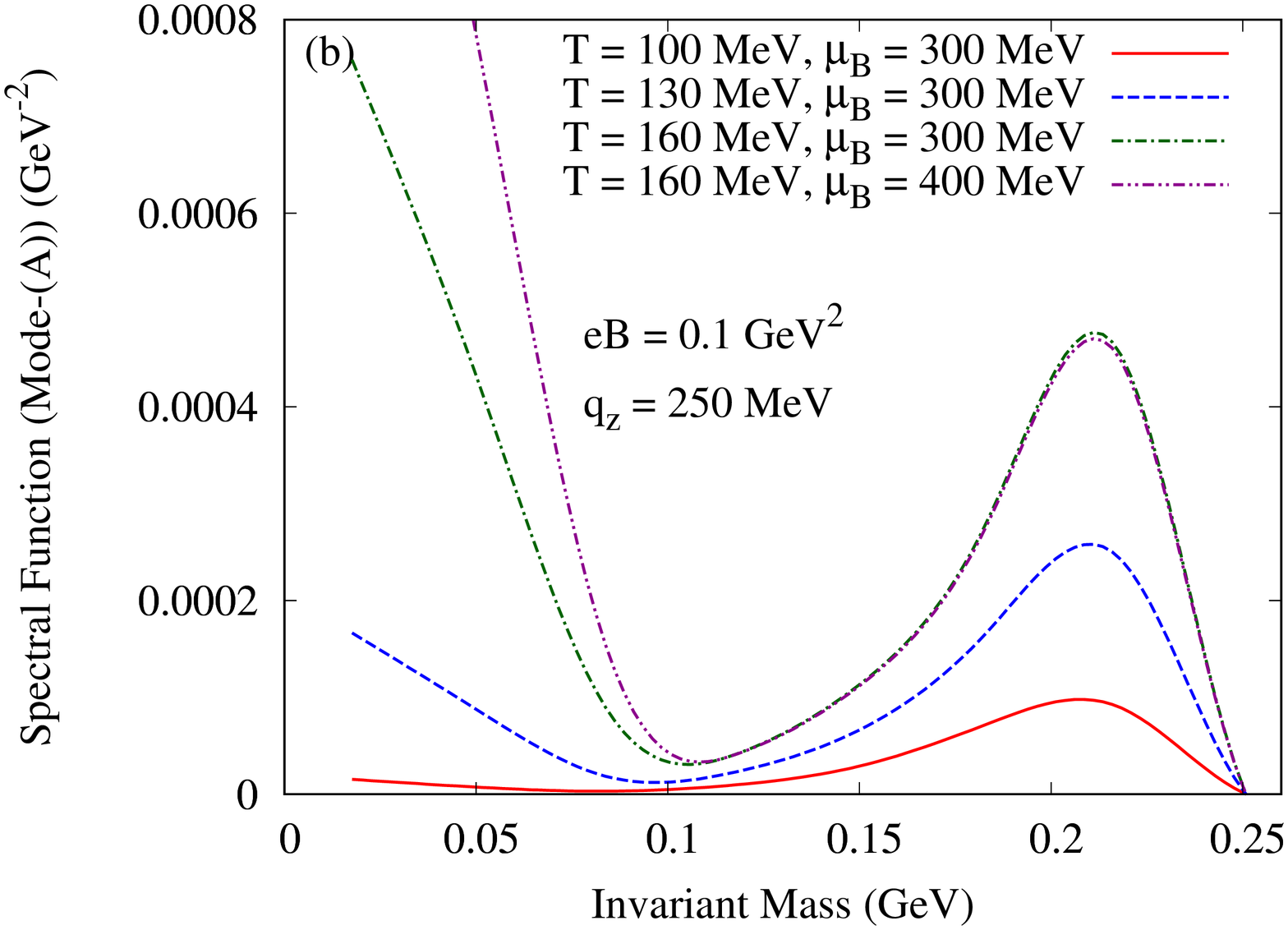}		
	\end{center}
	\caption{The in-medium spectral functions of $\rho^0$ for Mode-(A) as a function of invariant mass is shown in the low invariant mass region dominated by Landau cut contributions with $\rho^0$ longitudinal momentum $q_z=250$ MeV : (a) At temperature $T=$ 130 MeV and at baryon chemical potential $\mu_B=$ 300 MeV for three different values of magnetic field ($eB=$ 0.10, 0.15 and 0.20 GeV$^2$ respectively) and  (b) at magnetic field $eB=$ 0.10 GeV$^2$ for four different combinations of temperature and baryon chemical potential (($T=100$ MeV, $\mu_B=300$ MeV), ($T=130$ MeV, $\mu_B=300$ MeV), ($T=160$ MeV, $\mu_B=300$ MeV) and ($T=160$ MeV, $\mu_B=400$ MeV) respectively).}
	\label{fig.spectra.eB.2}
\end{figure}

We have already mentioned that, a non-trivial Landau cut in the physical kinematic region would appear in presence of the external magnetic field. In our case, the non-zero contribution to the Landau cut comes only from the form factor $\IM\Pi_\alpha$ which is reflected in the the spectral function of Mode-(A). In Fig.~\ref{fig.spectra.eB.2}, the spectral function as a function of $\rho^0$ invariant mass with $\rho^0$ longitudinal momentum $q_z=250$ MeV is shown in the low invariant mass region which is dominated by the Landau cut contribution. It can be observed that the magnitude of the spectral function in this region is much lower as compared to the Unitary cut regions. Fig.~\ref{fig.spectra.eB.2}(a) shows the spectral function at temperature $T=130$ MeV and at baryon chemical potential $\mu_B=300$ MeV for three different values of magnetic field ($eB=0.10$, 0.15 and 0.20 GeV$^2$ respectively). As can be seen in the graph, the threshold of the Landau cut moves towards the higher values of invariant mass with the increase in magnetic field as a consequence of similar behaviour of the Landau cut contribution to the imaginary part as shown in Fig.~\ref{fig.impi.eB.3}. Also the height of the spectral function is enhanced with the increase in $eB$. 
Fig.~\ref{fig.spectra.eB.2}(b) shows the corresponding plots of spectral function at magnetic field $eB=0.10$ GeV$^2$ for four different combinations of temperature and baryon chemical potential (($T=100$ MeV, $\mu_B=300$ MeV), ($T=130$ MeV, $\mu_B=300$ MeV), ($T=160$ MeV, $\mu_B=300$ MeV) and ($T=160$ MeV, $\mu_B=400$ MeV) respectively). As can be seen in the graph, the height of the spectral function increases with the increase in temperature and density owing to an enhancement of the corresponding scattering processes in presence of external magnetic field.
\begin{figure}[h]
	\begin{center}
		\includegraphics[angle=-90,scale=0.35]{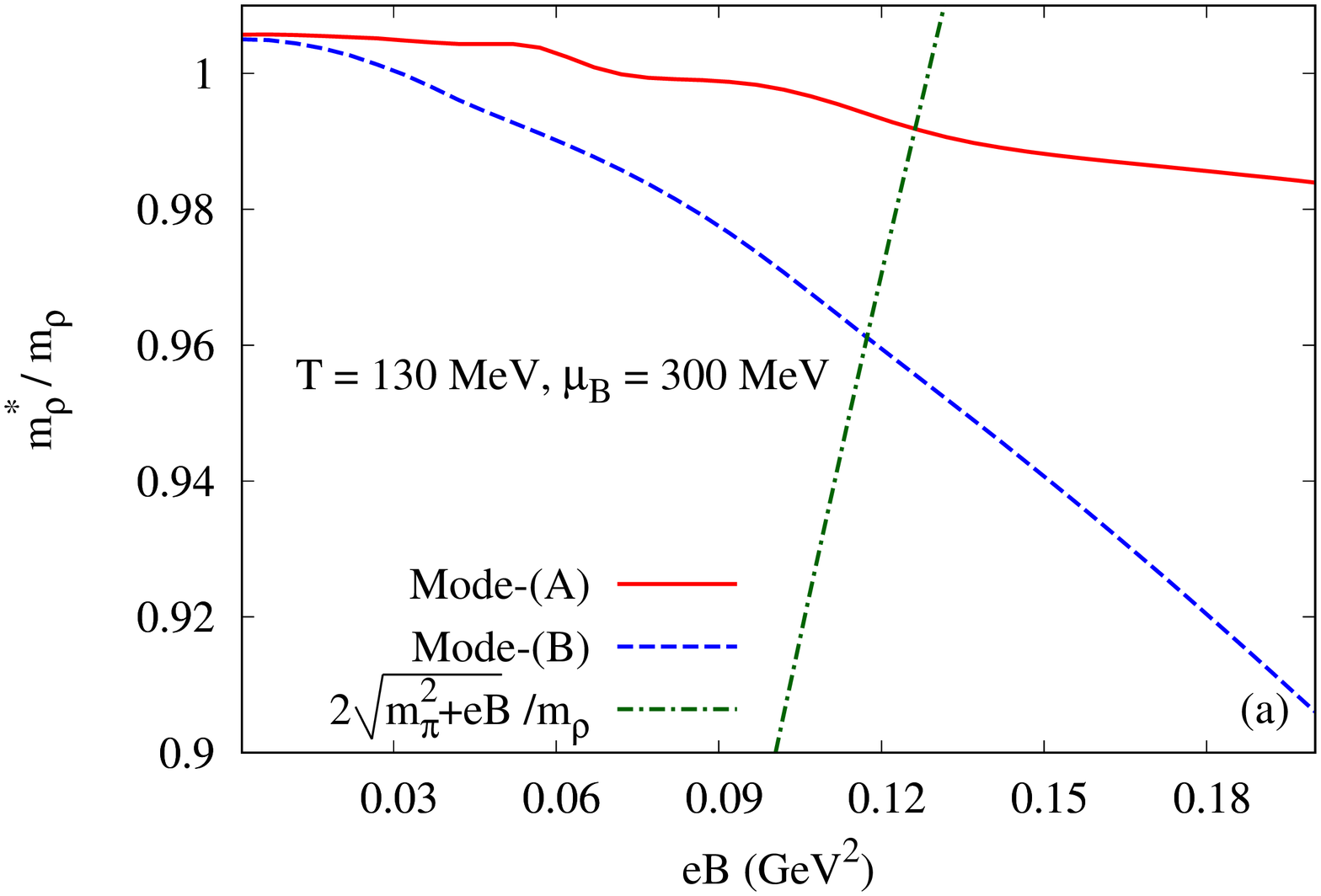}		
		\includegraphics[angle=-90,scale=0.35]{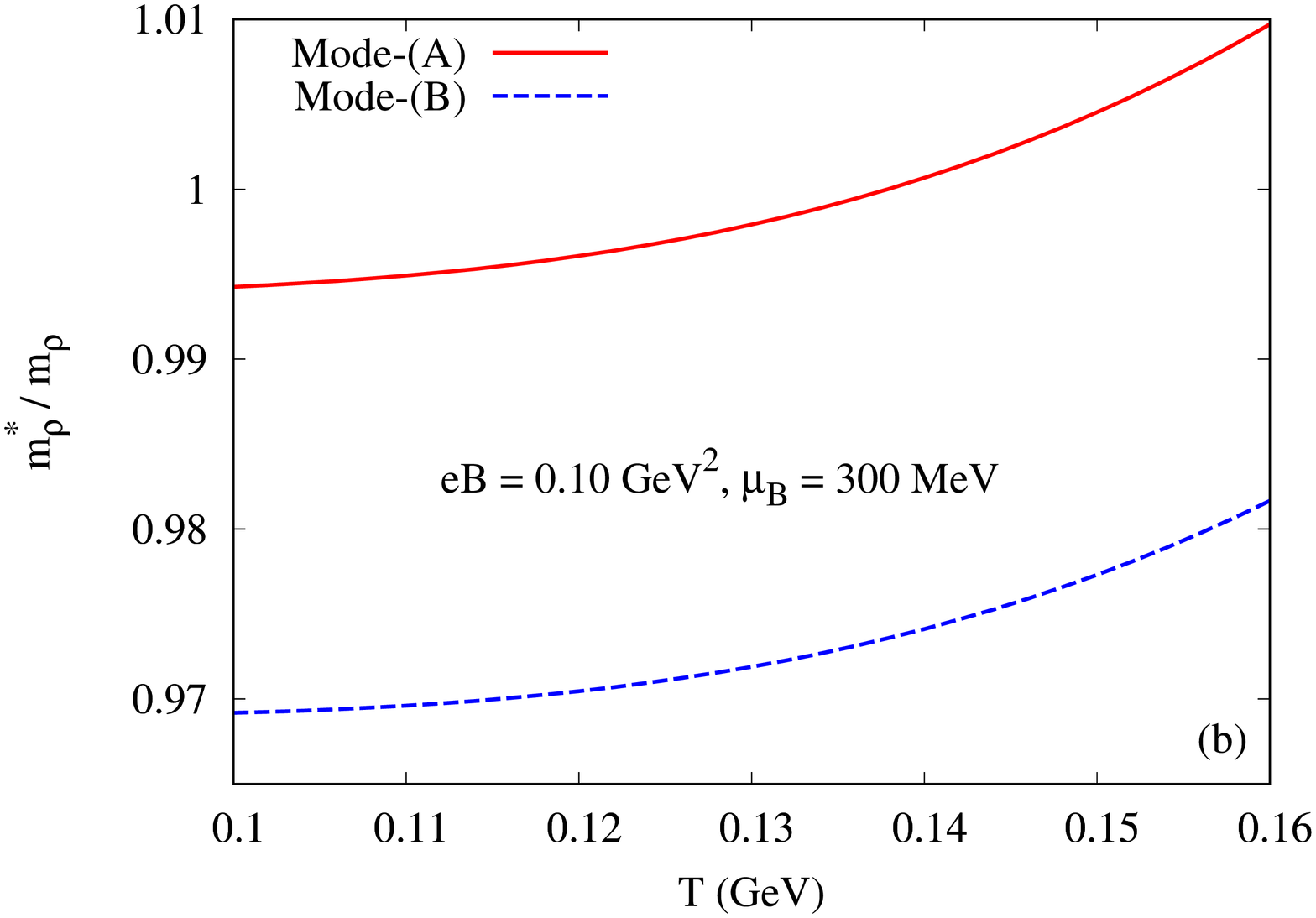} \\
		\includegraphics[angle=-90,scale=0.35]{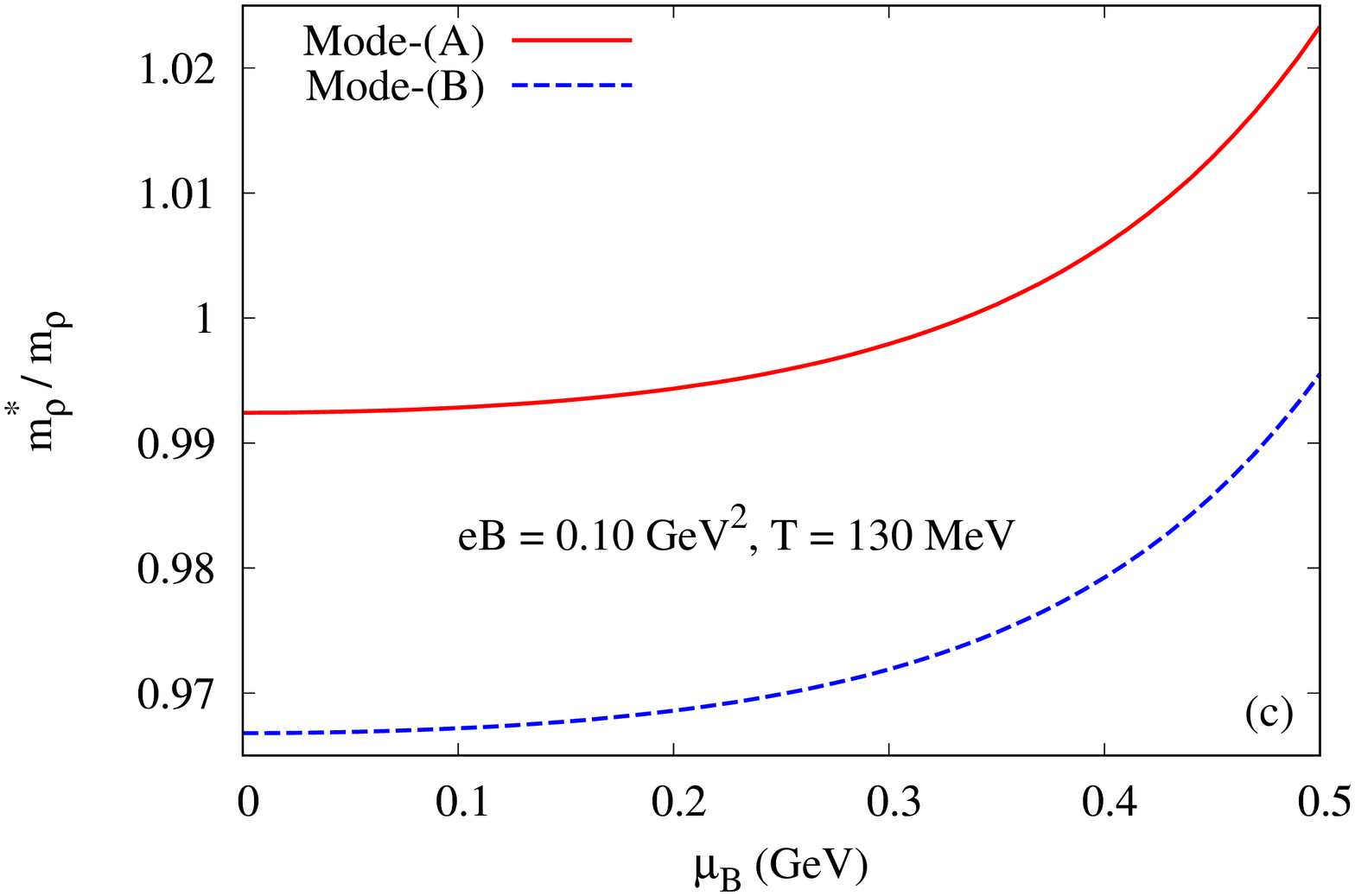}			
	\end{center}
	\caption{The ratio of effective mass of $\rho^0$ to its vacuum mass for different modes (a) as a function of magnetic field at temperature $T=$ 130 MeV and at baryon chemical potential $\mu_B=$ 300 MeV (b) as a function of temperature at magnetic field $eB=$ 0.10 GeV$^2$ and at baryon chemical potential $\mu_B=$ 300 MeV and (c) as a function of baryon chemical potential at temperature $T=$ 130 MeV and at magnetic field $eB=$ 0.10 GeV$^2$. The green dash-dotted curve in (a) corresponds to the Unitary cut threshold for decay of $\rho^0\rightarrow\pi^+\pi^-$. Here $m_\rho=770$ MeV.}
	\label{fig.mstar}
\end{figure}

We now proceed to obtain the effective mass and dispersion relation of the $\rho^0$ in a magnetized medium. They follow from the pole of the complete $\rho^0$ propagator given in Eq.~\eqref{eq.complete.prop.3} which are obtained by solving the following transcendental equations
\begin{eqnarray}
\omega^2 - q_z^2-m_\rho^2+ \RE\Pi_\alpha(q^0=\omega, q_z,eB,T,\mu_B) &=& 0 \\
\omega^2 - q_z^2-m_\rho^2+ \RE\Pi_\beta(q^0=\omega, q_z,eB,T,\mu_B) &=& 0 
\end{eqnarray}
whose numerical solutions $\omega = \omega(q_z,eB,T,\mu_B)$ represent the dispersion relations for the Mode-(A) and (B) corresponding to $\rho^0$ propagation in the magnetized medium. The effective mass $m_\rho^*$ of $\rho^0$ is obtained from the dispersion relation by setting $q_z=0$ i.e. $m_\rho^*(eB,T,\mu_B)=\omega(q_z=0,eB,T,\mu_B)$.

Fig.~\ref{fig.mstar}(a) depicts the variation of $m_\rho^*/m_\rho$ as a function of magnetic field at a temperature $T=130$ MeV and at a baryon chemical potential $\mu_B=300$ MeV. The effective mass for the two modes starts from the same value arround $eB=0$ and with the increase in magnetic field, they get separated. For both the modes, the effective $\rho^0$ mass decreases with the increase in the magnetic field which is due to the strong positive contribution coming from the dominating $eB$-dependent vacuum part. The effect of magnetic field is found to be more in Mode-(B) as compared to Mode-(A). At a magnetic field value $eB=0.20$ GeV$^2$, the effective $\rho^0$ mass in Mode-(A) decreases by about 2\% whereas for the Mode-(B) it decreases by about 10\%. 
Fig.~\ref{fig.mstar}(b) depicts the corresponding variation of effective mass with temperature at a magnetic field $eB=0.10$ GeV$^2$ and at a baryon chemical potential $\mu_B=300$ MeV. We find that, for both the modes effective mass of $\rho^0$ get enhanced by a small amount with the increase in temperature. Even at $T=160$ MeV the change in effective mass is less than 2\%.
In Fig.~\ref{fig.mstar}(c), the variation of effective $\rho^0$ mass is shown as a function of baryon chemical potential at  a magnetic field $eB=0.10$ GeV$^2$ and at a temperature $T=130$ MeV. In this case also, we observe an enhancement of the effective mass for both the modes with the increase in baryon density. Though the effect of $\mu_B$ on effective mass is more at a higher value of $\mu_B$  the change in the effective mass remains less than 2\% even at $\mu_B=500$ MeV.

\begin{figure}[h]
	\begin{center}
		\includegraphics[angle=-90,scale=0.35]{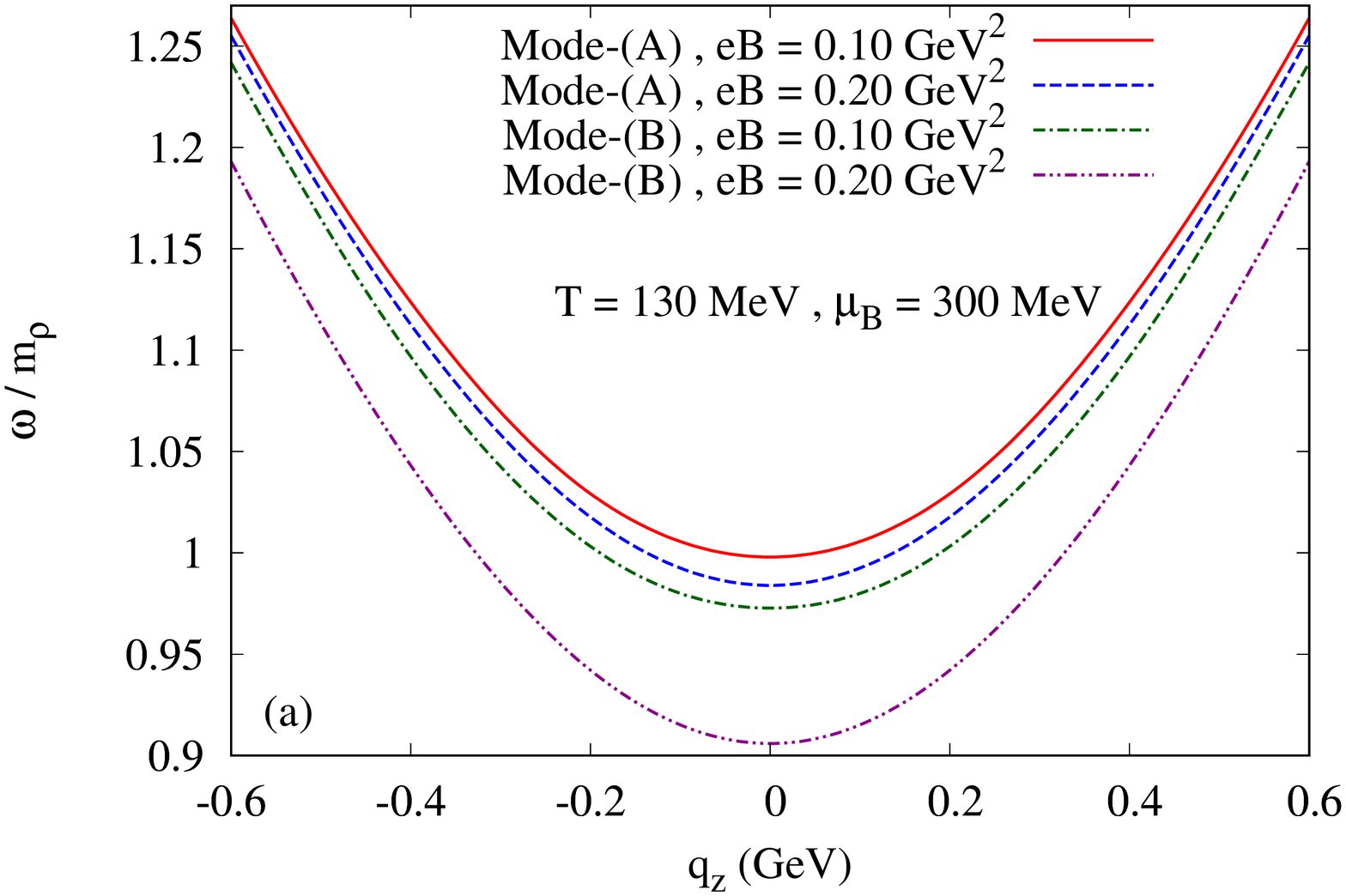}		
		\includegraphics[angle=-90,scale=0.35]{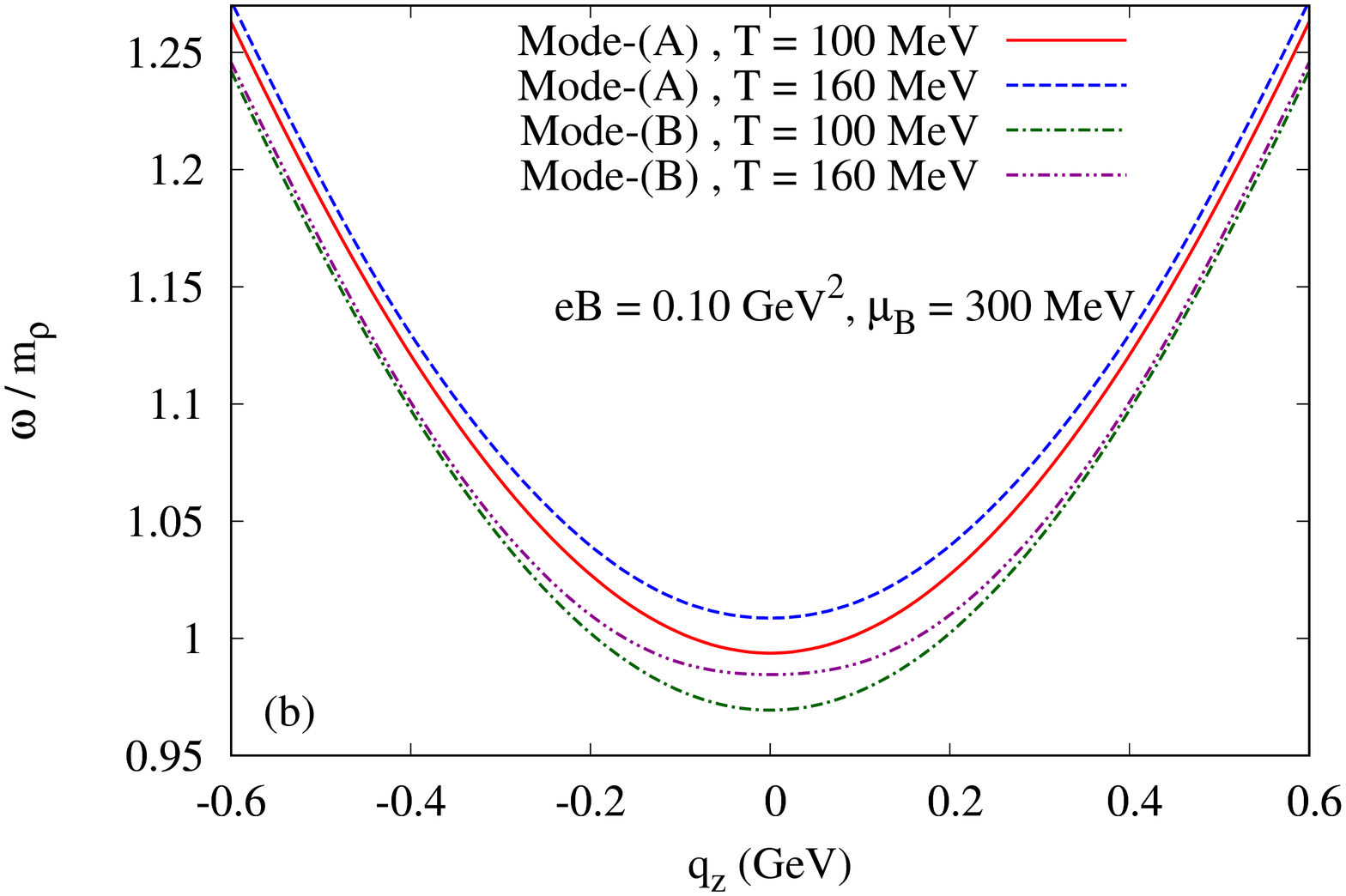} \\
		\includegraphics[angle=-90,scale=0.35]{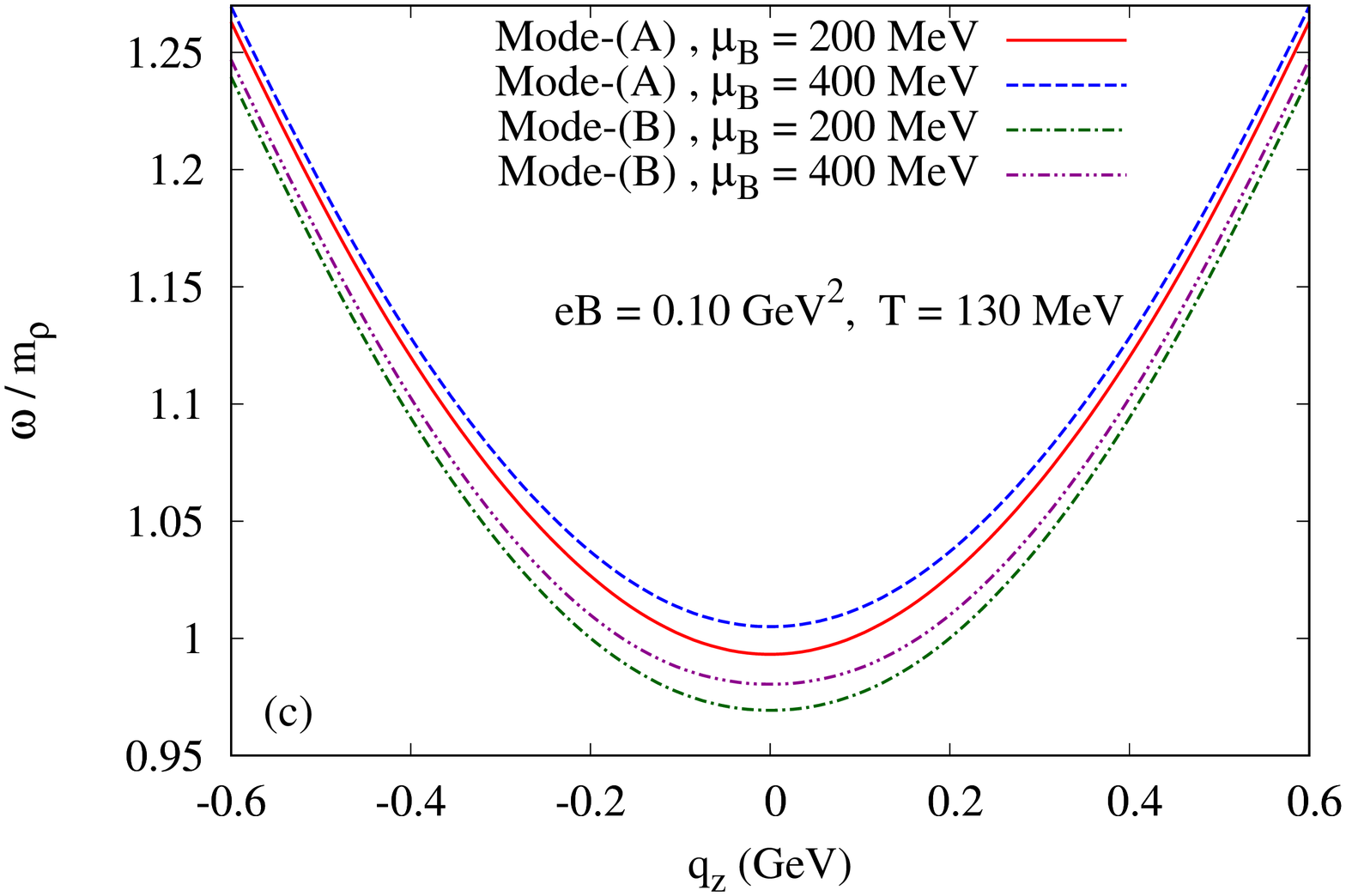}			
	\end{center}
	\caption{The dispersion relations of $\rho^0$ for different modes: (a) At temperature $T=$ 130 MeV and at baryon chemical potential $\mu_B=$ 300 MeV for two different values of magnetic field ($eB=$ 0.10 and 0.20 GeV$^2$ respectively), (b) at magnetic field $eB=$ 0.10 GeV$^2$ and at baryon chemical potential $\mu_B=$ 300 MeV for two different temperatures ($T=$ 100 and 160 MeV respectively) and (c) at magnetic field $eB=$ 0.10 GeV$^2$ and at temperature $T=$ 130 MeV for two different values of baryon chemical potential ($\mu_B=$ 200 and 400 MeV respectively). }
	\label{fig.disp}
\end{figure}

Next, we present the dispersion curves of $\rho^0$ propagation in magnetized medium for both the modes in Fig.~\ref{fig.disp}. We have plotted the energy $\omega$ of the $\rho^0$ scaled with the inverse of the vacuum rho mass $m_\rho=770$ MeV as a function of the longitudinal momentum of $\rho^0$. Fig.~\ref{fig.disp}(a) depicts the dispersion curves at temperature $T=$ 130 MeV and at baryon chemical potential $\mu_B=$ 300 MeV for two different values of magnetic field ($eB=$ 0.10 and 0.20 GeV$^2$ respectively). Fig.~\ref{fig.disp}(b) shows the same at a magnetic field $eB=$ 0.10 GeV$^2$ and  baryon chemical potential $\mu_B=$ 300 MeV for two different temperatures ($T=$ 100 and 160 MeV respectively). Finally, Fig.~\ref{fig.disp}(c) shows the corresponding graphs at a magnetic field $eB=$ 0.10 GeV$^2$ and at  a temperature $T=$ 130 MeV for two different values of baryon chemical potential ($\mu_B=$ 200 and 400 MeV respectively). In all the cases, the dispersion curves are well separated from each other at lower transverse momentum. With the increase in $q_z$, the loop correction becomes subleading with respect to the kinetic energy of $\rho^0$ and thus, it approaches to a light-like dispersion.
\begin{figure}[h]
	\begin{center}
		\includegraphics[angle=-90,scale=0.35]{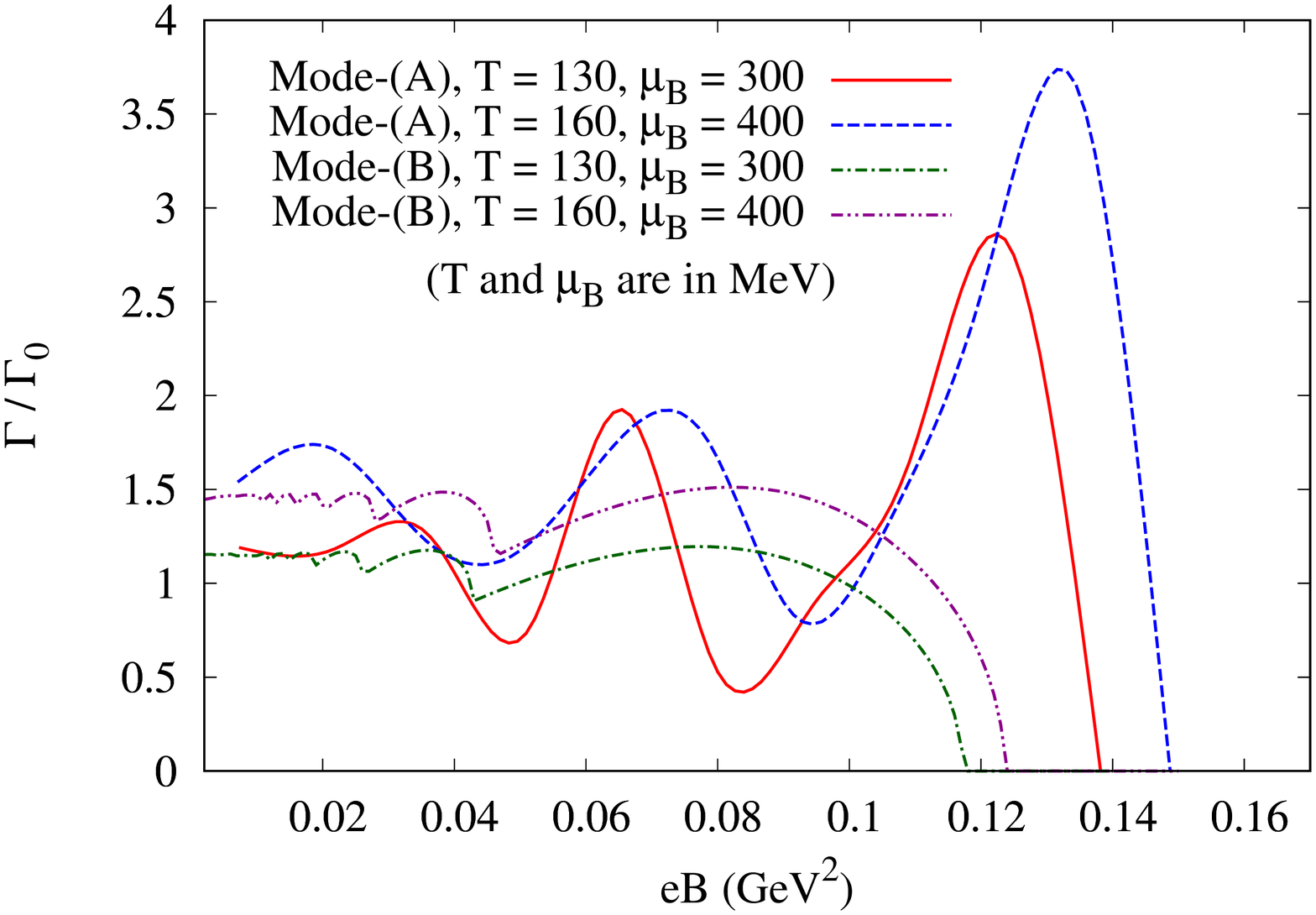}			
	\end{center}
	\caption{The ratio of the decay width of $\rho^0$ to its vacuum width as a function of magnetic field for different modes with two different combinations of temperature and baryon chemical potential (($T=130$ MeV, $\mu_B=300$ MeV) and ($T=160$ MeV, $\mu_B=400$ MeV) respectively). Here $\Gamma_0 = 156$ MeV. }
	\label{fig.decay}
\end{figure}

Finally we calculate the decay width of $\rho^0$ for the decay into charged pions which is defined for the two modes as
\begin{eqnarray}
\Gamma^{(A)} (eB,T,\mu_B) &=& \frac{\IM\Pi_\alpha(q^0=m_\rho^*,q_z=0,eB,T,\mu_B)}{m_\rho^*(eB,T,\mu_B)} \\
\Gamma^{(B)} (eB,T,\mu_B) &=& \frac{\IM\Pi_\beta(q^0=m_\rho^*,q_z=0,eB,T,\mu_B)}{m_\rho^*(eB,T,\mu_B)}~.
\end{eqnarray}
In Fig.~\ref{fig.decay}, the variation of the decay width $\Gamma$ of $\rho^0$ scaled with inverse of its vacuum width ($\Gamma_0=156$ MeV) 
for the two modes is shown as a function of magnetic field. Note that the vacuum decay width is obtained from the imaginary part of the vacuum self energy as 
\begin{eqnarray}
\Gamma_0 = \frac{\IM\Pi_\text{pure-vac}(q^0=m_\rho,\vec{q}=\vec{0})}{m_\rho} = 156 ~\text{MeV}~.
\end{eqnarray}
Results are presented for two different combinations of temperature and baryon chemical potential (($T=130$ MeV, $\mu_B=300$ MeV) and ($T=160$ MeV, $\mu_B=400$ MeV) respectively). Because of the presence of threshold singularity in $\IM\Pi_\alpha$, $\Gamma^{(A)}$ also suffers from the presence of threshold singularity for which it needs to be coarse grained. However, $\IM\Pi_\beta$ and hence $\Gamma^{(B)}$ is finite and free from the singularities. As can be seen from the figure, the ratio $\Gamma/\Gamma_0$ starts from a value greater than unity near $eB=0$ which is due to the enhancement of the decay width over its vacuum value due to the effect of finite temperature and density. Also for a particular value of magnetic field, larger  decay width is observed at higher temperature and density. Near $eB=0$, the two modes have almost the same decay widths which begin to differ from each other with the increase in the magnetic field.  An oscillatory behaviour of the decay width can be clearly seen throughout the 
magnetic field range. One should also notice that, for both the modes,  the oscillation amplitude increases whereas oscillation frequency decreases with $eB$. Finally at a critical value of the magnetic field, the decay width becomes zero. This is because of fact that, the eB-dependent Unitary cut threshold for the $\pi\pi$ loop has to  satisfy
\begin{eqnarray}
m_\rho^*(eB)> 2\sqrt{m_\pi^2+eB}
\end{eqnarray} 
 for a kinematically favorable decay of $\rho^0\rightarrow\pi^+\pi^-$. But, with the increase in magnetic field, the RHS of the above equation increases, whereas  $m_\rho^*$ in the LHS decreases so that at some critical value of magnetic field, the above inequality is violated and the decay width becomes zero. Physically it means, that $\rho^0$ becomes stable against the decay into $\pi^+\pi^-$ pair. This critical value of the field  may be considered as the critical value of the magnetic field required for the ``melting" of the spectral function of $\rho^0$.
\begin{figure}[h]
	\begin{center}
		\includegraphics[angle=-90,scale=0.35]{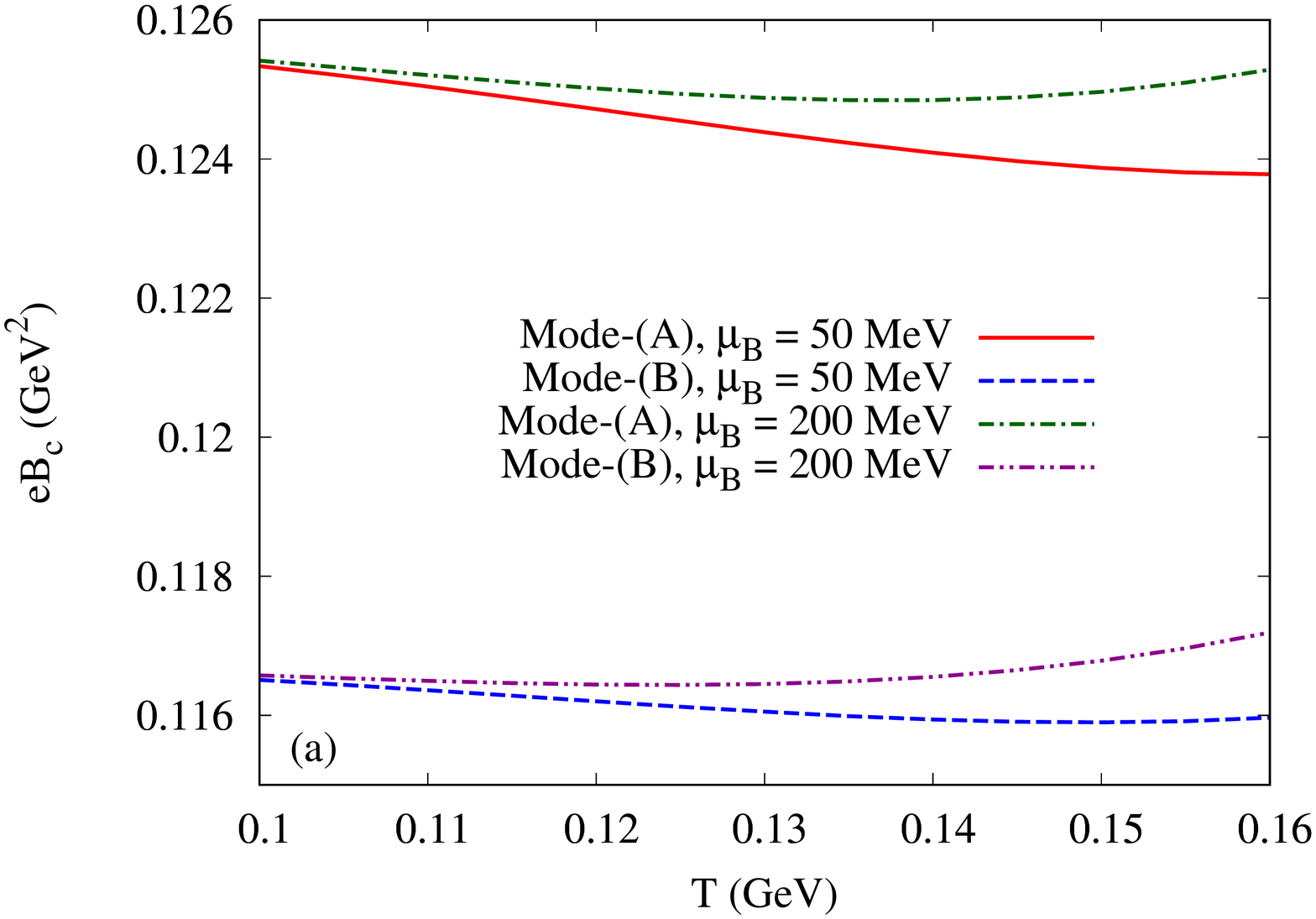}
		\includegraphics[angle=-90,scale=0.35]{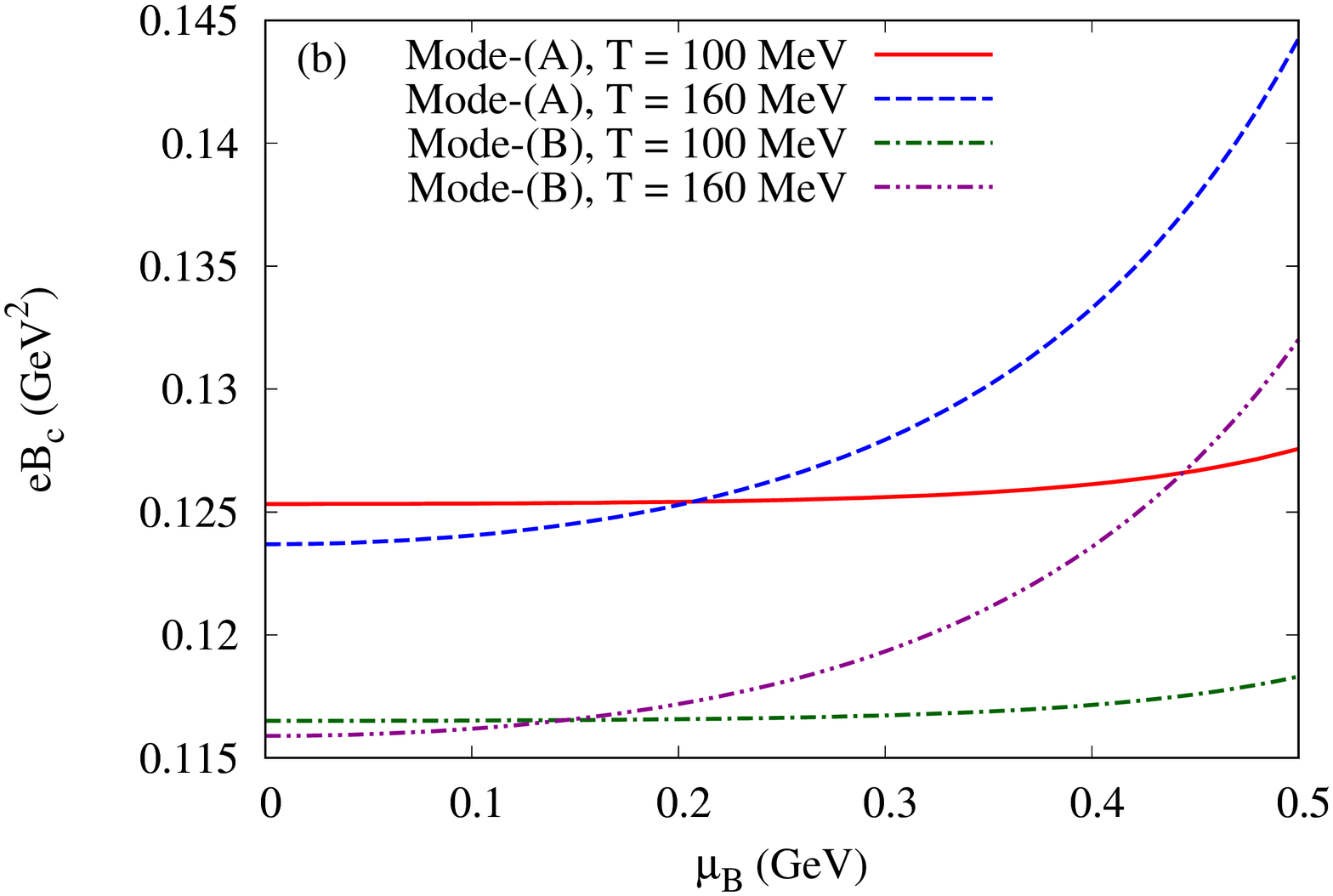}					
	\end{center}
	\caption{The variation of the critical value of magnetic field for stopping the decay of $\rho^0$ into $\pi^+\pi^-$ 
	pair for different modes as a function of (a) temperature at two different values of baryon chemical
	potential ($\mu_B=$ 50 and 200  MeV respectively) and (b) baryon chemical potential at two different values of temperature ($T=$ 100 and 160 MeV respectively). }
	\label{fig.eBc}
\end{figure}

In order to calculate the critical value of the magnetic field $eB_c$ for a given temperature $T$ and baryon chemical potential $\mu_B$, we need to solve the transcendental equation
\begin{eqnarray}
m_\rho^*(eB_c,T,\mu_B)= 2\sqrt{m_\pi^2+eB_c}~.
\end{eqnarray} 
The green dash-dotted curve in Fig.~\ref{fig.mstar}(a) corresponds to $m_\rho^*/m_\rho = 2\sqrt{m_\pi^2+eB}$ so that, the intersection of this curve with the $m_\rho^* = m_\rho^*(eB)$ represents the solution of the above equation. In Fig.~\ref{fig.eBc},
we  show the variation of the  critical  magnetic field $eB_c$ for the two decay modes.
Fig.~\ref{fig.eBc}(a) depicts  $eB_c$ as a function of temperature for two different values of baryon chemical potential ($\mu_B=50$ and 200 MeV) whereas Fig.~\ref{fig.eBc}(b) shows the corresponding variation with baryon chemical potential at two different values of temperature ($T=100$ and 160 MeV). Although, with 
fixed temperature, the variation with respect to $\mu_B$ shows monotonically increasing trend, both  the plots suggests  non-monotonic variations of the  critical  magnetic field with respect to the temperature. More specifically,  there exists a maximum value of chemical potential
(see Fig.~\ref{fig.eBc}(b)) below which the critical field decreases with the  temperature there by requiring relatively weaker magnetic field to completely stop
the particular decay channel. However, for even larger values of $\mu_B$,  a significant increase with   temperature can be observed for both of the decay modes.

Few comments on the magnitude of the external magnetic field are in order. 
The analytical expressions provided in this paper are valid for any arbitrary value of the external magnetic field which is 
constant in space-time. In presenting numerical results, we have considered magnetic field values in the range 
$0\leq eB \leq 0.20$ GeV$^2$. It is worth noting that the magnetic field created in the HIC experiments is expected to  
decay rapidly with time \cite{PhysRevC.83.054911}. However, a non-zero electrical conductivity of the strongly interacting fireball could 
possibly sustain the external magnetic field a bit longer \cite{PhysRevC.82.034904,PhysRevC.83.017901,PhysRevC.93.014905} implying a slowly varying function of time during the entire life time of the QGP. The magnitude of the 
external magnetic field at the time of chemical freezeout (when the hadronic degrees of freedom manifests) 
is expected to  be small because of  the very small conductivity of the hadron gas.  The experimental  estimation of the same is not reported yet.  In order to understand the plasma properties from the experimental data one  
solves relativistic magnetohydrodynamics equation usually with the assumption of ideal QGP fluid in the background electromagnetic field \cite{PhysRevC.96.054909,PhysRevC.96.034902,refId0}. However, the ideal fluidity assumption can only be validated after knowing the transport coefficients   
at temperatures of  phenomenological interest which are not yet certain. Despite these uncertinities,  
it should be mentioned here that  the complete blocking of the neutral rho decay seems to be quite  
unlikely in the recent energy regimes of the HIC experiments. Though, one might expect a suppression in the $\rho^0\rightarrow\pi^+\pi^-$ channel. 
Being  the only possible strong decay channel of $\rho^0$ meson, its  suppression is expected to  lead to the enhancement of dilepton and 
photon productions from $\rho^0$ decay. For example  $\rho^0\rightarrow\pi^0\gamma$ channel is expected to possess 64\% branching ratio at 
the critical magnetic field of the order 10$^{15}$ T \cite{filip}. However, recent measurement \cite{ALBUQUERQUE2019823,Acharya:2018qnp} 
shows almost no  suppression in the strong decay channel of $\rho^0$ in 
peripheral $Pb-Pb$ collisions (case of non-zero
external magnetic field) at LHC energies. However, the observed suppression
in the central region (case of zero external magnetic field) is interpreted as the re-scattering mechanism in the hadronic
medium. Thus, it suggests that  the magnetic effects on the neutral $\rho$ decay, if exists,  is negligibly small in the current HIC scenario. 
On the other hand, such magnetic modifications of mesonic properties  can occur 	
in   situations present inside the  high density compact objects with strong magnetic field such as magnetars. The tools used
in the present work can be used to see the effects of the changes of haronic properties on the equation of state, symmetry energy,
mass-radius relationship, etc. after generalization to models appropriate for the description of hadronic matter at low temperature and at high
density supposed to be present in a magnetized neutron star or magnetar \cite{Yuan_1999,Wei:2005aga}. 

\section{Summary and Conclusions} \label{sec.summary}

In this work, the  spectral properties  of the neutral rho meson is studied at finite temperature and density in  
a constant external  magnetic field using the real time formalism of finite temperature field theory. 
The effective $\rho\pi\pi$ and $\rho NN$ interactions   are considered for the evaluation of the one loop self energy  of $\rho^0$. Accordingly, the magnetically modified
in-medium propagators for  pions  and protons  are used which contain  infinite sum over  the Landau levels implying no constraint on the strength   of the external 
magnetic field. From the self-energy, 
the  eB-dependent
vacuum part  is extracted by means of dimensional regularization in which the ultraviolet divergence corresponding 
to the pure vacuum self energy is isolated as the pole of gamma and Hurwitz zeta functions. It is shown  that 
the external magnetic field does not create additional divergences so that the vacuum counter terms required in absence of the background field
 remain sufficient to renormalize the theory at non zero magnetic field. 

The general Lorentz structure for the in-medium  massive vector boson self energy in presence of external magnetic field 
has been constructed with  four linearly independent basis tensors out of which three form a mutually orthogonal set. Thus, the  
  the extraction of the form factors from the self energy becomes considerably  simple. Moreover, it is shown that with vanishing 
perpendicular momentum of the external particle, one can arrive at new set of  constraint relations among the form factors which essentially 
leave only two form factors  to be determined from the  self energy. As a consistency check,  the numerical  $B\rightarrow 0$ limit of  the real as well as imaginary
parts of the form factors  are shown to  reproduce  the  zero  field results. Solving 
the Dyson-Schwinger equation  with the one loop self energy,  the complete interacting $\rho^0$ propagator is obtained.
Consequently, two distinct modes are observed in  the study of the effective mass,  dispersion relations and the 
spectral function  of $\rho^0$ where one of the modes ( Mode-A) possesses two fold degeneracy. It is known ~\cite{Ghosh:2017rjo,Ghosh:2018xhh} 
that  non trivial Landau cuts appear in presence of external magnetic field along with finite
temperature even if the loop particles are of equal mass which is  completely a magnetic field effect. However,
in contrast to Mode-A, the 
non-trivial Landau cut is found to be absent in case of  Mode-B.   Also, sharper decrease in the effective
mass is observed for the later which essentially stems from the dominant eB-dependent vacuum contribution in the real part of the corresponding form factor.

 Finally, the decay width for  $\rho^0\rightarrow\pi^+\pi^-$ channel is obtained for the two distinct
modes and is found to  become zero at certain critical values of magnetic field depending upon the temperature and baryon chemical potential.   The corresponding 
variation of the critical field with these  external parameters shows increasing trend for
large baryonic chemical potential. However, it is observed that, both  the distinct modes possess a maximum value of $\mu_B$ below which the temperature dependence 
gets reversed. Especially, at a given temperature (say  $T=160$ MeV) , $eB_c$ attains the lowest values (123 MeV$^2$ for Mode-A and 116 MeV$^2$ for Mode-B)
in case of   zero chemical potential.  In Ref.~\cite{1674-1137-40-2-023102},  charged rho meson condensation has been studied at finite temperature and density. For charged rho mesons, the 
critical field for which the vector meson mass vanishes is observed to lie in the range of 0.2-0.6 GeV$^2$ at zero density with temperature in the range 0.2-0.5 GeV.  However, in case of $\rho^0$, the absence of the trivial Landau shift in the energy eigenvalue results in much slower decrease in the effective mass. As a consequence, unrealistically high  magnetic field values are required to observe 
neutral rho condensation in presence of temperature and medium (see Fig.\ref{fig.mstar}). In this scenario, the suppression in the  $\rho^0\rightarrow\pi^+\pi^-$ channel
can serve as an important alternative.   Magnetic modification of rho meson properties studied in this work deals with effective hadronic interactions. Thus,  the  observable modification can only occur if the initial burst of magnetic field  survives  up to hadronization 
retaining an appreciable field strength.
	However, the  recent report~\cite{ALBUQUERQUE2019823,Acharya:2018qnp} argued  
	that the observed suppression in the branching ratio of 
	$\rho^0\rightarrow\pi^+\pi^-$ channel in the central collisions 
	($B \sim 0$) is due
	to the re-scattering mechanism in the hadronic medium implying that   the magnetic 
	field effects in the neutral $\rho$ decay is negligible in  HIC experiments. On the other hand, the present study can be 
	relevant in   situations present inside  magnetars.

\section*{Acknowledgements}

S.G. acknowledges the Indian Institute of Technology Gandhinagar for the post doctoral fellowship. We are highly grateful to the people of India for their generous support for the research in fundamental sciences. 

\clearpage

\appendix

\section{Useful Identities} \label{appendix.identities}

We have the following list of $d$-dimensional integrals in Minkowski space~\cite{Peskin:1995ev}:
\begin{eqnarray}
\int\frac{d^dk}{\FB{2\pi}^d}\frac{1}{\FB{k^2-\Delta}^n} &=& 
\frac{i\FB{-1}^n}{\FB{4\pi}^{d/2}}\frac{\Gamma\FB{n-d/2}}{\Gamma\FB{n}}\FB{\frac{1}{\Delta}}^{n-d/2} \\
\int\frac{d^dk}{\FB{2\pi}^d}\frac{k^2}{\FB{k^2-\Delta}^n} &=& 
\frac{i\FB{-1}^{n-1}}{\FB{4\pi}^{d/2}}\FB{\frac{d}{2}}\frac{\Gamma\FB{n-1-d/2}}{\Gamma\FB{n}}\FB{\frac{1}{\Delta}}^{n-1-d/2} \\
\int\frac{d^dk}{\FB{2\pi}^d}\frac{k^\mu k^\nu}{\FB{k^2-\Delta}^n} &=& 
\frac{i\FB{-1}^{n-1}}{\FB{4\pi}^{d/2}}\FB{\frac{g^\munu}{2}}\frac{\Gamma\FB{n-1-d/2}}{\Gamma\FB{n}}\FB{\frac{1}{\Delta}}^{n-1-d/2}~.
\end{eqnarray}

Using the orthogonality properties of the generalized Laguerre polynomials, one can derive the following identities
\begin{eqnarray}
&&\int\frac{d^2k_\perp}{(2\pi)^2}e^{-2\alpha_k}L_{l}(2\alpha_k)L_{n}(2\alpha_k)\kper^\mu\kper^\nu  
= -\gper^\munu\frac{(eB)^2}{32\pi}\TB{(2n+1)\delta^{n}_{l}-(n+1)\delta^{n+1}_{l}-n\delta^{n-1}_{l} }\\
&&\int\frac{d^2k_\perp}{(2\pi)^2}e^{-2\alpha_k}L_l(2\alpha_k)L_n(2\alpha_k) 
= \frac{eB}{8\pi}\delta^{n}_{l} \\
&&\int\frac{d^2k_\perp}{(2\pi)^2}e^{-2\alpha_k}L^1_{l-1}(2\alpha_k)L^1_{n-1}(2\alpha_k)\kper^\mu\kper^\nu  
= -\gper^\munu\frac{(eB)^2}{32\pi}n\delta_{l-1}^{n-1} \\
&&\int\frac{d^2k_\perp}{(2\pi)^2}e^{-2\alpha_k}L^1_{l-1}(2\alpha_k)L^1_{n-1}(2\alpha_k)\kper^2  
= -\frac{(eB)^2}{16\pi}n\delta_{l-1}^{n-1}
\end{eqnarray}
where, $\alpha_k=-k_\perp^2/eB$.


\section{Calculation of Vacuum Self Energy}\label{app.vac.self}
In order to evaluate the momentum integrals in Eqs.~\eqref{eq.Pi.vac.pi.1} and \eqref{eq.Pi.vac.N.1}, they are rewritten as 
\begin{eqnarray}
\FB{\Pi^\munu_\pi}_\text{pure-vac}(q) &=& i\int\frac{d^4k}{(2\pi)^4} \frac{\mathcal{N}_\pi^\munu(q,k)}
{(k^2-m_\pi^2+i\epsilon)((q+k)^2-m_\pi^2+i\epsilon)}
\label{eq.Pi.vac.pi.2}\\
\FB{\Pi^\munu_\text{N}}_\text{pure-vac}(q) &=& i\int\frac{d^4k}{(2\pi)^4}\frac{\mathcal{N}_\text{N}^\munu(q,k)}
{(k^2-m_N^2+i\epsilon)((q+k)^2-m_N^2+i\epsilon)}
\label{eq.Pi.vac.N.2}
\end{eqnarray}
where, $\mathcal{N}_\text{N}^\munu(q,k)$ contains the trace over Dirac matrices:
\begin{eqnarray}
\mathcal{N}_\text{N}^\munu(q,k) &=& -2g_\rhoNN^2 \Tr\TB{\Gamma^\nu(q)(\cancel{q}+\cancel{k}+m_N)\Gamma^\mu(-q)(\cancel{k}+m_N)\frac{}{}} \nn \\
&=& -8g_\rhoNN^2\TB{ \frac{}{} (m_N^2-k^2-k\cdot q)g^\munu + 2k^\mu k^\nu + (q^\mu k^\nu+q^\nu k^\mu)  + \kappa_\rho\FB{q^2g^\munu-q^\mu q^\nu} \right. \nn \\ && \left.
+ \frac{\kappa_\rho^2}{4m_N^2}\SB{ (m_N^2+k^2-k\cdot q)(q^2g^\munu-q^\mu q^\nu)-2q^2k^\mu k^\nu - 2(k\cdot q)^2g^\munu 
+2(k\cdot q)(q^\mu k^\nu+q^\nu k^\mu) \frac{}{}} }~.
\label{eq.Nmunu_N}
\end{eqnarray}
Applying standard Feynman paramerization, the denominators of Eqs.~\eqref{eq.Pi.vac.pi.2} and \eqref{eq.Pi.vac.N.2} are combined to get,
\begin{eqnarray}
\FB{\Pi^\munu_\pi}_\text{pure-vac}(q) &=& i\int_{0}^{1}dx\int\frac{d^dk}{(2\pi)^d}\Lambda_\pi^{2-d/2} \frac{\mathcal{N}_\pi^\munu(q,k)}
{\TB{(k+xq)^2-\Delta_\pi}^2}\Bigg|_{d\rightarrow 4}
\label{eq.Pi.vac.pi.3}\\
\FB{\Pi^\munu_\text{N}}_\text{pure-vac}(q) &=& i\int_{0}^{1}dx\int\frac{d^dk}{(2\pi)^d}\Lambda_\text{N}^{2-d/2}\frac{\mathcal{N}_\text{N}^\munu(q,k)}
{\TB{(k+xq)^2-\Delta_\text{N}}^2}\Bigg|_{d\rightarrow 4}
\label{eq.Pi.vac.N.3}
\end{eqnarray}
where, 
\begin{eqnarray}
\Delta_\pi = m_\pi^2 -x(1-x)q^2 - i\epsilon \label{eq.Delta.pi}\\
\Delta_\text{N} = m_N^2 -x(1-x)q^2 - i\epsilon \label{eq.Delta.N} 
\end{eqnarray}
and the space-time dimension has been changed from $4$ to $d$ in order to work with the dimensional regularization so that the additional scale parameters $\Lambda_\pi$ and $\Lambda_\text{N}$ of dimension GeV$^2$ have been introduced to keep the overall dimension of the self energy same. It is now straight forward to perform the momentum integrals of the above equations after a momentum shift $k\rightarrow (k-xq)$ using the identities provided in Appendix~\ref{appendix.identities}, so that, the vacuum self energies becomes
\begin{eqnarray}
\FB{\Pi^\munu_\pi}_\text{pure-vac}(q) &=& (q^2g^\munu-q^\mu q^\nu)\FB{\frac{g_\rhopipi^2q^2}{32\pi^2}}\int_{0}^{1}dx\Gamma(\varepsilon-1)\FB{\frac{\Delta_\pi}{4\pi\Lambda_\pi}}^{-\varepsilon}\Bigg|_{\varepsilon\rightarrow 0} \label{eq.Pi.vac.pi.4} \\
\FB{\Pi^\munu_\text{N}}_\text{pure-vac}(q) &=& (q^2g^\munu-q^\mu q^\nu)\FB{\frac{g_\rhoNN^2}{2\pi^2}}\int_{0}^{1}dx
\TB{ \SB{2x(1-x)+\kappa_\rho+\frac{\kappa_\rho^2}{2}}\Gamma(\varepsilon) + \frac{\kappa_\rho^2}{4m_N^2}\Delta_\text{N}\Gamma(\varepsilon-1) } \FB{\frac{\Delta_\text{N}}{4\pi\Lambda_\text{N}}}^{-\varepsilon}\Bigg|_{\varepsilon\rightarrow 0}
\label{eq.Pi.vac.N.4}
\end{eqnarray}
where $\varepsilon=\FB{2-d/2}$. Expanding the above equations about $\varepsilon=0$, we get
\begin{eqnarray}
\FB{\Pi^\munu_\pi}_\text{pure-vac}(q) &=& (q^2g^\munu-q^\mu q^\nu)\FB{\frac{-g_\rhopipi^2q^2}{32\pi^2}}\int_{0}^{1}dx\Delta_\pi
\TB{\frac{1}{\varepsilon}-\gamma_\text{E}+1-\ln\FB{\frac{\Delta_\pi}{4\pi\Lambda_\pi}}}\Bigg|_{\varepsilon\rightarrow 0} 
\label{eq.Pi.vac.pi.5} \\
\FB{\Pi^\munu_\text{N}}_\text{pure-vac}(q) &=& (q^2g^\munu-q^\mu q^\nu)\FB{\frac{g_\rhoNN^2}{2\pi^2}}\int_{0}^{1}dx
\TB{ \SB{2x(1-x)+\kappa_\rho+\frac{\kappa_\rho^2}{2} -\frac{\kappa_\rho^2}{4m_N^2}\Delta_\text{N} } \right. \nn \\
	&& \left. \SB{\frac{1}{\varepsilon}-\gamma_\text{E}-\ln\FB{\frac{\Delta_\tN}{4\pi\Lambda_\tN}}}
	-\frac{\kappa_\rho^2}{4m_N^2}\Delta_\text{N}} \Bigg|_{\varepsilon\rightarrow 0}
\label{eq.Pi.vac.N.5}
\end{eqnarray}
where, $\gamma_\text{E}$ is the Euler-Mascheroni constant. 


\section{Calculation of eB-dependent Vacuum Contribution for $\pi\pi$ Loop} \label{app.eb.vac.pi}

In this appendix, we sketch how to obtain Eqs.~\eqref{eq.ebvac.pi.10} and \eqref{eq.ebvac.pi.11}.
We rewrite Eq.~\eqref{eq.ebvac.pi.1} as
\begin{eqnarray}
\FB{\Pi^\munu_\pi}_\text{vac}(q,eB) = i\sum_{l=0}^{\infty}\sum_{n=0}^{\infty}\int\frac{d^2k_\parallel}{(2\pi)^2}\int\frac{d^2k_\perp}{(2\pi)^2}
\frac{\mathcal{N}^\munu_{\pi,nl}(q,k)}{(k_\parallel^2-m_l^2+i\epsilon)((q_\parallel+k_\parallel)^2-m_n^2+i\epsilon)}
\label{eq.ebvac.pi.2}
\end{eqnarray}
For the simplicity in analytic calculations, we take the transverse momentum of the $\rho^0$ to be zero i.e. $q_\perp=0$. This implies
that the $d^2k_\perp$ integration can be performed analytically using the orthogonality of the Laguerre polynomial details of which can be obtained from Appendix~\ref{app.d2k.integral}, so that the self energy becomes
\begin{eqnarray}
\FB{\Pi^\munu_\pi}_\text{vac}(q_\parallel,eB) = i\sum_{l=0}^{\infty}\sum_{n=0}^{\infty}\int\frac{d^2k_\parallel}{(2\pi)^2}
\frac{\tilde{\mathcal{N}}^\munu_{\pi,nl}(q_\parallel,k_\parallel)}{(k_\parallel^2-m_l^2+i\epsilon)((q_\parallel+k_\parallel)^2-m_n^2+i\epsilon)}
\label{eq.ebvac.pi.3}
\end{eqnarray}
where, $\tilde{\mathcal{N}}^\munu_{\pi,nl}(q_\parallel,k_\parallel)$ is given in Eq.~\eqref{eq.N.pi.5}.
Next, we use the standard Feynman parametrization technique to combine the denominators of Eq.~\eqref{eq.ebvac.pi.3} and change the reduced space-time dimension from $2$ to $d$ in order to apply the dimensional regularization for which a scale parameter 
$\Lambda_\pi$ of dimension GeV$^2$ has to be introduced in order to keep the overall dimension of the self energy same. This leads to 
\begin{eqnarray}
\FB{\Pi^\munu_\pi}_\text{vac}(q_\parallel,eB) = i\sum_{l=0}^{\infty}\sum_{n=0}^{\infty}\int_{0}^{1}dx\int\frac{d^dk_\parallel}{(2\pi)^d}\Lambda_\pi^{1-d/2}
\frac{\tilde{\mathcal{N}}^\munu_{\pi,nl}(q_\parallel,k_\parallel)}{\TB{(k_\parallel+xq_\parallel)^2-\Delta^\pi_{nl}}^2}\Bigg|_{d\rightarrow2}
\label{eq.ebvac.pi.4}
\end{eqnarray}
where, 
\begin{eqnarray}
\Delta^\pi_{nl} = \Delta_\pi(q_\perp=0) + 2eB\SB{l+\frac{1}{2}-x(l-n)}
\end{eqnarray}
with $\Delta_\pi$ is defined in Eq.~\eqref{eq.Delta.pi}. It is now trivial to perform the $d^dk_\parallel$ integration after a shift of momentum $\kpll\rightarrow(\kpll-x\qpll)$ using the identities provided in Appendix~\ref{appendix.identities}, so that the self energy becomes
\begin{eqnarray}
\FB{\Pi^\munu_\pi}_\text{vac}(q_\parallel,eB) &=& \frac{-g_\rhopipi^2q_\parallel^2}{16\pi^2}eB\int_{0}^{1}dx\sum_{n=0}^{\infty}~
\sum_{l=(n-1)}^{(n+1)}
(-1)^{n+l}\FB{4\pi\Lambda_\pi}^\varepsilon\TB{ -(q_\parallel^2g_\parallel^\munu-q_\parallel^\mu q_\parallel^\nu)\delta_l^n\Gamma(\varepsilon)\FB{\Delta_{nl}^\pi}^{-\varepsilon} \right. \nn \\ 
&& \left. -q_\parallel^2g_\perp^\munu\frac{eB}{2}\SB{(2n+1)\delta_l^n-(n+1)\delta_l^{n+1}-n\delta_l^{n-1}}\Gamma(\varepsilon+1)
\FB{\Delta_{nl}^\pi}^{-\varepsilon-1} } \Bigg|_{\varepsilon\rightarrow0}
\label{eq.ebvac.pi.5}
\end{eqnarray}
where $\varepsilon=(1-d/2)$ and the presence of Kronecker delta functions in Eq.~\eqref{eq.N.pi.5} has made the double sum into a single one or in other words the sum over index $l$ runs only from $(n-1)$ to $(n+1)$. The infinite sum in the above equations can be expressed in terms of Hurwitz zeta function so that we get after some simplifications
\begin{eqnarray}
\FB{\Pi^\munu_\pi}_\text{vac}(q_\parallel,eB) &=& \frac{-g_\rhopipi^2q_\parallel^2}{16\pi^2}eB\int_{0}^{1}dx
\FB{\frac{4\pi\Lambda_\pi}{2eB}}^\varepsilon
\TB{ -(q_\parallel^2g_\parallel^\munu-q_\parallel^\mu q_\parallel^\nu)\Gamma(\varepsilon)\zeta\FB{\varepsilon,z_\pi+\frac{1}{2}}
	-\frac{q_\parallel^2}{2}g_\perp^\munu\Gamma(\varepsilon+1) \times
	\right. \nn \\ && \left.
	\SB{ \zeta\FB{\varepsilon,z_\pi+\frac{1}{2}} + \zeta\FB{\varepsilon,z_\pi+x+\frac{1}{2}} - z_\pi\zeta\FB{\varepsilon+1,z_\pi+\frac{1}{2}} -z_\pi \zeta\FB{\varepsilon+1,z_\pi+x+\frac{1}{2}} }} 
 \Bigg|_{\varepsilon\rightarrow0}~.
\label{eq.ebvac.pi.6}
\end{eqnarray}
where, $z_\pi=\frac{\Delta_\pi(q_\perp=0)}{2eB}$. Expanding the above equation about $\varepsilon=0$, we get,
\begin{eqnarray}
\FB{\Pi^\munu_\pi}_\text{vac}(q_\parallel,eB) &=& \frac{-g_\rhopipi^2q_\parallel^2}{32\pi^2}\int_{0}^{1}dx
\TB{ \SB{\frac{1}{\varepsilon}-\gamma_\text{E} + \ln\FB{\frac{4\pi\Lambda_\pi}{2eB}} } \Delta_\pi(q_\perp=0)
	(q_\parallel^2g^\munu-q_\parallel^\mu q_\parallel^\nu) \right. \nn \\ && \left.
	- (q_\parallel^2g_\parallel^\munu-q_\parallel^\mu q_\parallel^\nu)2eB
	\SB{\ln\Gamma\FB{z_\pi+\frac{1}{2}}-\ln\sqrt{2\pi}} \right. \nn \\ && \left.
	+ q_\parallel^2g_\perp^\munu\SB{ \Delta_\pi(q_\perp=0)+\frac{eB}{2}-\frac{1}{2}\Delta_\pi(q_\perp=0)
	\SB{\psi\FB{z_\pi+\frac{1}{2}} + \psi\FB{z_\pi+x+\frac{1}{2}  } } } }
\Bigg|_{\varepsilon\rightarrow0}
\label{eq.ebvac.pi.7}
\end{eqnarray}
where, $\psi(z)$ is the digamma function. It is now trivial to check that, in the limit $eB\rightarrow0$, the above equation exactly boils down to the pure vacuum contribution 
given in Eq.~\eqref{eq.Pi.vac.pi.6}. Thus extracting the pure vacuum contribution from the above equation we get, 
\begin{eqnarray}
\FB{\Pi^\munu_\pi}_\text{vac}(q_\parallel,eB) = \FB{\Pi^\munu_\pi}_\text{pure-vac}(q_\parallel) + 
\FB{\Pi^\munu_\pi}_\text{eB-vac}(q_\parallel,eB)
\label{eq.ebvac.pi.8}
\end{eqnarray}
where,
\begin{eqnarray}
\FB{\Pi^\munu_\pi}_\text{eB-vac}(q_\parallel,eB) &=& \frac{-g_\rhopipi^2q_\parallel^2}{32\pi^2}\int_{0}^{1}dx
\TB{ \SB{\ln\FB{\frac{\Delta_\pi(q_\perp=0)}{2eB}}-1 } \Delta_\pi(q_\perp=0)
	(q_\parallel^2g^\munu-q_\parallel^\mu q_\parallel^\nu) \right. \nn \\ && \left.
	- (q_\parallel^2g_\parallel^\munu-q_\parallel^\mu q_\parallel^\nu)2eB
	\SB{\ln\Gamma\FB{z_\pi+\frac{1}{2}}-\ln\sqrt{2\pi}} \right. \nn \\ && \left.
	+ q_\parallel^2g_\perp^\munu\SB{ \Delta_\pi(q_\perp=0)+\frac{eB}{2}-\frac{1}{2}\Delta_\pi(q_\perp=0)
		\SB{\psi\FB{z_\pi+\frac{1}{2}} + \psi\FB{z_\pi+x+\frac{1}{2}  } } } }
\label{eq.ebvac.pi.9}
\end{eqnarray}
which is finite and independent of scale.


\section{Calculation of eB-dependent Vacuum Contribution for proton-proton Loop} \label{app.eb.vac.p}

In this appendix, we sketch how to obtain Eqs.~\eqref{eq.ebvac.p.10} and \eqref{eq.ebvac.p.11}
We rewrite Eq.~\eqref{eq.ebvac.p.1} as
\begin{eqnarray}
\FB{\Pi^\munu_\tp}_\text{vac}(q,eB) = i\sum_{l=0}^{\infty}\sum_{n=0}^{\infty}\int\frac{d^2k_\parallel}{(2\pi)^2}\int\frac{d^2k_\perp}{(2\pi)^2}
\frac{\mathcal{N}^\munu_{\tp,nl}(q,k)}{(k_\parallel^2-M_l^2+i\epsilon)((q_\parallel+k_\parallel)^2-M_n^2+i\epsilon)}
\label{eq.ebvac.p.2}
\end{eqnarray}
where, $\mathcal{N}^\munu_{\tp,nl}(q,k)$ is given in Eq.~\eqref{eq.N.p.2}.
For the simplicity in analytic calculations, we take the transverse momentum of the $\rho^0$ to be zero i.e. $q_\perp=0$. This implies
that the $d^2k_\perp$ integration can be performed analytically using the orthogonality of the Laguerre polynomial details of which can be obtained from Appendix~\ref{app.d2k.integral}, so that the self energy becomes
\begin{eqnarray}
\FB{\Pi^\munu_\tp}_\text{vac}(q_\parallel,eB) = i\sum_{l=0}^{\infty}\sum_{n=0}^{\infty}\int\frac{d^2k_\parallel}{(2\pi)^2}
\frac{\tilde{\mathcal{N}}^\munu_{\tp,nl}(q_\parallel,k_\parallel)}{(k_\parallel^2-M_l^2+i\epsilon)((q_\parallel+k_\parallel)^2-M_n^2+i\epsilon)}
\label{eq.ebvac.p.3}
\end{eqnarray}
where, $\tilde{\mathcal{N}}^\munu_{\tp,nl}(q_\parallel,k_\parallel)$ can be read off from Eq.~\eqref{eq.N.p.5}.
Next, we use the standard Feynman parametrization technique to combine the denominators of Eq.~\eqref{eq.ebvac.p.3} and change the reduced space-time dimension from $2$ to $d$ in order to apply the dimensional regularization for which a scale parameter 
$\Lambda_N$ of dimension GeV$^2$ has to be introduced in order to keep the overall dimension of the self energy same. This leads to 
\begin{eqnarray}
\FB{\Pi^\munu_\tp}_\text{vac}(q_\parallel,eB) = i\sum_{l=0}^{\infty}\sum_{n=0}^{\infty}\int_{0}^{1}dx\int\frac{d^dk_\parallel}{(2\pi)^d}\Lambda_N^{1-d/2}
\frac{\tilde{\mathcal{N}}^\munu_{\tp,nl}(q_\parallel,k_\parallel)}{\TB{(k_\parallel+xq_\parallel)^2-\Delta^\tp_{nl}}^2}\Bigg|_{d\rightarrow2}
\label{eq.ebvac.p.4}
\end{eqnarray}
where, 
\begin{eqnarray}
\Delta^\tp_{nl} = \Delta_N(q_\perp=0) + 2eB\SB{l-x(l-n)}
\end{eqnarray}
with $\Delta_N$ is defined in Eq.~\eqref{eq.Delta.N}. It is now trivial to perform the $d^dk_\parallel$ integration after a shift of momentum $\kpll\rightarrow(\kpll-x\qpll)$ using the identities provided in Appendix~\ref{appendix.identities}, so that the self energy becomes
\begin{eqnarray}
\FB{\Pi^\munu_\tp}_\text{vac}(q_\parallel,eB) &=& \frac{g_\rhoNN^2}{4\pi^2}eB\int_{0}^{1}dx\sum_{n=0}^{\infty}~
\sum_{l=(n-1)}^{(n+1)}(-1)^{n+l}\FB{4\pi\Lambda_\pi}^\varepsilon \Bigg[\Bigg. \TB{ 4eB\gpll^\munu n\delta_{l-1}^{n-1} 
 \right. \nn \\ && \left.
+ \SB{ (m_N^2+x(1-x)\qpll^2)\gpll^\munu-2x(1-x)\qpll^\mu\qpll^\nu }
\FB{\delta_{l-1}^{n-1}+\delta_{l}^{n}} - (m_N^2+x(1-x)\qpll^2)\gper^\munu\FB{\delta_{l-1}^{n}+\delta_{l}^{n-1}} } \times 
\nn \\ &&
\Gamma(\varepsilon+1)\FB{\Delta^\tp_{nl}}^{-\varepsilon-1} - \SB{ \gpll^\munu\FB{\delta_{l-1}^{n-1}+\delta_{l}^{n}}\varepsilon 
	+\gper^\munu\FB{\delta_{l-1}^{n}+\delta_{l}^{n-1}}(-\varepsilon+1) } \Gamma(\varepsilon)\FB{\Delta^\tp_{nl}}^{-\varepsilon} \nn \\
&& +\kappa_\rho \SB{ (\qpll^2\gpll^\munu-\qpll^\mu\qpll^\nu)\FB{\delta_{l-1}^{n-1}+\delta_{l}^{n}} -\qpll^2\gper^\munu
\FB{\delta_{l-1}^{n}+\delta_{l}^{n-1}}}\Gamma(\varepsilon+1)\FB{\Delta^\tp_{nl}}^{-\varepsilon-1} \nn \\ 
&& +\frac{\kappa_\rho^2}{4m_N^2}\TB{ \SB{-4eBn\delta_{l-1}^{n-1} +(m_N^2+x(1-x)\qpll^2)
\FB{\delta_{l-1}^{n-1}+\delta_{l}^{n}} }(\qpll^2\gpll^\munu-\qpll^\mu\qpll^\nu) \right. \nn \\ && \left.
-\qpll^2(m_N^2+x(1+x)\qpll^2)\gper^\munu \FB{\delta_{l-1}^{n}+\delta_{l}^{n-1}} }\Gamma(\varepsilon+1)\FB{\Delta^\tp_{nl}}^{-\varepsilon-1} 
\nn \\ &&
- \frac{\kappa_\rho^2}{4m_N^2}\SB{ (\qpll^2\gpll^\munu-\qpll^\mu\qpll^\nu)(-\varepsilon-1)\FB{\delta_{l-1}^{n-1}+\delta_{l}^{n}}
+\qpll^2\gper^\munu\FB{\delta_{l-1}^{n}+\delta_{l}^{n-1}}\varepsilon }\Gamma(\varepsilon)\FB{\Delta^\tp_{nl}}^{-\varepsilon}\Bigg.\Bigg]
\Bigg|_{\varepsilon\rightarrow0}
\label{eq.ebvac.p.5}
\end{eqnarray}
where $\varepsilon=(1-d/2)$ and the presence of Kronecker delta functions in Eq.~\eqref{eq.N.p.5} has made the double sum into a single one or in other words the sum over index $l$ runs only from $(n-1)$ to $(n+1)$. The infinite sum in the above equations can be expressed in terms of Hurwitz zeta function so that we get after some simplifications
\begin{eqnarray}
\FB{\Pi^\munu_\tp}_\text{vac}(q_\parallel,eB) &=& \frac{g_\rhopipi^2}{4\pi^2}\int_{0}^{1}dx
\FB{\frac{4\pi\Lambda_N}{2eB}}^\varepsilon \Bigg[\Bigg. \TB{2eB\gpll^\munu\SB{\zeta(\varepsilon,z_N)-z_N\zeta(\varepsilon+1,z_N)} 
	\right. \nn \\ && \left.
+\SB{(m_N^2+x(1-x)\qpll^2)\gpll^\munu-2x(1-x)\qpll^\mu\qpll^\nu} \SB{\zeta(\varepsilon+1,z_N)-\frac{1}{2}z_N^{-\varepsilon-1}} 
\right. \nn \\ && \left.
+(m_N^2+x(1-x)\qpll^2)\gper^\munu\zeta(\varepsilon+1,z_N+x) } \Gamma(\varepsilon+1) 
-2eB \SB{ \gpll^\munu\varepsilon\FB{\zeta(\varepsilon,z_N)-\frac{1}{2}z_N^{-\varepsilon}} \right. \nn \\ && \left.
+\gper^\munu(\varepsilon-1)\zeta(\varepsilon,z_N+x) \frac{}{}}\Gamma(\varepsilon) 
+ \kappa_\rho\SB{ (\qpll^2\gpll^\munu-\qpll^\mu\qpll^\nu)\FB{ \zeta(1+\varepsilon,z_N)-\frac{1}{2}z_N^{-\varepsilon-1}}
\right. \nn \\ && \left. +\qpll^2\gper^\munu\zeta(\varepsilon+1,z_N+x) }\Gamma(\varepsilon+1) + \frac{\kappa_\rho^2}{4m_N^2}
\TB{ \SB{-2eB\FB{ \zeta(\varepsilon,z_N)-z_N\zeta(\varepsilon+1,z_N)\frac{}{}} \right.\right.\nn \\ && \left. \left.
+ (m_N^2+x(1-x)\qpll^2)
\FB{ \zeta(\varepsilon+1,z_N)-\frac{1}{2}z_N^{-\varepsilon-1}}}(\qpll^2\gpll^\munu-\qpll^\mu\qpll^\nu) \right. \nn \\ && \left.
+ \qpll^2\gper^\munu(m_N^2+x(1-x)\qpll^2)\zeta(\varepsilon+1,z_N+x) }\Gamma(\varepsilon+1) + \frac{\kappa_\rho^2}{4m_N^2} 2eB
\SB{(\qpll^2\gpll^\munu-\qpll^\mu\qpll^\nu)(\varepsilon+1) \times \right. \nn \\ && \left.
\FB{\zeta(\varepsilon,z_N)-\frac{1}{2}z_N^{-\varepsilon}}+\qpll^2\gper^\munu\varepsilon\zeta(\varepsilon,z_N+x)}\Gamma(\varepsilon)
\Bigg.\Bigg]
\Bigg|_{\varepsilon\rightarrow0}~,
\label{eq.ebvac.p.6}
\end{eqnarray}
where, $z_N=\frac{\Delta_N(q_\perp=0)}{2eB}$. Expanding the above equation about $\varepsilon=0$, we get,
\begin{eqnarray}
\FB{\Pi^\munu_\tp}_\text{vac}(q_\parallel,eB) &=& \frac{g_\rhoNN^2}{4\pi^2}\int_{0}^{1}dx \Bigg[\Bigg.
 \SB{\frac{1}{\varepsilon}-\gamma_\text{E} + \ln\FB{\frac{4\pi\Lambda_N}{2eB}} } \SB{2x(1-x)+\kappa_\rho+\frac{\kappa_\rho^2}{2}-\frac{\kappa_\rho^2}{4m_N^2}\Delta_N(\qper=0)} 
 (q_\parallel^2g^\munu-q_\parallel^\mu q_\parallel^\nu) \nn \\
&& -2x(1-x)\FB{ \psi(z_N)+\frac{1}{2z_N}}(q_\parallel^2\gpll^\munu-q_\parallel^\mu q_\parallel^\nu) 
+ 2eB \gper^\munu \SB{ \FB{z_N-\frac{m_N^2}{eB}}\psi(z_N+x)+z_N \right. \nn \\ && \left.
	+ \ln\Gamma(z+x)-\ln\sqrt{2\pi}\frac{}{}} - \kappa_\rho \SB{ (q_\parallel^2\gpll^\munu-q_\parallel^\mu q_\parallel^\nu) 
 \FB{\psi(z_N)+\frac{1}{2z_N}}  + \qpll^2\gper^\munu\psi(z+x) } \nn \\ &&
+ \frac{\kappa_\rho^2}{4m_N^2}2eB\TB{ (q_\parallel^2\gpll^\munu-q_\parallel^\mu q_\parallel^\nu)
	\SB{ -\frac{m_N^2}{eB}\FB{\psi(z_N)+\frac{1}{2z_N}} + \frac{1}{2}\ln(z_N)+\ln\Gamma(z_N)-\ln\sqrt{2\pi} } \right. \nn \\  && \left.
-\qpll^2\gper^\munu\SB{ \FB{\frac{m_N^2}{eB}-z_N}\psi(z_N+x)+\Delta_N(\qper=0) }}
\Bigg|_{\varepsilon\rightarrow0}~.
\label{eq.ebvac.p.7}
\end{eqnarray}
It is now trivial to check that, in the limit $eB\rightarrow0$, the above equation exactly boils down to the $\frac{1}{2}$ times pure vacuum contribution 
given in Eq.~\eqref{eq.Pi.vac.N.6}. Thus extracting the pure vacuum contribution from the above equation we get, 
\begin{eqnarray}
\FB{\Pi^\munu_\tp}_\text{vac}(q_\parallel,eB) = \frac{1}{2}\FB{\Pi^\munu_\tN}_\text{pure-vac}(q_\parallel) + 
\FB{\Pi^\munu_\tp}_\text{eB-vac}(q_\parallel,eB)
\label{eq.ebvac.p.8}
\end{eqnarray}
where,
\begin{eqnarray}
\FB{\Pi^\munu_\tp}_\text{eB-vac}(q_\parallel,eB) &=& \frac{g_\rhoNN^2}{4\pi^2}\int_{0}^{1}dx \Bigg[\Bigg.
 \ln\FB{\frac{\Delta_N(\qper=0)}{2eB}} \SB{2x(1-x)+\kappa_\rho+\frac{\kappa_\rho^2}{2}-\frac{\kappa_\rho^2}{4m_N^2}\Delta_N(\qper=0)} 
(q_\parallel^2g^\munu-q_\parallel^\mu q_\parallel^\nu) \nn \\
&& -2x(1-x)\FB{ \psi(z_N)+\frac{1}{2z_N}}(q_\parallel^2\gpll^\munu-q_\parallel^\mu q_\parallel^\nu) 
+ 2eB \gper^\munu \SB{ \FB{z_N-\frac{m_N^2}{eB}}\psi(z_N+x)+z_N \right. \nn \\ && \left.
	+ \ln\Gamma(z+x)-\ln\sqrt{2\pi}\frac{}{}} - \kappa_\rho \SB{ (q_\parallel^2\gpll^\munu-q_\parallel^\mu q_\parallel^\nu) 
	\FB{\psi(z_N)+\frac{1}{2z_N}}  + \qpll^2\gper^\munu\psi(z+x) } \nn \\ &&
+ \frac{\kappa_\rho^2}{4m_N^2}2eB\TB{ (q_\parallel^2\gpll^\munu-q_\parallel^\mu q_\parallel^\nu)
	\SB{ -\frac{m_N^2}{eB}\FB{\psi(z_N)+\frac{1}{2z_N}} + \frac{1}{2}\ln(z_N)+\ln\Gamma(z_N)-\ln\sqrt{2\pi} } \right. \nn \\  && \left.
	-\qpll^2\gper^\munu\SB{ \FB{\frac{m_N^2}{eB}-z_N}\psi(z_N+x)+\Delta_N(\qper=0) } 
	+ \frac{\kappa_\rho^2}{4m_N^2}(\qpll^2g^\munu-\qpll^\mu\qpll^\nu)\Delta_N(\qper=0) }
\label{eq.ebvac.p.9}
\end{eqnarray}
which is finite and independent of scale.


\section{Analytic Evaluation of $d^2k_\perp$ Integral for $\qper=0$ } \label{app.d2k.integral}

In this appendix we will calculate the quantities
\begin{eqnarray}
\tilde{\mathcal{N}}^\munu_{\pi,nl}(q_\parallel,k_\parallel) &=& \int\frac{d^2\kper}{(2\pi)^2}
\mathcal{N}^\munu_{\pi,nl}(q_\parallel,q_\perp=0,k) \\
\tilde{\mathcal{N}}^\munu_{\tp,nl}(q_\parallel,k_\parallel) &=& \int\frac{d^2\kper}{(2\pi)^2}
\mathcal{N}^\munu_{\tp,nl}(q_\parallel,q_\perp=0,k)~.
\end{eqnarray}
We have the expression for $\mathcal{N}^\munu_{\pi,nl}(q,k)$ from Eqs.~\eqref{eq.N.pi.2} and \eqref{eq.N.pi.1} as
\begin{eqnarray}
\mathcal{N}^\munu_{\pi,nl}(q,k) &=& 4g_\rhopipi^2(-1)^{n+l}e^{-\alpha_k-\alpha_p}L_l(2\alpha_k)L_n(2\alpha_p)\TB{q^4k^\mu k^\nu + (q\cdot k)^2q^\mu q^\nu - q^2(q\cdot k)(q^\mu k^\nu+q^\nu k^\mu) \frac{}{}}~.
\label{eq.N.pi.3}
\end{eqnarray} 
which for $q_\perp=0$ becomes
\begin{eqnarray}
\mathcal{N}^\munu_{\pi,nl}(q_\parallel,k) &=& 4g_\rhopipi^2(-1)^{n+l}e^{-2\alpha_k}L_l(2\alpha_k)L_n(2\alpha_k)\TB{q_\parallel^4k^\mu k^\nu + (q_\parallel\cdot k_\parallel)^2q_\parallel^\mu q_\parallel^\nu - q_\parallel^2(q_\parallel\cdot k_\parallel)(q_\parallel^\mu k^\nu+q_\parallel^\nu k^\mu) \frac{}{}}~.
\label{eq.N.pi.4}
\end{eqnarray} 
We now perform the $d^2k_\perp$ integration using the orthogonality of the Laguerre polynomial (identities provided in Appendix~\ref{appendix.identities}) to obtain 
\begin{eqnarray}
\tilde{\mathcal{N}}^\munu_{\pi,nl}(q_\parallel,k_\parallel) &=&  4g_\rhopipi^2(-1)^{n+l}\frac{eB}{8\pi}
\TB{\SB{q_\parallel^4k_\parallel^\mu k_\parallel^\nu + (q_\parallel\cdot k_\parallel)^2q_\parallel^\mu q_\parallel^\nu - q_\parallel^2(q_\parallel\cdot k_\parallel)(q_\parallel^\mu k_\parallel^\nu+q_\parallel^\nu k_\parallel^\mu)}\delta_l^n \frac{}{} \right. \nn \\ 
	&& \left. - q_\parallel^4 g_\perp^\munu \frac{eB}{4}\SB{(2n+1)\delta_l^n-(n+1)\delta_l^{n+1}-n\delta_l^{n-1}} \frac{}{}}
\label{eq.N.pi.5}
\end{eqnarray} 

Similarly, $\mathcal{N}^\munu_{\tp,nl}(q,k)$ is obtained from Eq.~\eqref{eq.N.p.2} as
\begin{eqnarray}
\mathcal{N}^\munu_{\tp,nl}(q,k) &=& -g_\rhoNN^2(-1)^{n+l}e^{-\alpha_k-\alpha_p}
\Tr\TB{\Gamma^\nu(q)\mathcal{D}_n(q+k)\Gamma^\mu(-q)\mathcal{D}_l(k)}
\label{eq.N.p.3}
\end{eqnarray} 
Evaluating the trace over the Dirac matrices in the above equation, we get for $q_\perp=0$ (considering the Lorentz symmetric part since the self energy should be symmetric in the two Lorentz indices)
\begin{eqnarray}
\mathcal{N}^\munu_{\tp,nl}(q_\parallel,k) &=& -8g_\rhoNN^2(-1)^{n+l}e^{-2\alpha_k} \Bigg[\Bigg. 8(2k_\perp^\mu k_\perp^\nu-k_\perp^2g^\munu)
L_{l-1}^1(2\alpha_k)L_{n-1}^1(2\alpha_k)  \nn \\ && 
+ \SB{  (m_N^2-\kpll^2-\kpll\cdot\qpll)\gpll^\munu + 2\kpll^\mu\kpll^\nu + (\qpll^\mu\kpll^\nu+\qpll^\nu\kpll^\mu) }
\SB{L_{l-1}(2\alpha_k)L_{n-1}(2\alpha_k) + L_{l}(2\alpha_k)L_{n}(2\alpha_k) \frac{}{}}  \nn \\ && 
-(m_N^2-\kpll^2-\kpll\cdot\qpll)\gper^\munu \SB{L_{l}(2\alpha_k)L_{n-1}(2\alpha_k) + L_{l-1}(2\alpha_k)L_{n}(2\alpha_k)\frac{}{}} 
\nn \\ && 
+ \kappa_\rho \TB{ (\qpll^2\gpll^\munu-\qpll^\mu\qpll^\nu)\SB{L_{l-1}(2\alpha_k)L_{n-1}(2\alpha_k) + L_{l}(2\alpha_k)L_{n}(2\alpha_k) \frac{}{}} \right. \nn \\ && \left.  
	-\qpll^2\gper^\munu \SB{L_{l}(2\alpha_k)L_{n-1}(2\alpha_k) + L_{l-1}(2\alpha_k)L_{n}(2\alpha_k)\frac{}{}} } 
\nn \\ && 
+ \frac{\kappa_\rho^2}{4m_N^2} \TB{8\SB{\kper^2(\qpll^2g^\munu-\qpll^\mu\qpll^\nu) - \qpll^2\gper^\munu \SB{L_{l}(2\alpha_k)L_{n-1}(2\alpha_k) + L_{l-1}(2\alpha_k)L_{n}(2\alpha_k)\frac{}{}}} 
	\right. \nn \\ && \left.
	-  \SB{ 2(\kpll\cdot\qpll)^2\gpll^\munu + 2\qpll^2\kpll^\mu\kpll^\nu -2(\kpll\cdot\qpll)(\qpll^\mu\kpll^\nu+\qpll^\nu\kpll^\mu) 
		-(m_N^2+\kpll^2-\kpll\cdot\qpll)(\qpll^2\gpll^\munu-\qpll^\mu\qpll^\nu) } \times
	\right. \nn \\ && \left.
	\SB{L_{l-1}(2\alpha_k)L_{n-1}(2\alpha_k) + L_{l}(2\alpha_k)L_{n}(2\alpha_k) \frac{}{}}
	-\SB{ \qpll^2(m_N^2+\kpll^2-\kpll\cdot\qpll)-2(\kpll\cdot\qpll)^2 }\gper^\munu \times
	\right. \nn \\ && \left. 
	\SB{L_{l}(2\alpha_k)L_{n-1}(2\alpha_k) + L_{l-1}(2\alpha_k)L_{n}(2\alpha_k)\frac{}{}}  		
} \Bigg.\Bigg],
\label{eq.N.p.4}
\end{eqnarray} 
where the terms involving odd powers of $\kper^\mu$ are discarded as they will vanish while integrating over $d^2\kper$.

We now perform the $d^2k_\perp$ integration using the orthogonality of the Laguerre polynomial (identities provided in Appendix~\ref{appendix.identities}) to obtain, 
\begin{eqnarray}
\tilde{\mathcal{N}}^\munu_{\tp,nl}(q_\parallel,k_\parallel) &=&  -g_\rhoNN^2(-1)^{n+l}\frac{eB}{\pi} \Bigg[\Bigg. 4eB\gpll^\munu n\delta_{l-1}^{n-1} + \SB{  (m_N^2-\kpll^2-\kpll\cdot\qpll)\gpll^\munu + 2\kpll^\mu\kpll^\nu + (\qpll^\mu\kpll^\nu+\qpll^\nu\kpll^\mu) }
\FB{\delta_{l-1}^{n-1}+\delta_{l}^{n}}  \nn \\ && 
-(m_N^2-\kpll^2-\kpll\cdot\qpll)\gper^\munu \FB{\delta_{l-1}^{n}+\delta_{l}^{n-1}}
+ \kappa_\rho \TB{ (\qpll^2\gpll^\munu-\qpll^\mu\qpll^\nu)\FB{\delta_{l-1}^{n-1}+\delta_{l}^{n}}  
	-\qpll^2\gper^\munu \FB{ \delta_{l-1}^{n}+\delta_{l}^{n-1} }  } 
\nn \\ && 
+ \frac{\kappa_\rho^2}{4m_N^2} \TB{ -4eB(\qpll^2\gpll^\munu-\qpll^\mu\qpll^\nu)n\delta_{l-1}^{n-1}  
	-  \SB{ 2(\kpll\cdot\qpll)^2\gpll^\munu + 2\qpll^2\kpll^\mu\kpll^\nu -2(\kpll\cdot\qpll)(\qpll^\mu\kpll^\nu+\qpll^\nu\kpll^\mu) 
		\right.\right. \nn \\ && \hspace{-2.3cm} \left. \left.		
		-(m_N^2+\kpll^2-\kpll\cdot\qpll)(\qpll^2\gpll^\munu-\qpll^\mu\qpll^\nu) } 
	\FB{\delta_{l-1}^{n-1}+\delta_{l}^{n}}
	-\SB{ \qpll^2(m_N^2+\kpll^2-\kpll\cdot\qpll)-2(\kpll\cdot\qpll)^2 }\gper^\munu
	\FB{\delta_{l-1}^{n}+\delta_{l}^{n-1}}  		
} \Bigg.\Bigg]
\label{eq.N.p.5}
\end{eqnarray} 
It is to be noted that, a Kronecker delta with -ve index is zero which comes from our constraint on the Laguerre polynomials $L^a_{-1}=0$.

\section{Details of $\mathcal{N}^\mu_{~\mu}$ and $\mathcal{N}^{00}$ for different loop } \label{app.Nmumu00}

In this appendix, we list the explicit forms of $\mathcal{N}^\mu_{~\mu}$ and $\mathcal{N}^{00}$ for all the different loops. 
For the zero magnetic field case, we have for the $\pi\pi$ Loop
\begin{eqnarray}
g_\munu\mathcal{N}_\pi^\munu(q,k) &=& g_\rhopipi^2 \TB{q^4k^\mu k^\nu + (q\cdot k)^2q^2 - q^2(q\cdot k)2q\cdot k \frac{}{}} \label{eq.N.pi.vac.mumu}\\
\mathcal{N}_\pi^{00}(q,k) &=& g_\rhopipi^2 \TB{q^4k_0^2 + (q\cdot k)^2q_0^2 - q^2(q\cdot k)2q^0k^0 \frac{}{}} \label{eq.N.pi.vac.00}\\
\end{eqnarray}
and for the NN-Loop,
\begin{eqnarray}
g_\munu\mathcal{N}_\text{N}^\munu(q,k) &=& -8g_\rhoNN^2\TB{ \frac{}{} (m_N^2-k^2-k\cdot q)4 + 2k^2 + q\cdot k  + \kappa_\rho3q^2 \right. \nn \\ && \left.
	+ \frac{\kappa_\rho^2}{4m_N^2}\SB{ (m_N^2+k^2-k\cdot q)3q^2-2q^2k^2 - 2(k\cdot q)^24 
		+4(k\cdot q)^2 \frac{}{}} }~.
\label{eq.Nmunu_Nmumu} \\
\mathcal{N}_\text{N}^{00}(q,k) &=& -8g_\rhoNN^2\TB{ \frac{}{} (m_N^2-k^2-k\cdot q) + 2k_0^2 + 2q^0k^0  
	- \kappa_\rho\vec{q}^2 \right. \nn \\ && \left.
	+ \frac{\kappa_\rho^2}{4m_N^2}\SB{ -(m_N^2+k^2-k\cdot q)\vec{q}^2-2q^2k_0^2 - 2(k\cdot q)^2 
		+4(k\cdot q)q^0k^0 \frac{}{}} }~.
\label{eq.Nmunu_N00}
\end{eqnarray}

The corresponding expressions for $\pi\pi$ loop for finite magnetic field case are given by
\begin{eqnarray}
g_\munu\tilde{\mathcal{N}}^\munu_{\pi,nl}(q_\parallel,k_\parallel) &=&  4g_\rhopipi^2(-1)^{n+l}\frac{eB}{8\pi}
\TB{\SB{q_\parallel^4\kpll^2 + (q_\parallel\cdot k_\parallel)^2\qpll^2 - q_\parallel^2(q_\parallel\cdot k_\parallel)2\qpll\cdot\kpll}\delta_l^n \frac{}{} \right. \nn \\ 
	&& \left. - q_\parallel^4 \frac{eB}{2}\SB{(2n+1)\delta_l^n-(n+1)\delta_l^{n+1}-n\delta_l^{n-1}} \frac{}{}}
\label{eq.N.pi.mumu} \\
\tilde{\mathcal{N}}^{00}_{\pi,nl}(q_\parallel,k_\parallel) &=&  4g_\rhopipi^2(-1)^{n+l}\frac{eB}{8\pi}
\TB{\frac{}{}q_\parallel^4k_0^2 + (q_\parallel\cdot k_\parallel)^2q_0^2 - q_\parallel^2(q_\parallel\cdot k_\parallel)2q^0k^0}\delta_l^n 
\label{eq.N.pi.00}
\end{eqnarray}

whereas the same for proton-proton loop are
\begin{eqnarray}
g_\munu\tilde{\mathcal{N}}^\munu_{\tp,nl}(q_\parallel,k_\parallel) &=&  -g_\rhoNN^2(-1)^{n+l}\frac{eB}{\pi} \Bigg[\Bigg. 8eB n\delta_{l-1}^{n-1} + \SB{  (m_N^2-\kpll^2-\kpll\cdot\qpll)2 + 2\kpll^2 + 2\qpll\cdot\kpll }
\FB{\delta_{l-1}^{n-1}+\delta_{l}^{n}}  \nn \\ && 
-(m_N^2-\kpll^2-\kpll\cdot\qpll)2 \FB{\delta_{l-1}^{n}+\delta_{l}^{n-1}}
+ \kappa_\rho \TB{ \qpll^2\FB{\delta_{l-1}^{n-1}+\delta_{l}^{n}}  
	-\qpll^22 \FB{ \delta_{l-1}^{n}+\delta_{l}^{n-1} }  } 
\nn \\ && 
+ \frac{\kappa_\rho^2}{4m_N^2} \TB{ -4eB\qpll^2n\delta_{l-1}^{n-1}  
	-  \SB{ 2(\kpll\cdot\qpll)^22 + 2\qpll^2\kpll^2 - 2(\kpll\cdot\qpll)2\qpll\cdot\kpll 
		\right.\right. \nn \\ && \hspace{-2.3cm} \left. \left.		
		-(m_N^2+\kpll^2-\kpll\cdot\qpll)\qpll^2 } 
	\FB{\delta_{l-1}^{n-1}+\delta_{l}^{n}}
	-\SB{ \qpll^2(m_N^2+\kpll^2-\kpll\cdot\qpll)-2(\kpll\cdot\qpll)^2 }2
	\FB{\delta_{l-1}^{n}+\delta_{l}^{n-1}}  		
} \Bigg.\Bigg] 
\label{eq.N.p.mumu} \\
\tilde{\mathcal{N}}^{00}_{\tp,nl}(q_\parallel,k_\parallel) &=&  -g_\rhoNN^2(-1)^{n+l}\frac{eB}{\pi} \Bigg[\Bigg. 4eB n\delta_{l-1}^{n-1} + \SB{  (m_N^2-\kpll^2-\kpll\cdot\qpll) + 2k_0^2 + 2q^0k^0 }
\FB{\delta_{l-1}^{n-1}+\delta_{l}^{n}}  \nn \\ && 
+ \kappa_\rho \TB{ -q_z^2\FB{\delta_{l-1}^{n-1}+\delta_{l}^{n}}    } 
+ \frac{\kappa_\rho^2}{4m_N^2} \TB{ 4eBq_z^2n\delta_{l-1}^{n-1}  
	-  \SB{ 2(\kpll\cdot\qpll)^2 + 2\qpll^2k_0^2 - 2(\kpll\cdot\qpll)2q^0k^0 
		\right.\right. \nn \\ && \hspace{0cm} \left. \left.		
		+(m_N^2+\kpll^2-\kpll\cdot\qpll)q_z^2 } 
	\FB{\delta_{l-1}^{n-1}+\delta_{l}^{n}}  		
} \Bigg.\Bigg]
\label{eq.N.p.00}~. 
\end{eqnarray}

\bibliographystyle{apsrev4-1}
\bibliography{snigdha}

\begin{thebibliography}{60}%
\makeatletter
\providecommand \@ifxundefined [1]{%
 \@ifx{#1\undefined}
}%
\providecommand \@ifnum [1]{%
 \ifnum #1\expandafter \@firstoftwo
 \else \expandafter \@secondoftwo
 \fi
}%
\providecommand \@ifx [1]{%
 \ifx #1\expandafter \@firstoftwo
 \else \expandafter \@secondoftwo
 \fi
}%
\providecommand \natexlab [1]{#1}%
\providecommand \enquote  [1]{``#1''}%
\providecommand \bibnamefont  [1]{#1}%
\providecommand \bibfnamefont [1]{#1}%
\providecommand \citenamefont [1]{#1}%
\providecommand \href@noop [0]{\@secondoftwo}%
\providecommand \href [0]{\begingroup \@sanitize@url \@href}%
\providecommand \@href[1]{\@@startlink{#1}\@@href}%
\providecommand \@@href[1]{\endgroup#1\@@endlink}%
\providecommand \@sanitize@url [0]{\catcode `\\12\catcode `\$12\catcode
  `\&12\catcode `\#12\catcode `\^12\catcode `\_12\catcode `\%12\relax}%
\providecommand \@@startlink[1]{}%
\providecommand \@@endlink[0]{}%
\providecommand \url  [0]{\begingroup\@sanitize@url \@url }%
\providecommand \@url [1]{\endgroup\@href {#1}{\urlprefix }}%
\providecommand \urlprefix  [0]{URL }%
\providecommand \Eprint [0]{\href }%
\providecommand \doibase [0]{http://dx.doi.org/}%
\providecommand \selectlanguage [0]{\@gobble}%
\providecommand \bibinfo  [0]{\@secondoftwo}%
\providecommand \bibfield  [0]{\@secondoftwo}%
\providecommand \translation [1]{[#1]}%
\providecommand \BibitemOpen [0]{}%
\providecommand \bibitemStop [0]{}%
\providecommand \bibitemNoStop [0]{.\EOS\space}%
\providecommand \EOS [0]{\spacefactor3000\relax}%
\providecommand \BibitemShut  [1]{\csname bibitem#1\endcsname}%
\let\auto@bib@innerbib\@empty
\bibitem [{\citenamefont {Skokov}\ \emph {et~al.}(2009)\citenamefont {Skokov},
  \citenamefont {Illarionov},\ and\ \citenamefont {Toneev}}]{Skokov:2009qp}%
  \BibitemOpen
  \bibfield  {author} {\bibinfo {author} {\bibfnamefont {V.}~\bibnamefont
  {Skokov}}, \bibinfo {author} {\bibfnamefont {A.~{\relax Yu}.}\ \bibnamefont
  {Illarionov}}, \ and\ \bibinfo {author} {\bibfnamefont {V.}~\bibnamefont
  {Toneev}},\ }\href {\doibase 10.1142/S0217751X09047570} {\bibfield  {journal}
  {\bibinfo  {journal} {Int. J. Mod. Phys.}\ }\textbf {\bibinfo {volume}
  {A24}},\ \bibinfo {pages} {5925} (\bibinfo {year} {2009})},\ \Eprint
  {http://arxiv.org/abs/0907.1396} {arXiv:0907.1396 [nucl-th]} \BibitemShut
  {NoStop}%
\bibitem [{\citenamefont {Bzdak}\ and\ \citenamefont
  {Skokov}(2013)}]{PhysRevLett.110.192301}%
  \BibitemOpen
  \bibfield  {author} {\bibinfo {author} {\bibfnamefont {A.}~\bibnamefont
  {Bzdak}}\ and\ \bibinfo {author} {\bibfnamefont {V.}~\bibnamefont {Skokov}},\
  }\href {\doibase 10.1103/PhysRevLett.110.192301} {\bibfield  {journal}
  {\bibinfo  {journal} {Phys. Rev. Lett.}\ }\textbf {\bibinfo {volume} {110}},\
  \bibinfo {pages} {192301} (\bibinfo {year} {2013})}\BibitemShut {NoStop}%
\bibitem [{\citenamefont {McLerran}\ and\ \citenamefont
  {Skokov}(2014)}]{MCLERRAN2014184}%
  \BibitemOpen
  \bibfield  {author} {\bibinfo {author} {\bibfnamefont {L.}~\bibnamefont
  {McLerran}}\ and\ \bibinfo {author} {\bibfnamefont {V.}~\bibnamefont
  {Skokov}},\ }\href {\doibase https://doi.org/10.1016/j.nuclphysa.2014.05.008}
  {\bibfield  {journal} {\bibinfo  {journal} {Nuclear Physics A}\ }\textbf
  {\bibinfo {volume} {929}},\ \bibinfo {pages} {184 } (\bibinfo {year}
  {2014})}\BibitemShut {NoStop}%
\bibitem [{\citenamefont {Tuchin}(2013{\natexlab{a}})}]{PhysRevC.88.024910}%
  \BibitemOpen
  \bibfield  {author} {\bibinfo {author} {\bibfnamefont {K.}~\bibnamefont
  {Tuchin}},\ }\href {\doibase 10.1103/PhysRevC.88.024910} {\bibfield
  {journal} {\bibinfo  {journal} {Phys. Rev. C}\ }\textbf {\bibinfo {volume}
  {88}},\ \bibinfo {pages} {024910} (\bibinfo {year}
  {2013}{\natexlab{a}})}\BibitemShut {NoStop}%
\bibitem [{\citenamefont {Tuchin}(2013{\natexlab{b}})}]{PhysRevC.87.024912}%
  \BibitemOpen
  \bibfield  {author} {\bibinfo {author} {\bibfnamefont {K.}~\bibnamefont
  {Tuchin}},\ }\href {\doibase 10.1103/PhysRevC.87.024912} {\bibfield
  {journal} {\bibinfo  {journal} {Phys. Rev. C}\ }\textbf {\bibinfo {volume}
  {87}},\ \bibinfo {pages} {024912} (\bibinfo {year}
  {2013}{\natexlab{b}})}\BibitemShut {NoStop}%
\bibitem [{\citenamefont {Tuchin}(2013{\natexlab{c}})}]{AdvHighEnergyPhys2013}%
  \BibitemOpen
  \bibfield  {author} {\bibinfo {author} {\bibfnamefont {K.}~\bibnamefont
  {Tuchin}},\ }\href {\doibase 10.1155/2013/490495} {\bibfield  {journal}
  {\bibinfo  {journal} {Advances in High Energy Physics}\ }\textbf {\bibinfo
  {volume} {2013}},\ \bibinfo {pages} {490495} (\bibinfo {year}
  {2013}{\natexlab{c}})}\BibitemShut {NoStop}%
\bibitem [{\citenamefont {Vachaspati}(1991)}]{VACHASPATI1991258}%
  \BibitemOpen
  \bibfield  {author} {\bibinfo {author} {\bibfnamefont {T.}~\bibnamefont
  {Vachaspati}},\ }\href {\doibase
  https://doi.org/10.1016/0370-2693(91)90051-Q} {\bibfield  {journal} {\bibinfo
   {journal} {Physics Letters B}\ }\textbf {\bibinfo {volume} {265}},\ \bibinfo
  {pages} {258 } (\bibinfo {year} {1991})}\BibitemShut {NoStop}%
\bibitem [{\citenamefont {Grasso}\ and\ \citenamefont
  {Rubinstein}(2001)}]{GRASSO2001163}%
  \BibitemOpen
  \bibfield  {author} {\bibinfo {author} {\bibfnamefont {D.}~\bibnamefont
  {Grasso}}\ and\ \bibinfo {author} {\bibfnamefont {H.~R.}\ \bibnamefont
  {Rubinstein}},\ }\href {\doibase
  https://doi.org/10.1016/S0370-1573(00)00110-1} {\bibfield  {journal}
  {\bibinfo  {journal} {Physics Reports}\ }\textbf {\bibinfo {volume} {348}},\
  \bibinfo {pages} {163 } (\bibinfo {year} {2001})}\BibitemShut {NoStop}%
\bibitem [{\citenamefont {GIOVANNINI}(2004)}]{S0218271804004530}%
  \BibitemOpen
  \bibfield  {author} {\bibinfo {author} {\bibfnamefont {M.}~\bibnamefont
  {GIOVANNINI}},\ }\href {\doibase 10.1142/S0218271804004530} {\bibfield
  {journal} {\bibinfo  {journal} {International Journal of Modern Physics D}\
  }\textbf {\bibinfo {volume} {13}},\ \bibinfo {pages} {391} (\bibinfo {year}
  {2004})},\ \Eprint
  {http://arxiv.org/abs/https://doi.org/10.1142/S0218271804004530}
  {https://doi.org/10.1142/S0218271804004530} \BibitemShut {NoStop}%
\bibitem [{\citenamefont {Giovannini}(2018)}]{0264-9381-35-8-084003}%
  \BibitemOpen
  \bibfield  {author} {\bibinfo {author} {\bibfnamefont {M.}~\bibnamefont
  {Giovannini}},\ }\href {http://stacks.iop.org/0264-9381/35/i=8/a=084003}
  {\bibfield  {journal} {\bibinfo  {journal} {Classical and Quantum Gravity}\
  }\textbf {\bibinfo {volume} {35}},\ \bibinfo {pages} {084003} (\bibinfo
  {year} {2018})}\BibitemShut {NoStop}%
\bibitem [{\citenamefont {Kaspi}\ and\ \citenamefont
  {Beloborodov}(2017)}]{annurev-astro}%
  \BibitemOpen
  \bibfield  {author} {\bibinfo {author} {\bibfnamefont {V.~M.}\ \bibnamefont
  {Kaspi}}\ and\ \bibinfo {author} {\bibfnamefont {A.~M.}\ \bibnamefont
  {Beloborodov}},\ }\href {\doibase 10.1146/annurev-astro-081915-023329}
  {\bibfield  {journal} {\bibinfo  {journal} {Annual Review of Astronomy and
  Astrophysics}\ }\textbf {\bibinfo {volume} {55}},\ \bibinfo {pages} {261}
  (\bibinfo {year} {2017})}\BibitemShut {NoStop}%
\bibitem [{\citenamefont {Ferrer}\ \emph {et~al.}(2005)\citenamefont {Ferrer},
  \citenamefont {de~la Incera},\ and\ \citenamefont {Manuel}}]{Ferrer:2005vd}%
  \BibitemOpen
  \bibfield  {author} {\bibinfo {author} {\bibfnamefont {E.~J.}\ \bibnamefont
  {Ferrer}}, \bibinfo {author} {\bibfnamefont {V.}~\bibnamefont {de~la
  Incera}}, \ and\ \bibinfo {author} {\bibfnamefont {C.}~\bibnamefont
  {Manuel}},\ }\href {\doibase 10.1103/PhysRevLett.95.152002} {\bibfield
  {journal} {\bibinfo  {journal} {Phys. Rev. Lett.}\ }\textbf {\bibinfo
  {volume} {95}},\ \bibinfo {pages} {152002} (\bibinfo {year} {2005})},\
  \Eprint {http://arxiv.org/abs/hep-ph/0503162} {arXiv:hep-ph/0503162 [hep-ph]}
  \BibitemShut {NoStop}%
\bibitem [{\citenamefont {Ferrer}\ \emph {et~al.}(2006)\citenamefont {Ferrer},
  \citenamefont {de~la Incera},\ and\ \citenamefont {Manuel}}]{Ferrer:2006vw}%
  \BibitemOpen
  \bibfield  {author} {\bibinfo {author} {\bibfnamefont {E.~J.}\ \bibnamefont
  {Ferrer}}, \bibinfo {author} {\bibfnamefont {V.}~\bibnamefont {de~la
  Incera}}, \ and\ \bibinfo {author} {\bibfnamefont {C.}~\bibnamefont
  {Manuel}},\ }\href {\doibase 10.1016/j.nuclphysb.2006.04.013} {\bibfield
  {journal} {\bibinfo  {journal} {Nucl. Phys.}\ }\textbf {\bibinfo {volume}
  {B747}},\ \bibinfo {pages} {88} (\bibinfo {year} {2006})},\ \Eprint
  {http://arxiv.org/abs/hep-ph/0603233} {arXiv:hep-ph/0603233 [hep-ph]}
  \BibitemShut {NoStop}%
\bibitem [{\citenamefont {Ferrer}\ and\ \citenamefont {de~la
  Incera}(2007)}]{Ferrer:2007iw}%
  \BibitemOpen
  \bibfield  {author} {\bibinfo {author} {\bibfnamefont {E.~J.}\ \bibnamefont
  {Ferrer}}\ and\ \bibinfo {author} {\bibfnamefont {V.}~\bibnamefont {de~la
  Incera}},\ }\href {\doibase 10.1103/PhysRevD.76.045011} {\bibfield  {journal}
  {\bibinfo  {journal} {Phys. Rev.}\ }\textbf {\bibinfo {volume} {D76}},\
  \bibinfo {pages} {045011} (\bibinfo {year} {2007})},\ \Eprint
  {http://arxiv.org/abs/nucl-th/0703034} {arXiv:nucl-th/0703034 [NUCL-TH]}
  \BibitemShut {NoStop}%
\bibitem [{\citenamefont {Fukushima}\ and\ \citenamefont
  {Warringa}(2008)}]{Fukushima:2007fc}%
  \BibitemOpen
  \bibfield  {author} {\bibinfo {author} {\bibfnamefont {K.}~\bibnamefont
  {Fukushima}}\ and\ \bibinfo {author} {\bibfnamefont {H.~J.}\ \bibnamefont
  {Warringa}},\ }\href {\doibase 10.1103/PhysRevLett.100.032007} {\bibfield
  {journal} {\bibinfo  {journal} {Phys. Rev. Lett.}\ }\textbf {\bibinfo
  {volume} {100}},\ \bibinfo {pages} {032007} (\bibinfo {year} {2008})},\
  \Eprint {http://arxiv.org/abs/0707.3785} {arXiv:0707.3785 [hep-ph]}
  \BibitemShut {NoStop}%
\bibitem [{\citenamefont {Feng}\ \emph {et~al.}(2010)\citenamefont {Feng},
  \citenamefont {Hou}, \citenamefont {Ren},\ and\ \citenamefont
  {Wu}}]{Feng:2009vt}%
  \BibitemOpen
  \bibfield  {author} {\bibinfo {author} {\bibfnamefont {B.}~\bibnamefont
  {Feng}}, \bibinfo {author} {\bibfnamefont {D.}~\bibnamefont {Hou}}, \bibinfo
  {author} {\bibfnamefont {H.-c.}\ \bibnamefont {Ren}}, \ and\ \bibinfo
  {author} {\bibfnamefont {P.-p.}\ \bibnamefont {Wu}},\ }\href {\doibase
  10.1103/PhysRevLett.105.042001} {\bibfield  {journal} {\bibinfo  {journal}
  {Phys. Rev. Lett.}\ }\textbf {\bibinfo {volume} {105}},\ \bibinfo {pages}
  {042001} (\bibinfo {year} {2010})},\ \Eprint {http://arxiv.org/abs/0911.4997}
  {arXiv:0911.4997 [hep-ph]} \BibitemShut {NoStop}%
\bibitem [{\citenamefont {Fayazbakhsh}\ and\ \citenamefont
  {Sadooghi}(2010)}]{Fayazbakhsh:2010gc}%
  \BibitemOpen
  \bibfield  {author} {\bibinfo {author} {\bibfnamefont {S.}~\bibnamefont
  {Fayazbakhsh}}\ and\ \bibinfo {author} {\bibfnamefont {N.}~\bibnamefont
  {Sadooghi}},\ }\href {\doibase 10.1103/PhysRevD.82.045010} {\bibfield
  {journal} {\bibinfo  {journal} {Phys. Rev.}\ }\textbf {\bibinfo {volume}
  {D82}},\ \bibinfo {pages} {045010} (\bibinfo {year} {2010})},\ \Eprint
  {http://arxiv.org/abs/1005.5022} {arXiv:1005.5022 [hep-ph]} \BibitemShut
  {NoStop}%
\bibitem [{\citenamefont {Fayazbakhsh}\ and\ \citenamefont
  {Sadooghi}(2011)}]{Fayazbakhsh:2010bh}%
  \BibitemOpen
  \bibfield  {author} {\bibinfo {author} {\bibfnamefont {S.}~\bibnamefont
  {Fayazbakhsh}}\ and\ \bibinfo {author} {\bibfnamefont {N.}~\bibnamefont
  {Sadooghi}},\ }\href {\doibase 10.1103/PhysRevD.83.025026} {\bibfield
  {journal} {\bibinfo  {journal} {Phys. Rev.}\ }\textbf {\bibinfo {volume}
  {D83}},\ \bibinfo {pages} {025026} (\bibinfo {year} {2011})},\ \Eprint
  {http://arxiv.org/abs/1009.6125} {arXiv:1009.6125 [hep-ph]} \BibitemShut
  {NoStop}%
\bibitem [{\citenamefont {Chernodub}(2010)}]{Chernodub:2010qx}%
  \BibitemOpen
  \bibfield  {author} {\bibinfo {author} {\bibfnamefont {M.~N.}\ \bibnamefont
  {Chernodub}},\ }\href {\doibase 10.1103/PhysRevD.82.085011} {\bibfield
  {journal} {\bibinfo  {journal} {Phys. Rev.}\ }\textbf {\bibinfo {volume}
  {D82}},\ \bibinfo {pages} {085011} (\bibinfo {year} {2010})},\ \Eprint
  {http://arxiv.org/abs/1008.1055} {arXiv:1008.1055 [hep-ph]} \BibitemShut
  {NoStop}%
\bibitem [{\citenamefont {Chernodub}(2013)}]{Chernodub:2012tf}%
  \BibitemOpen
  \bibfield  {author} {\bibinfo {author} {\bibfnamefont {M.~N.}\ \bibnamefont
  {Chernodub}},\ }\href {\doibase 10.1007/978-3-642-37305-3_6} {\bibfield
  {journal} {\bibinfo  {journal} {Lect. Notes Phys.}\ }\textbf {\bibinfo
  {volume} {871}},\ \bibinfo {pages} {143} (\bibinfo {year} {2013})},\ \Eprint
  {http://arxiv.org/abs/1208.5025} {arXiv:1208.5025 [hep-ph]} \BibitemShut
  {NoStop}%
\bibitem [{\citenamefont {Vafa}\ and\ \citenamefont
  {Witten}(1984)}]{VAFA1984173}%
  \BibitemOpen
  \bibfield  {author} {\bibinfo {author} {\bibfnamefont {C.}~\bibnamefont
  {Vafa}}\ and\ \bibinfo {author} {\bibfnamefont {E.}~\bibnamefont {Witten}},\
  }\href {\doibase https://doi.org/10.1016/0550-3213(84)90230-X} {\bibfield
  {journal} {\bibinfo  {journal} {Nuclear Physics B}\ }\textbf {\bibinfo
  {volume} {234}},\ \bibinfo {pages} {173 } (\bibinfo {year}
  {1984})}\BibitemShut {NoStop}%
\bibitem [{\citenamefont {Chernodub}(2012)}]{PhysRevD.86.107703}%
  \BibitemOpen
  \bibfield  {author} {\bibinfo {author} {\bibfnamefont {M.~N.}\ \bibnamefont
  {Chernodub}},\ }\href {\doibase 10.1103/PhysRevD.86.107703} {\bibfield
  {journal} {\bibinfo  {journal} {Phys. Rev. D}\ }\textbf {\bibinfo {volume}
  {86}},\ \bibinfo {pages} {107703} (\bibinfo {year} {2012})}\BibitemShut
  {NoStop}%
\bibitem [{\citenamefont {Li}\ and\ \citenamefont {Wang}(2013)}]{LI2013141}%
  \BibitemOpen
  \bibfield  {author} {\bibinfo {author} {\bibfnamefont {C.}~\bibnamefont
  {Li}}\ and\ \bibinfo {author} {\bibfnamefont {Q.}~\bibnamefont {Wang}},\
  }\href {\doibase https://doi.org/10.1016/j.physletb.2013.02.050} {\bibfield
  {journal} {\bibinfo  {journal} {Physics Letters B}\ }\textbf {\bibinfo
  {volume} {721}},\ \bibinfo {pages} {141 } (\bibinfo {year}
  {2013})}\BibitemShut {NoStop}%
\bibitem [{\citenamefont {Hidaka}\ and\ \citenamefont
  {Yamamoto}(2013)}]{PhysRevD.87.094502}%
  \BibitemOpen
  \bibfield  {author} {\bibinfo {author} {\bibfnamefont {Y.}~\bibnamefont
  {Hidaka}}\ and\ \bibinfo {author} {\bibfnamefont {A.}~\bibnamefont
  {Yamamoto}},\ }\href {\doibase 10.1103/PhysRevD.87.094502} {\bibfield
  {journal} {\bibinfo  {journal} {Phys. Rev. D}\ }\textbf {\bibinfo {volume}
  {87}},\ \bibinfo {pages} {094502} (\bibinfo {year} {2013})}\BibitemShut
  {NoStop}%
\bibitem [{\citenamefont {Liu}\ \emph {et~al.}(2015)\citenamefont {Liu},
  \citenamefont {Yu},\ and\ \citenamefont {Huang}}]{PhysRevD.91.014017}%
  \BibitemOpen
  \bibfield  {author} {\bibinfo {author} {\bibfnamefont {H.}~\bibnamefont
  {Liu}}, \bibinfo {author} {\bibfnamefont {L.}~\bibnamefont {Yu}}, \ and\
  \bibinfo {author} {\bibfnamefont {M.}~\bibnamefont {Huang}},\ }\href
  {\doibase 10.1103/PhysRevD.91.014017} {\bibfield  {journal} {\bibinfo
  {journal} {Phys. Rev. D}\ }\textbf {\bibinfo {volume} {91}},\ \bibinfo
  {pages} {014017} (\bibinfo {year} {2015})}\BibitemShut {NoStop}%
\bibitem [{\citenamefont {Kawaguchi}\ and\ \citenamefont
  {Matsuzaki}(2016)}]{PhysRevD.93.125027}%
  \BibitemOpen
  \bibfield  {author} {\bibinfo {author} {\bibfnamefont {M.}~\bibnamefont
  {Kawaguchi}}\ and\ \bibinfo {author} {\bibfnamefont {S.}~\bibnamefont
  {Matsuzaki}},\ }\href {\doibase 10.1103/PhysRevD.93.125027} {\bibfield
  {journal} {\bibinfo  {journal} {Phys. Rev. D}\ }\textbf {\bibinfo {volume}
  {93}},\ \bibinfo {pages} {125027} (\bibinfo {year} {2016})}\BibitemShut
  {NoStop}%
\bibitem [{\citenamefont {Liu}\ \emph {et~al.}(2016)\citenamefont {Liu},
  \citenamefont {Yu},\ and\ \citenamefont {Huang}}]{1674-1137-40-2-023102}%
  \BibitemOpen
  \bibfield  {author} {\bibinfo {author} {\bibfnamefont {H.}~\bibnamefont
  {Liu}}, \bibinfo {author} {\bibfnamefont {L.}~\bibnamefont {Yu}}, \ and\
  \bibinfo {author} {\bibfnamefont {M.}~\bibnamefont {Huang}},\ }\href
  {http://stacks.iop.org/1674-1137/40/i=2/a=023102} {\bibfield  {journal}
  {\bibinfo  {journal} {Chinese Physics C}\ }\textbf {\bibinfo {volume} {40}},\
  \bibinfo {pages} {023102} (\bibinfo {year} {2016})}\BibitemShut {NoStop}%
\bibitem [{\citenamefont {Ghosh}\ \emph {et~al.}(2016)\citenamefont {Ghosh},
  \citenamefont {Mukherjee}, \citenamefont {Mandal}, \citenamefont {Sarkar},\
  and\ \citenamefont {Roy}}]{Ghosh:2016evc}%
  \BibitemOpen
  \bibfield  {author} {\bibinfo {author} {\bibfnamefont {S.}~\bibnamefont
  {Ghosh}}, \bibinfo {author} {\bibfnamefont {A.}~\bibnamefont {Mukherjee}},
  \bibinfo {author} {\bibfnamefont {M.}~\bibnamefont {Mandal}}, \bibinfo
  {author} {\bibfnamefont {S.}~\bibnamefont {Sarkar}}, \ and\ \bibinfo {author}
  {\bibfnamefont {P.}~\bibnamefont {Roy}},\ }\href {\doibase
  10.1103/PhysRevD.94.094043} {\bibfield  {journal} {\bibinfo  {journal} {Phys.
  Rev.}\ }\textbf {\bibinfo {volume} {D94}},\ \bibinfo {pages} {094043}
  (\bibinfo {year} {2016})},\ \Eprint {http://arxiv.org/abs/1612.02966}
  {arXiv:1612.02966 [nucl-th]} \BibitemShut {NoStop}%
\bibitem [{\citenamefont {{Bandyopadhyay, Aritra}}\ and\ \citenamefont
  {{Mallik, S.}}(2017)}]{aritra_weak}%
  \BibitemOpen
  \bibfield  {author} {\bibinfo {author} {\bibnamefont {{Bandyopadhyay,
  Aritra}}}\ and\ \bibinfo {author} {\bibnamefont {{Mallik, S.}}},\ }\href
  {\doibase 10.1140/epjc/s10052-017-5357-9} {\bibfield  {journal} {\bibinfo
  {journal} {Eur. Phys. J. C}\ }\textbf {\bibinfo {volume} {77}},\ \bibinfo
  {pages} {771} (\bibinfo {year} {2017})}\BibitemShut {NoStop}%
\bibitem [{\citenamefont {{Kawaguchi, Mamiya}}\ and\ \citenamefont {{Matsuzaki,
  Shinya}}(2017)}]{kawaguchi_strong}%
  \BibitemOpen
  \bibfield  {author} {\bibinfo {author} {\bibnamefont {{Kawaguchi, Mamiya}}}\
  and\ \bibinfo {author} {\bibnamefont {{Matsuzaki, Shinya}}},\ }\href
  {\doibase 10.1140/epja/i2017-12254-1} {\bibfield  {journal} {\bibinfo
  {journal} {Eur. Phys. J. A}\ }\textbf {\bibinfo {volume} {53}},\ \bibinfo
  {pages} {68} (\bibinfo {year} {2017})}\BibitemShut {NoStop}%
\bibitem [{\citenamefont {Ghosh}\ \emph {et~al.}(2017)\citenamefont {Ghosh},
  \citenamefont {Mukherjee}, \citenamefont {Mandal}, \citenamefont {Sarkar},\
  and\ \citenamefont {Roy}}]{Ghosh:2017rjo}%
  \BibitemOpen
  \bibfield  {author} {\bibinfo {author} {\bibfnamefont {S.}~\bibnamefont
  {Ghosh}}, \bibinfo {author} {\bibfnamefont {A.}~\bibnamefont {Mukherjee}},
  \bibinfo {author} {\bibfnamefont {M.}~\bibnamefont {Mandal}}, \bibinfo
  {author} {\bibfnamefont {S.}~\bibnamefont {Sarkar}}, \ and\ \bibinfo {author}
  {\bibfnamefont {P.}~\bibnamefont {Roy}},\ }\href {\doibase
  10.1103/PhysRevD.96.116020} {\bibfield  {journal} {\bibinfo  {journal} {Phys.
  Rev.}\ }\textbf {\bibinfo {volume} {D96}},\ \bibinfo {pages} {116020}
  (\bibinfo {year} {2017})},\ \Eprint {http://arxiv.org/abs/1704.05319}
  {arXiv:1704.05319 [hep-ph]} \BibitemShut {NoStop}%
\bibitem [{\citenamefont {Elizalde}\ \emph {et~al.}(1994)\citenamefont
  {Elizalde}, \citenamefont {Odintsov}, \citenamefont {Romeo}, \citenamefont
  {Bytsenko},\ and\ \citenamefont {Zerbini}}]{doi:10.1142/2065}%
  \BibitemOpen
  \bibfield  {author} {\bibinfo {author} {\bibfnamefont {E.}~\bibnamefont
  {Elizalde}}, \bibinfo {author} {\bibfnamefont {S.~D.}\ \bibnamefont
  {Odintsov}}, \bibinfo {author} {\bibfnamefont {A.}~\bibnamefont {Romeo}},
  \bibinfo {author} {\bibfnamefont {A.~A.}\ \bibnamefont {Bytsenko}}, \ and\
  \bibinfo {author} {\bibfnamefont {S.}~\bibnamefont {Zerbini}},\ }\href
  {\doibase 10.1142/2065} {\emph {\bibinfo {title} {Zeta Regularization
  Techniques with Applications}}}\ (\bibinfo  {publisher} {WORLD SCIENTIFIC},\
  \bibinfo {year} {1994})\ \Eprint
  {http://arxiv.org/abs/https://www.worldscientific.com/doi/pdf/10.1142/2065}
  {https://www.worldscientific.com/doi/pdf/10.1142/2065} \BibitemShut {NoStop}%
\bibitem [{\citenamefont {Krehl}\ \emph {et~al.}(2000)\citenamefont {Krehl},
  \citenamefont {Hanhart}, \citenamefont {Krewald},\ and\ \citenamefont
  {Speth}}]{Krehl:1999km}%
  \BibitemOpen
  \bibfield  {author} {\bibinfo {author} {\bibfnamefont {O.}~\bibnamefont
  {Krehl}}, \bibinfo {author} {\bibfnamefont {C.}~\bibnamefont {Hanhart}},
  \bibinfo {author} {\bibfnamefont {S.}~\bibnamefont {Krewald}}, \ and\
  \bibinfo {author} {\bibfnamefont {J.}~\bibnamefont {Speth}},\ }\href
  {\doibase 10.1103/PhysRevC.62.025207} {\bibfield  {journal} {\bibinfo
  {journal} {Phys. Rev.}\ }\textbf {\bibinfo {volume} {C62}},\ \bibinfo {pages}
  {025207} (\bibinfo {year} {2000})},\ \Eprint
  {http://arxiv.org/abs/nucl-th/9911080} {arXiv:nucl-th/9911080 [nucl-th]}
  \BibitemShut {NoStop}%
\bibitem [{\citenamefont {Bellac}(2011)}]{Bellac:2011kqa}%
  \BibitemOpen
  \bibfield  {author} {\bibinfo {author} {\bibfnamefont {M.~L.}\ \bibnamefont
  {Bellac}},\ }\href {\doibase 10.1017/CBO9780511721700} {\emph {\bibinfo
  {title} {{Thermal Field Theory}}}},\ Cambridge Monographs on Mathematical
  Physics\ (\bibinfo  {publisher} {Cambridge University Press},\ \bibinfo
  {year} {2011})\BibitemShut {NoStop}%
\bibitem [{\citenamefont {Mallik}\ and\ \citenamefont
  {Sarkar}(2016)}]{Mallik:2016anp}%
  \BibitemOpen
  \bibfield  {author} {\bibinfo {author} {\bibfnamefont {S.}~\bibnamefont
  {Mallik}}\ and\ \bibinfo {author} {\bibfnamefont {S.}~\bibnamefont
  {Sarkar}},\ }\href {\doibase 10.1017/9781316535585} {\emph {\bibinfo {title}
  {{Hadrons at Finite Temperature}}}}\ (\bibinfo  {publisher} {Cambridge
  University Press},\ \bibinfo {address} {Cambridge},\ \bibinfo {year}
  {2016})\BibitemShut {NoStop}%
\bibitem [{\citenamefont {Schwinger}(1951)}]{Schwinger:1951nm}%
  \BibitemOpen
  \bibfield  {author} {\bibinfo {author} {\bibfnamefont {J.~S.}\ \bibnamefont
  {Schwinger}},\ }\href {\doibase 10.1103/PhysRev.82.664} {\bibfield  {journal}
  {\bibinfo  {journal} {Phys. Rev.}\ }\textbf {\bibinfo {volume} {82}},\
  \bibinfo {pages} {664} (\bibinfo {year} {1951})}\BibitemShut {NoStop}%
\bibitem [{\citenamefont {Ayala}\ \emph {et~al.}(2005)\citenamefont {Ayala},
  \citenamefont {Sanchez}, \citenamefont {Piccinelli},\ and\ \citenamefont
  {Sahu}}]{Ayala:2004dx}%
  \BibitemOpen
  \bibfield  {author} {\bibinfo {author} {\bibfnamefont {A.}~\bibnamefont
  {Ayala}}, \bibinfo {author} {\bibfnamefont {A.}~\bibnamefont {Sanchez}},
  \bibinfo {author} {\bibfnamefont {G.}~\bibnamefont {Piccinelli}}, \ and\
  \bibinfo {author} {\bibfnamefont {S.}~\bibnamefont {Sahu}},\ }\href {\doibase
  10.1103/PhysRevD.71.023004} {\bibfield  {journal} {\bibinfo  {journal} {Phys.
  Rev.}\ }\textbf {\bibinfo {volume} {D71}},\ \bibinfo {pages} {023004}
  (\bibinfo {year} {2005})},\ \Eprint {http://arxiv.org/abs/hep-ph/0412135}
  {arXiv:hep-ph/0412135 [hep-ph]} \BibitemShut {NoStop}%
\bibitem [{\citenamefont {Ghosh}\ and\ \citenamefont
  {Chandra}(2018)}]{Ghosh:2018xhh}%
  \BibitemOpen
  \bibfield  {author} {\bibinfo {author} {\bibfnamefont {S.}~\bibnamefont
  {Ghosh}}\ and\ \bibinfo {author} {\bibfnamefont {V.}~\bibnamefont
  {Chandra}},\ }\href {\doibase 10.1103/PhysRevD.98.076006} {\bibfield
  {journal} {\bibinfo  {journal} {Phys. Rev.}\ }\textbf {\bibinfo {volume}
  {D98}},\ \bibinfo {pages} {076006} (\bibinfo {year} {2018})},\ \Eprint
  {http://arxiv.org/abs/1808.05176} {arXiv:1808.05176 [hep-ph]} \BibitemShut
  {NoStop}%
\bibitem [{\citenamefont {Chakraborty}(2017)}]{Chakraborty:2017vvg}%
  \BibitemOpen
  \bibfield  {author} {\bibinfo {author} {\bibfnamefont {P.}~\bibnamefont
  {Chakraborty}},\ }\href@noop {} {\  (\bibinfo {year} {2017})},\ \Eprint
  {http://arxiv.org/abs/1711.04404} {arXiv:1711.04404 [nucl-th]} \BibitemShut
  {NoStop}%
\bibitem [{\citenamefont {Gorban}(2006)}]{Gorban}%
  \BibitemOpen
  \bibfield  {author} {\bibinfo {author} {\bibfnamefont {A.~N.}\ \bibnamefont
  {Gorban}},\ }\href@noop {} {\bibfield  {journal} {\bibinfo  {journal} {Model
  Reduction and Coarse--Graining Approaches for Multiscale Phenomena,
  Springer}\ ,\ \bibinfo {pages} {117}} (\bibinfo {year} {2006})},\ \Eprint
  {http://arxiv.org/abs/cond-mat/0602024} {arXiv:cond-mat/0602024 [cond-mat]}
  \BibitemShut {NoStop}%
\bibitem [{\citenamefont {P.~Ehrenfest}(2002)}]{Ehrenfest}%
  \BibitemOpen
  \bibfield  {author} {\bibinfo {author} {\bibfnamefont {T.~E.-A.}\
  \bibnamefont {P.~Ehrenfest}},\ }\href@noop {} {\emph {\bibinfo {title} {The
  Conceptual Foundations of the Statistical Approach in Mechanics}}}\ (\bibinfo
   {publisher} {Dover Phoneix},\ \bibinfo {year} {2002})\BibitemShut {NoStop}%
\bibitem [{\citenamefont {Dey}\ \emph {et~al.}(1990)\citenamefont {Dey},
  \citenamefont {Eletsky},\ and\ \citenamefont {Ioffe}}]{Dey:1990ba}%
  \BibitemOpen
  \bibfield  {author} {\bibinfo {author} {\bibfnamefont {M.}~\bibnamefont
  {Dey}}, \bibinfo {author} {\bibfnamefont {V.~L.}\ \bibnamefont {Eletsky}}, \
  and\ \bibinfo {author} {\bibfnamefont {B.~L.}\ \bibnamefont {Ioffe}},\
  }\bibfield  {booktitle} {\emph {\bibinfo {booktitle} {{Phys. Lett. B252
  (1990) 620-624}}},\ }\href {\doibase 10.1016/0370-2693(90)90495-R} {\bibfield
   {journal} {\bibinfo  {journal} {Phys. Lett.}\ }\textbf {\bibinfo {volume}
  {B252}},\ \bibinfo {pages} {620} (\bibinfo {year} {1990})}\BibitemShut
  {NoStop}%
\bibitem [{\citenamefont {Kapusta}\ and\ \citenamefont
  {Shuryak}(1994)}]{Kapusta:1993hq}%
  \BibitemOpen
  \bibfield  {author} {\bibinfo {author} {\bibfnamefont {J.~I.}\ \bibnamefont
  {Kapusta}}\ and\ \bibinfo {author} {\bibfnamefont {E.~V.}\ \bibnamefont
  {Shuryak}},\ }\href {\doibase 10.1103/PhysRevD.49.4694} {\bibfield  {journal}
  {\bibinfo  {journal} {Phys. Rev.}\ }\textbf {\bibinfo {volume} {D49}},\
  \bibinfo {pages} {4694} (\bibinfo {year} {1994})},\ \Eprint
  {http://arxiv.org/abs/hep-ph/9312245} {arXiv:hep-ph/9312245 [hep-ph]}
  \BibitemShut {NoStop}%
\bibitem [{\citenamefont {Hohler}\ and\ \citenamefont
  {Rapp}(2014)}]{Hohler:2013eba}%
  \BibitemOpen
  \bibfield  {author} {\bibinfo {author} {\bibfnamefont {P.~M.}\ \bibnamefont
  {Hohler}}\ and\ \bibinfo {author} {\bibfnamefont {R.}~\bibnamefont {Rapp}},\
  }\href {\doibase https://doi.org/10.1016/j.physletb.2014.02.021} {\bibfield
  {journal} {\bibinfo  {journal} {Physics Letters B}\ }\textbf {\bibinfo
  {volume} {731}},\ \bibinfo {pages} {103 } (\bibinfo {year}
  {2014})}\BibitemShut {NoStop}%
\bibitem [{\citenamefont {Brown}\ and\ \citenamefont
  {Rho}(1991)}]{Brown:1991kk}%
  \BibitemOpen
  \bibfield  {author} {\bibinfo {author} {\bibfnamefont {G.~E.}\ \bibnamefont
  {Brown}}\ and\ \bibinfo {author} {\bibfnamefont {M.}~\bibnamefont {Rho}},\
  }\href {\doibase 10.1103/PhysRevLett.66.2720} {\bibfield  {journal} {\bibinfo
   {journal} {Phys. Rev. Lett.}\ }\textbf {\bibinfo {volume} {66}},\ \bibinfo
  {pages} {2720} (\bibinfo {year} {1991})}\BibitemShut {NoStop}%
\bibitem [{\citenamefont {Alam}\ \emph {et~al.}(2001)\citenamefont {Alam},
  \citenamefont {Sarkar}, \citenamefont {Roy}, \citenamefont {Hatsuda},\ and\
  \citenamefont {Sinha}}]{Alam:1999sc}%
  \BibitemOpen
  \bibfield  {author} {\bibinfo {author} {\bibfnamefont {J.}~\bibnamefont
  {Alam}}, \bibinfo {author} {\bibfnamefont {S.}~\bibnamefont {Sarkar}},
  \bibinfo {author} {\bibfnamefont {P.}~\bibnamefont {Roy}}, \bibinfo {author}
  {\bibfnamefont {T.}~\bibnamefont {Hatsuda}}, \ and\ \bibinfo {author}
  {\bibfnamefont {B.}~\bibnamefont {Sinha}},\ }\href {\doibase
  10.1006/aphy.2000.6091} {\bibfield  {journal} {\bibinfo  {journal} {Annals
  Phys.}\ }\textbf {\bibinfo {volume} {286}},\ \bibinfo {pages} {159} (\bibinfo
  {year} {2001})},\ \Eprint {http://arxiv.org/abs/hep-ph/9909267}
  {arXiv:hep-ph/9909267 [hep-ph]} \BibitemShut {NoStop}%
\bibitem [{\citenamefont {Leupold}\ \emph {et~al.}(2010)\citenamefont
  {Leupold}, \citenamefont {Metag},\ and\ \citenamefont
  {Mosel}}]{Leupold:2009kz}%
  \BibitemOpen
  \bibfield  {author} {\bibinfo {author} {\bibfnamefont {S.}~\bibnamefont
  {Leupold}}, \bibinfo {author} {\bibfnamefont {V.}~\bibnamefont {Metag}}, \
  and\ \bibinfo {author} {\bibfnamefont {U.}~\bibnamefont {Mosel}},\ }\href
  {\doibase 10.1142/S0218301310014728} {\bibfield  {journal} {\bibinfo
  {journal} {Int. J. Mod. Phys.}\ }\textbf {\bibinfo {volume} {E19}},\ \bibinfo
  {pages} {147} (\bibinfo {year} {2010})},\ \Eprint
  {http://arxiv.org/abs/0907.2388} {arXiv:0907.2388 [nucl-th]} \BibitemShut
  {NoStop}%
\bibitem [{\citenamefont {Voronyuk}\ \emph {et~al.}(2011)\citenamefont
  {Voronyuk}, \citenamefont {Toneev}, \citenamefont {Cassing}, \citenamefont
  {Bratkovskaya}, \citenamefont {Konchakovski},\ and\ \citenamefont
  {Voloshin}}]{PhysRevC.83.054911}%
  \BibitemOpen
  \bibfield  {author} {\bibinfo {author} {\bibfnamefont {V.}~\bibnamefont
  {Voronyuk}}, \bibinfo {author} {\bibfnamefont {V.~D.}\ \bibnamefont
  {Toneev}}, \bibinfo {author} {\bibfnamefont {W.}~\bibnamefont {Cassing}},
  \bibinfo {author} {\bibfnamefont {E.~L.}\ \bibnamefont {Bratkovskaya}},
  \bibinfo {author} {\bibfnamefont {V.~P.}\ \bibnamefont {Konchakovski}}, \
  and\ \bibinfo {author} {\bibfnamefont {S.~A.}\ \bibnamefont {Voloshin}},\
  }\href {\doibase 10.1103/PhysRevC.83.054911} {\bibfield  {journal} {\bibinfo
  {journal} {Phys. Rev. C}\ }\textbf {\bibinfo {volume} {83}},\ \bibinfo
  {pages} {054911} (\bibinfo {year} {2011})}\BibitemShut {NoStop}%
\bibitem [{\citenamefont {Tuchin}(2010)}]{PhysRevC.82.034904}%
  \BibitemOpen
  \bibfield  {author} {\bibinfo {author} {\bibfnamefont {K.}~\bibnamefont
  {Tuchin}},\ }\href {\doibase 10.1103/PhysRevC.82.034904} {\bibfield
  {journal} {\bibinfo  {journal} {Phys. Rev. C}\ }\textbf {\bibinfo {volume}
  {82}},\ \bibinfo {pages} {034904} (\bibinfo {year} {2010})}\BibitemShut
  {NoStop}%
\bibitem [{\citenamefont {Tuchin}(2011)}]{PhysRevC.83.017901}%
  \BibitemOpen
  \bibfield  {author} {\bibinfo {author} {\bibfnamefont {K.}~\bibnamefont
  {Tuchin}},\ }\href {\doibase 10.1103/PhysRevC.83.017901} {\bibfield
  {journal} {\bibinfo  {journal} {Phys. Rev. C}\ }\textbf {\bibinfo {volume}
  {83}},\ \bibinfo {pages} {017901} (\bibinfo {year} {2011})}\BibitemShut
  {NoStop}%
\bibitem [{\citenamefont {Tuchin}(2016)}]{PhysRevC.93.014905}%
  \BibitemOpen
  \bibfield  {author} {\bibinfo {author} {\bibfnamefont {K.}~\bibnamefont
  {Tuchin}},\ }\href {\doibase 10.1103/PhysRevC.93.014905} {\bibfield
  {journal} {\bibinfo  {journal} {Phys. Rev. C}\ }\textbf {\bibinfo {volume}
  {93}},\ \bibinfo {pages} {014905} (\bibinfo {year} {2016})}\BibitemShut
  {NoStop}%
\bibitem [{\citenamefont {Roy}\ \emph {et~al.}(2017)\citenamefont {Roy},
  \citenamefont {Pu}, \citenamefont {Rezzolla},\ and\ \citenamefont
  {Rischke}}]{PhysRevC.96.054909}%
  \BibitemOpen
  \bibfield  {author} {\bibinfo {author} {\bibfnamefont {V.}~\bibnamefont
  {Roy}}, \bibinfo {author} {\bibfnamefont {S.}~\bibnamefont {Pu}}, \bibinfo
  {author} {\bibfnamefont {L.}~\bibnamefont {Rezzolla}}, \ and\ \bibinfo
  {author} {\bibfnamefont {D.~H.}\ \bibnamefont {Rischke}},\ }\href {\doibase
  10.1103/PhysRevC.96.054909} {\bibfield  {journal} {\bibinfo  {journal} {Phys.
  Rev. C}\ }\textbf {\bibinfo {volume} {96}},\ \bibinfo {pages} {054909}
  (\bibinfo {year} {2017})}\BibitemShut {NoStop}%
\bibitem [{\citenamefont {Das}\ \emph {et~al.}(2017)\citenamefont {Das},
  \citenamefont {Dave}, \citenamefont {Saumia},\ and\ \citenamefont
  {Srivastava}}]{PhysRevC.96.034902}%
  \BibitemOpen
  \bibfield  {author} {\bibinfo {author} {\bibfnamefont {A.}~\bibnamefont
  {Das}}, \bibinfo {author} {\bibfnamefont {S.~S.}\ \bibnamefont {Dave}},
  \bibinfo {author} {\bibfnamefont {P.~S.}\ \bibnamefont {Saumia}}, \ and\
  \bibinfo {author} {\bibfnamefont {A.~M.}\ \bibnamefont {Srivastava}},\ }\href
  {\doibase 10.1103/PhysRevC.96.034902} {\bibfield  {journal} {\bibinfo
  {journal} {Phys. Rev. C}\ }\textbf {\bibinfo {volume} {96}},\ \bibinfo
  {pages} {034902} (\bibinfo {year} {2017})}\BibitemShut {NoStop}%
\bibitem [{\citenamefont {{Inghirami, Gabriele}}\ \emph
  {et~al.}(2016)\citenamefont {{Inghirami, Gabriele}}, \citenamefont {{Del
  Zanna, Luca}}, \citenamefont {{Beraudo, Andrea}}, \citenamefont {{Moghaddam,
  Mohsen Haddadi}}, \citenamefont {{Becattini, Francesco}},\ and\ \citenamefont
  {{Bleicher, Marcus}}}]{refId0}%
  \BibitemOpen
  \bibfield  {author} {\bibinfo {author} {\bibnamefont {{Inghirami,
  Gabriele}}}, \bibinfo {author} {\bibnamefont {{Del Zanna, Luca}}}, \bibinfo
  {author} {\bibnamefont {{Beraudo, Andrea}}}, \bibinfo {author} {\bibnamefont
  {{Moghaddam, Mohsen Haddadi}}}, \bibinfo {author} {\bibnamefont {{Becattini,
  Francesco}}}, \ and\ \bibinfo {author} {\bibnamefont {{Bleicher, Marcus}}},\
  }\href {\doibase 10.1140/epjc/s10052-016-4516-8} {\bibfield  {journal}
  {\bibinfo  {journal} {Eur. Phys. J. C}\ }\textbf {\bibinfo {volume} {76}},\
  \bibinfo {pages} {659} (\bibinfo {year} {2016})}\BibitemShut {NoStop}%
\bibitem [{\citenamefont {Filip}(2015)}]{filip}%
  \BibitemOpen
  \bibfield  {author} {\bibinfo {author} {\bibfnamefont {P.}~\bibnamefont
  {Filip}},\ }\href {\doibase 10.1088/1742-6596/636/1/012013} {\bibfield
  {journal} {\bibinfo  {journal} {Journal of Physics: Conference Series 636
  012013}\ } (\bibinfo {year} {2015}),\ 10.1088/1742-6596/636/1/012013},\
  \Eprint {http://arxiv.org/abs/1504.07008} {arXiv:1504.07008 [hep-ph]}
  \BibitemShut {NoStop}%
\bibitem [{\citenamefont {Albuquerque}(2019)}]{ALBUQUERQUE2019823}%
  \BibitemOpen
  \bibfield  {author} {\bibinfo {author} {\bibfnamefont {D.}~\bibnamefont
  {Albuquerque}},\ }\href {\doibase
  https://doi.org/10.1016/j.nuclphysa.2018.08.033} {\bibfield  {journal}
  {\bibinfo  {journal} {Nuclear Physics A}\ }\textbf {\bibinfo {volume}
  {982}},\ \bibinfo {pages} {823 } (\bibinfo {year} {2019})},\ \bibinfo {note}
  {the 27th International Conference on Ultrarelativistic Nucleus-Nucleus
  Collisions: Quark Matter 2018}\BibitemShut {NoStop}%
\bibitem [{\citenamefont {Acharya}\ \emph {et~al.}(2018)\citenamefont {Acharya}
  \emph {et~al.}}]{Acharya:2018qnp}%
  \BibitemOpen
  \bibfield  {author} {\bibinfo {author} {\bibfnamefont {S.}~\bibnamefont
  {Acharya}} \emph {et~al.} (\bibinfo {collaboration} {ALICE}),\ }\href@noop {}
  {\  (\bibinfo {year} {2018})},\ \Eprint {http://arxiv.org/abs/1805.04365}
  {arXiv:1805.04365 [nucl-ex]} \BibitemShut {NoStop}%
\bibitem [{\citenamefont {Yuan}\ and\ \citenamefont {Zhang}(1999)}]{Yuan_1999}%
  \BibitemOpen
  \bibfield  {author} {\bibinfo {author} {\bibfnamefont {Y.~F.}\ \bibnamefont
  {Yuan}}\ and\ \bibinfo {author} {\bibfnamefont {J.~L.}\ \bibnamefont
  {Zhang}},\ }\href {\doibase 10.1086/307921} {\bibfield  {journal} {\bibinfo
  {journal} {The Astrophysical Journal}\ }\textbf {\bibinfo {volume} {525}},\
  \bibinfo {pages} {950} (\bibinfo {year} {1999})}\BibitemShut {NoStop}%
\bibitem [{\citenamefont {Wei}\ \emph {et~al.}(2006)\citenamefont {Wei},
  \citenamefont {Mao}, \citenamefont {Ko}, \citenamefont {Kisslinger},
  \citenamefont {Stoecker},\ and\ \citenamefont {Greiner}}]{Wei:2005aga}%
  \BibitemOpen
  \bibfield  {author} {\bibinfo {author} {\bibfnamefont {F.~X.}\ \bibnamefont
  {Wei}}, \bibinfo {author} {\bibfnamefont {G.~J.}\ \bibnamefont {Mao}},
  \bibinfo {author} {\bibfnamefont {C.~M.}\ \bibnamefont {Ko}}, \bibinfo
  {author} {\bibfnamefont {L.~S.}\ \bibnamefont {Kisslinger}}, \bibinfo
  {author} {\bibfnamefont {H.}~\bibnamefont {Stoecker}}, \ and\ \bibinfo
  {author} {\bibfnamefont {W.}~\bibnamefont {Greiner}},\ }\href {\doibase
  10.1088/0954-3899/32/1/005} {\bibfield  {journal} {\bibinfo  {journal} {J.
  Phys.}\ }\textbf {\bibinfo {volume} {G32}},\ \bibinfo {pages} {47} (\bibinfo
  {year} {2006})},\ \Eprint {http://arxiv.org/abs/nucl-th/0508065}
  {arXiv:nucl-th/0508065 [nucl-th]} \BibitemShut {NoStop}%
\bibitem [{\citenamefont {Peskin}\ and\ \citenamefont
  {Schroeder}(1995)}]{Peskin:1995ev}%
  \BibitemOpen
  \bibfield  {author} {\bibinfo {author} {\bibfnamefont {M.~E.}\ \bibnamefont
  {Peskin}}\ and\ \bibinfo {author} {\bibfnamefont {D.~V.}\ \bibnamefont
  {Schroeder}},\ }\href {http://www.slac.stanford.edu/~mpeskin/QFT.html} {\emph
  {\bibinfo {title} {{An Introduction to quantum field theory}}}}\ (\bibinfo
  {publisher} {Addison-Wesley},\ \bibinfo {address} {Reading, USA},\ \bibinfo
  {year} {1995})\BibitemShut {NoStop}%
\end{thebibliography}%

\end{document}